
\hsize=5.104truein
\hoffset=.25in
\vsize=8.5416truein
\voffset=-.25in
\font\sevenrm=cmr7
\font\eightrm=cmr8
\font\cs=cmcsc10
\def\ov{\overline}
\nopagenumbers
\input epsf

{}~~~~~
\vskip.417truein

\noindent {\bf PHYSICS OF THE QUARK-GLUON PLASMA}
\vskip.417truein

\hskip30pt Berndt M\"uller
\bigskip

\hskip30pt Department of Physics

\hskip30pt Duke University

\hskip30pt Durham, NC 27708-0305
\vskip.417truein

\centerline{\bf Abstract}
\bigskip

{\narrower
Central nuclear collisions at energies far above 1
GeV/nucleon may provide for conditions, where the transition from
highly excited hadronic matter into quark matter or quark-gluon plasma
can be probed.  Here I review our current understanding of the
physical properties of a quark-gluon plasma and review ideas about the
nature of, and signals for, the deconfinement transition.  I also give
a detailed presentation of recent progress in the treatment of the
formation of a thermalized state at the parton level.
\bigskip\medskip}

\noindent {\bf INTRODUCTION}
\bigskip\medskip

\noindent {\bf Overview}
\bigskip

After a collision between a $^{32}$S nucleus of 6.4 TeV total energy and a
heavy target nucleus several hundred charged particles are emitted.
No one in his or her right mind would care to study such events,
unless there existed a compelling reason for doing so.  The current
interest in nuclear collisions at very high energies (far above 1
GeV/u in the c.m. system) is fueled by the expectation that a {\it
quark gluon plasma} may be created temporarily in these events$^{1-3}$.
Whereas there is general consensus among theorists that QCD at
thermodynamic equilibrium exhibits a phase transition from the normal
color-confined phase of hadronic matter with broken chiral symmetry to
a deconfined, chirally symmetric phase at sufficiently high energy
density, many aspects of this transition are still a matter of
debate.  Such ``details'' are, e.g., the order of the phase
transition, the precise value of the critical energy density, the
nature of experimentally observable signatures of the transition, and
how fast thermal equilibrium conditions are established in nuclear
collisions over a sufficiently large space-time volume.  These
questions require much further theoretical and experimental study.
Here we are concerned with an up-to-date survey of (mostly) theoretical
aspects.  More detailed discussions of selected theoretical topics, as
well as reviews of experimental results, can be found in other
lectures presented at this school.
\vfill\eject

\noindent {\bf The ``Cosmic'' Connection}
\bigskip

Perhaps the most compelling reason why we should attempt to study the
quark-gluon plasma transition in laboratory experiments is that it
must have occurred in the early universe$^4$.  The history of the
thermal evolution of our universe is depicted in Fig. 1.  The relation
between temperature $T$ and cosmic time $t$ is approximately given
by$^5$:

$$T_{\hbox{\sevenrm MeV}} \simeq (5.75 N_f(T))^{{1\over 4}}
t_{\hbox{\sevenrm sec}}^{-1/2},$$
where $N_f(T)$ describes the number of particle degrees of freedom
that act as effectively massless modes at a given temperature, and the
subscripts indicate the units in which $T$ and $t$ are measured.  From
the present temperature of the cosmic background radiation (2.7 K) we
extrapolate back to a temperature of about $2\times 10^{12}$K $\simeq$
200 MeV at about 20$\mu s$ after the ``big bang''.  This is the
temperature above which, as we will discuss in detail below, hadrons
dissolve and their constituent quanta, quarks and gluons, are
liberated.
\bigskip

\centerline{\epsfbox{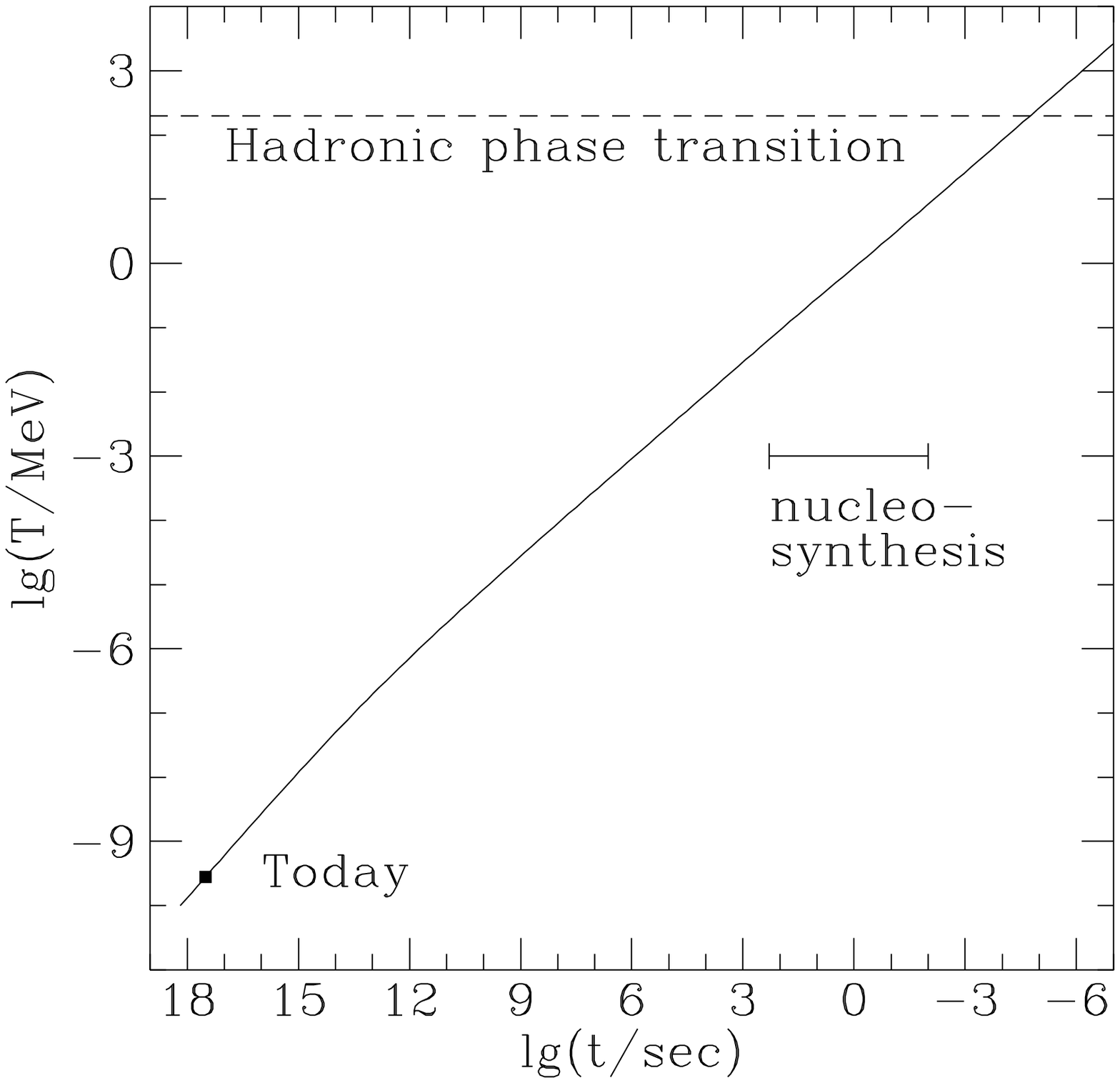}}
\centerline{\eightrm Figure 1. Thermal history of the Universe.}
\bigskip

Tracing the history of our universe backward, this is only the first
phase transition involving fundamental quantum fields that we
encounter.  Most likely, more transformations of a similar nature have
occurred at even earlier times.  If our current ideas of the origin of
electroweak symmetry breaking are correct, a phase transition in the
Higgs vacuum took place at $t\approx 10^{-11}$s, when the temperature
was around 250 GeV.  A similar phase transition at much earlier times
$(t\approx 10^{-35}$s) associated with the ``grand'' unification of
electroweak and strong interactions may have led to exponential
inflation of cosmic scales, due to the anti-gravitational pressure
exerted by an unstable vacuum state$^{6,7}$.

Why should we care about events occurring at such unimaginably short
moments after the creation of our universe?  The reason is that some
of the unsolved problems of cosmology are, most likely, associated
with events in the era before about $10^{-5}$s:

\item{$\bullet$} The observed {\it baryon number asymmetry} in the
universe is probably due to baryogenesis by topologically nontrivial
field configurations during the electroweak phase transition.

\item{$\bullet$} The nature of {\it dark matter} may be associated
with properties of the Higgs or the QCD vacuum.

\item{$\bullet$} The near {\it isotropy of the background radiation},
and the deviations from homogeneity underlying the observed {\it
large-scale structure} of the universe were, according to current
thinking, determined during the phase transition causing inflation.
\medskip

Understanding of the dynamical nature of phase transitions in
elementary quantum field theories is thus essential to further
progress toward the solution of these puzzles.  Since the QCD phase
transition is the only one of these that appears accessible to
laboratory experiments, we must study it with high priority.
\bigskip\medskip

\noindent {\bf Quark Stars}
\bigskip

In addition to events in the very early universe, quark matter may
also play a role in the internal structure of collapsed stars.  At the
high densities reached in the core of neutron stars nucleons may well
dissolve into their constituents, forming baryon-rich cold quark
matter.  This would not lead to greater stability of neutron stars,
quite to the contrary:  since quark matter would allow for a higher
central density of the star at a given total star mass, its formation
would actually facilitate collapse to a black hole.  Neutron stars with
a quark core have a lower value of their upper mass limit, probably
somewhere between 1.5 and 2 solar masses$^8$.  Stars with a quark core
would be more compact and hence could sustain higher rotation
rates$^9$.  This observation would be of practical interest, if
pulsars with periods in the sub-millisecond range are eventually
detected.

The problem with quantitative predictions here is that dense
baryon-rich nuclear matter is expected to contain a large strangeness
fraction, because the inclusion of strange quarks can lower the Fermi
energy.  This holds true for baryonic matter as well as quark matter.
Unfortunately the equation of state of baryonic matter containing
hyperons is poorly known.  The scalar coupling strength of
$\Lambda$-hyperons, which is not well determined experimentally has a
large influence on the central density of hyperon-rich neutron stars.
This uncertainty can easily mask the formation of a quark matter
core$^8$.  Better understanding of the equation of state of hyperon
interactions in dense baryon-rich matter, which can only come from
high-energy nuclear collisions, is therefore essential for further
progress towards the understanding of neutron star structure.
\bigskip\medskip

\noindent {\bf THE EQUATION OF STATE OF HADRONIC MATTER}
\bigskip

There are two approaches that have been widely followed to the problem
of the equation of state of strongly interacting matter:

\item{(a)} Consider color-singlet hadrons, i.e. baryons and mesons,
only and see how far one gets with taking into account their known
interactions and excitation spectra.  This approach was pioneered by
R. Hagedorn, and later studied by Hagedorn and Rafelski$^{10}$,
Walecka$^{11}$, Gasser and Leutwyler$^{12}$, and many others.

\item{(b)} Consider the fundamental constituents of hadrons, i.e.
quarks and gluons, and study the equation of state on the basis of
quantum chromodynamics.  This approach, anticipated in pre-QCD days
by P. Carruthers$^{13}$, was pioneered by Collins and Perry$^{14}$,
Baym and Chin$^{15}$, McLerran et al.$^{16}$ and by Shuryak$^{17}$.
\medskip

Here we will start by examining the approach based on hadrons and
their interactions, because its results are important in their own
right.  Any quark-gluon plasma formed in a nuclear reaction will
eventually hadronize and evolve through a hadron-dominated break-up
phase.  In addition, potential signatures for quark-gluon plasma
formation must always be compared with predictions of a scenario that
makes no reference to a phase transition from color-singlet hadrons to
deconfined quarks and gluons.
\bigskip\medskip

\noindent {\bf Hot and Dense Hadronic Matter}
\bigskip

How far do we get in explaining the equation of state of hadronic
matter by considering hadrons alone?  Experimental observations of
baryon and meson resonances indicate that the mass density of hadronic
states grows exponentially,
$$\rho (m) \simeq m^a \exp(m/m_0) \eqno(1)$$
as originally postulated by Hagedorn$^{18}$.  It is now understood
that an exponential mass spectrum is the natural consequence of quark
confinement, and it is found in even the simplest hadron models that
incorporate the confinement concept, such as the string model$^{19}$
or the MIT-bag model$^{20}$.  A numerical simulation of the spectrum
of excited string modes agrees very nicely with the observed hadron
spectrum even at low energies$^{21}$.

Now consider highly excited hadronic matter, characterized by a
temperature $T$.  The energy density of excited states is then given
by (2.1) integrated over momentum space, i.e.
$$\eqalignno{ n(E) &\simeq \exp (-E/T) \int_0^E dm\; \rho(m)\; E
\sqrt{E^2-m^2} \cr
&\simeq E^{a+2} \exp \left[ -E \left({1\over T} - {1\over
m_0}\right)\right]. &(2)\cr}$$
Obviously, $n(E)$ is only integrable as long as the factor in the
exponent remains positive, i.e. for $T<T_c\equiv m_0$.  From
experimental data we know that $m_0\approx$ 200 MeV, i.e. the
temperature of a (noninteracting) hadronic resonance gas cannot exceed
the Hagedorn temperature $T_c\sim 200$ MeV.  Cabibbo and Parisi
pointed out in 1975 that the distribution (2) remains integrable at
$T=T_c$ when $a<-4$, and then corresponds to a finite energy density
$\varepsilon = \int EdE\; n(E)$ at the critical point$^{22}$.  This is
reminiscent of the singular behavior of thermodynamical quantities at
a first-order phase transition, and they speculated that
it signals the transition to a color-deconfined quark-gluon plasma.

Although Hagedorn has stressed that the inclusion of hadronic
resonances effectively accounts for a large part of the interactions
among hadrons, not all interactions are taken into account in this
simple manner, especially if one neglects the resonance widths.  When
Hagedorn and Rafelski$^{23}$ studied the effect of the finite internal
size of hadrons by utilizing the finite excluded volume approximation,
they found that the equation of state developed a singularity at
finite energy density at a slightly lower critical temperature $T_c
< m_0$, independent of other details of the hadronic mass spectrum
(2.1).  We conclude from all this that hadronic matter as composed of
individual color-singlet hadrons ceases to exist at temperatures
exceeding $T_c \simeq$ 150-200 MeV.
\vfill\eject

\noindent {\bf Medium Modifications of Hadron Structure}
\bigskip

It is useful to sharpen the question concerning the influence of
interactions on the hadronic equation of state and ask whether hadron
masses themselves depend on temperature and density.  As this interesting
problem has been approached from several different angles, it is quite
enlightening to discuss a few of these.
\medskip

\noindent (a) {\it Sigma-condensate models:}
\medskip

\noindent In these models the nucleon mass $M$ is generated by the
coupling to a scalar field $\sigma$:
$${\cal L}_N = \ov{\psi}i\gamma\cdot \partial\psi -
g\ov{\psi}\psi\sigma, \eqno(3)$$
which develops a nonvanishing vacuum expectation value
$$\sigma_0 \equiv \langle 0\vert\sigma\vert 0\rangle = M/g.
\eqno(4)$$
However, since the $\sigma$-field interacts with hadrons, via an
equation of the form
$$(\partial^2 + \mu_{\sigma}^2)(\sigma - \sigma_0) =
-g\langle\ov{\psi}\psi\rangle, \eqno(5)$$
its expectation value is shifted in dense nuclear matter according to
the equation
$$M^* = g\sigma = M-{g^2\over \mu_{\sigma}^2}
\langle\ov{\psi}\psi\rangle. \eqno(6)$$
Here
$$\langle\ov{\psi}\psi\rangle = 4 \int {d^3k\over (2\pi)^3} \;
{M^*\over E(k)} (n(k) + \ov{n}(k)) \eqno(7)$$
is the scalar baryon density, $E(k)^2 = k^2 + M^{*2}$, and $n(k),
\ov{n}(k)$ denote the momentum space density of baryons and
antibaryons, respectively.  The numerical evaluation of eq. (2.6)
shows that $M^*$ drops rather suddenly to a very small value at
$T\simeq$ 200 MeV at zero net baryon density $\rho^{11}$.  The decrease
with $\rho$ at $T=0$ is more gradual, already corresponding to
$M^*/M \simeq 0.6$ at nuclear matter saturation density $\rho_0$.
$M^*$ becomes very small at $\rho/\rho_0 \sim 3-4$.
\bigskip

\centerline{\epsfbox{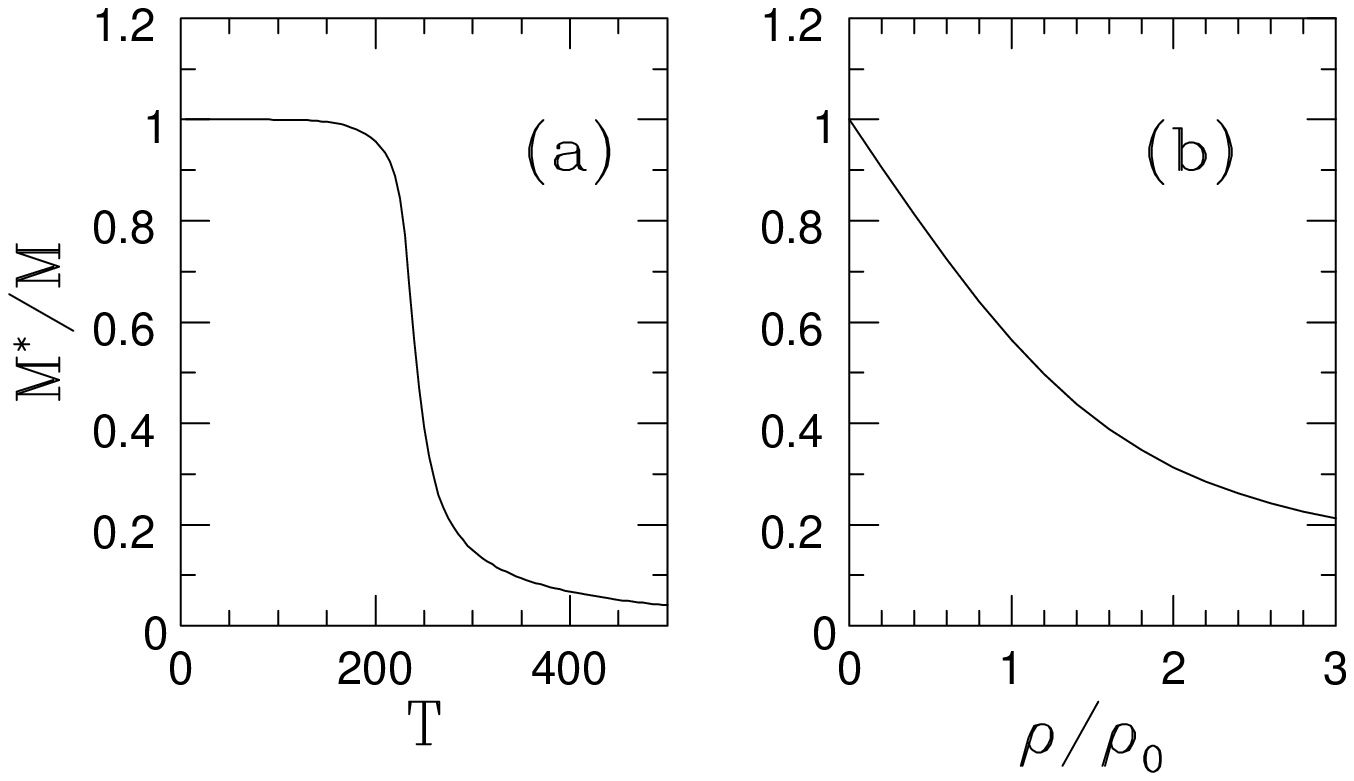}}
\noindent {\eightrm Figure 2. Effective mass of the nucleon in the
$\scriptstyle{\sigma-\omega}$ model:  (a) as function temperature
$\scriptstyle{T}$ at $\scriptstyle{\rho = 0}$, (b) as function of
the baryon density $\scriptstyle{\rho}$ at $\scriptstyle{T}$ = 0.}
\bigskip

\noindent (b) {\it Dispersion relations:}
\medskip

A second method of approach makes use of available data on hadron-hadron
scattering via dispersion relations.  The propagator of a hadron, say
a pion or kaon, in medium is modified due to interactions with other
particles.  In many cases this interaction is dominated by resonance
scattering, schematically illustrated in Figure 3.  This is the case,
e.g., for pions which interact with other pions from the medium to
form a short-lived rho-meson, and with kaons that form a $K^*$
resonance.  This also occurs with $K^-$ mesons interacting with
nucleons via the $\Sigma^*$ resonance at 1385 MeV, but {\it not} with
$K^+$ mesons which have a dominantly nonresonant interaction with
nucleons.
\bigskip

\centerline{\epsfbox{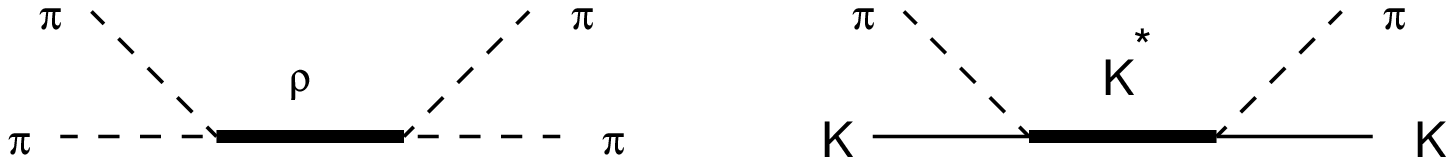}}
\noindent {\eightrm Figure 3. Resonance scattering on pions from the medium
leads to a medium-dependent mass shift of (a) pions, (b) kaons. }
\bigskip

The in-medium propagator of a (pseudo-) scalar meson has the form
$$D(k) = (k^2-m_i^2-\Pi_i(k))^{-1} \equiv (k^2-m_i^*(k)^2)^{-1}
\eqno(8)$$
where $i$ denotes the meson species and the polarization function
$\Pi_i(k)$ can be expressed as the sum of the contributions from
interactions with all other hadrons contained in the medium$^{24}$:
$$\Pi_i(k) = \sum_j \int {d^3p\over (2\pi)^3 E_p}{8\pi\over \sqrt{s}}
f_{ij}(0,s) n_j(p). \eqno(9)$$
Here $f_{ij}(0,s)$ is the forward scattering amplitude at energy
$s=(p+k)^2$ and $n_j(p)$ denotes the momentum space density of hadrons
of species $j$ in the medium.  Numerical evaluation of expression
(8) for pions and kaons reveals rather small mass shifts $\Delta m_i =
m_i^*(\vec k=0) - m_i$ of order 10-30 MeV in baryon-symmetric hadronic
matter up to temperatures around 150 MeV, beyond which the
approximations entering into (9) become unreliable.

However, it turns out that the effect on kaons in {\it baryon-rich}
matter can be substantial.  As shown in Figure 4, the mass shift of kaons
at $T=160$ MeV and baryon density $\rho = 4\rho_0$ is of the order of
100 MeV, being positive for $K^+$ and negative for $K^-$.  The reason
for the opposite behavior is that the interaction of $K^-$-mesons with
nucleons is resonance-dominated and attractive, whereas the
interaction of $K^+$-mesons is nonresonant and repulsive.  An
immediate consequence of the different mass shift is that the relative
abundance of charged kaons in dense baryon-rich matter is
modified$^{25}$.  In fact, the mass shift acts counter to the effect
of the nonvanishing baryochemical potential and limits the ratio of
abundances $K^+/K^-$ to about 5 at $T\simeq$ 150 MeV even for extreme
baryon density.  Notably this is just the ratio observed in heavy ion
collisions at the AGS$^{26}$.  Is this a coincidence?   Although kaons
in these reactions are probably emitted long after the moment of
highest density in the fireball, the $K^+/K^-$ ratio may be determined
by the reactions occurring at or before the moment of highest
compression.  The abundances probably get out of equilibrium and
remain frozen as the fireball expands and breaks up rapidly.
\bigskip

\centerline{\epsfbox{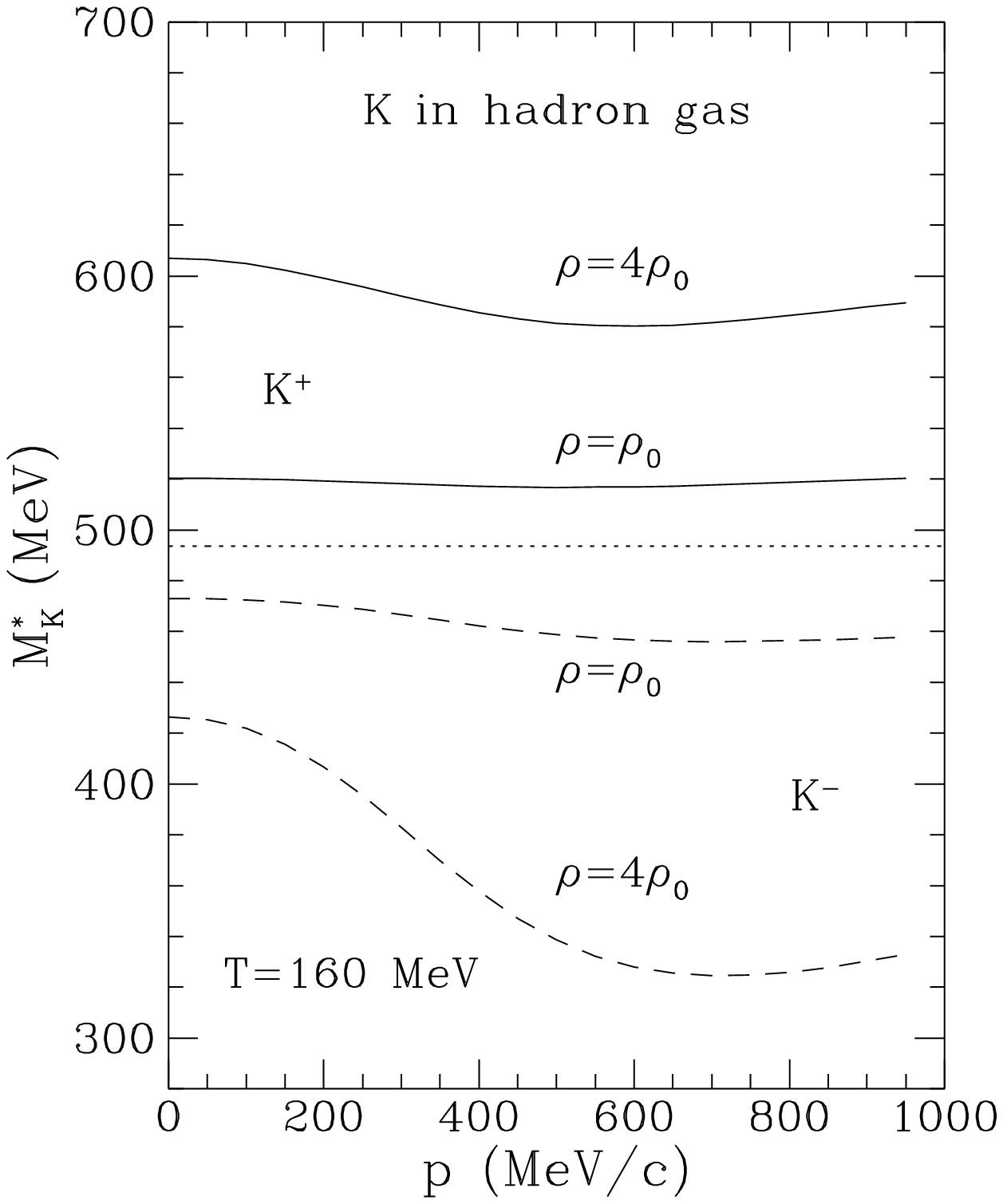}}
\noindent {\eightrm Figure 4. Mass shift of kaons in baryon-rich
hadronic matter.  The figure shows the medium-dependent change in the
energy of positive kaons (solid lines) and negative kaons (dashed
lines) as function of momentum for $\scriptstyle{T}$ = 160 MeV
and twodifferent baryon densities.}
\bigskip

\noindent (c) {\it QCD sum rules:}
\medskip

A third avenue of approach to medium modifications of hadronic
properties touches base with the underlying QCD dynamics by identifying
the sigma condensate $\sigma_0$ with the scalar quark-antiquark
condensate $\langle \ov{q} q\rangle$ in QCD.  This permits one to
advance the argument$^{27}$ that {\it all} hadron masses should be
modified in the {\it same} way, according to
$${m_i^*\over m_i} = {\langle \hbox{\sevenrm med}\vert \ov{q}q\vert
\hbox{\sevenrm med}\rangle \over \langle 0\vert\ov{q}q\vert 0\rangle}.
\eqno(10)$$
A qualitative estimate of the change in hadron masses can then be
obtained from any estimate of the scalar quark density in the medium,
as compared with the vacuum condensate$^{28}$
$$\langle 0\vert\ov{q}q\vert 0\rangle = -(225 \pm 25 \hbox{MeV})^3.
\eqno(11)$$
A simple way to estimate the medium contribution to
$\langle\ov{q}q\rangle$ is to calculate the scalar quark density
contained in the thermal pions
$$\delta\langle\ov{q}q\rangle_{\pi} = 3 \int {d^3k\over (2\pi)^3}
\langle\pi (k) \vert\ov{q}q\vert \pi(k)\rangle n_{\pi}(k),
\eqno(12a)$$
or, for baryon-rich matter, in the nucleons
$$\delta\langle\ov{q}q\rangle_N \simeq 4 \int {d^3k\over (2\pi)^3}
\langle N(k) \vert\ov{q}q\vert N(k)\rangle n_N(k). \eqno(12b)$$
Eq. (12a) may be conveniently evaluated in the framework of chiral
perturbation theory$^{12}$.  Results are shown in Figure~5a.  More
model-independent estimates for (12a,b) can be obtained by the
so-called QCD sum-rule approach$^{29-31}$.  Figure 5b shows the
predictions of this technique for the medium modifications of the
condensates of light quarks,  strange quarks and gluons in the QCD
vacuum as function of baryon density (at $T=0$).  Obviously, the
condensate of light quarks is most strongly affected in both
approaches, indicating a dramatic change in hadron structure at
$T\approx 200$ MeV or $\rho\simeq 4\rho_0$.
\bigskip

\centerline{\epsfbox{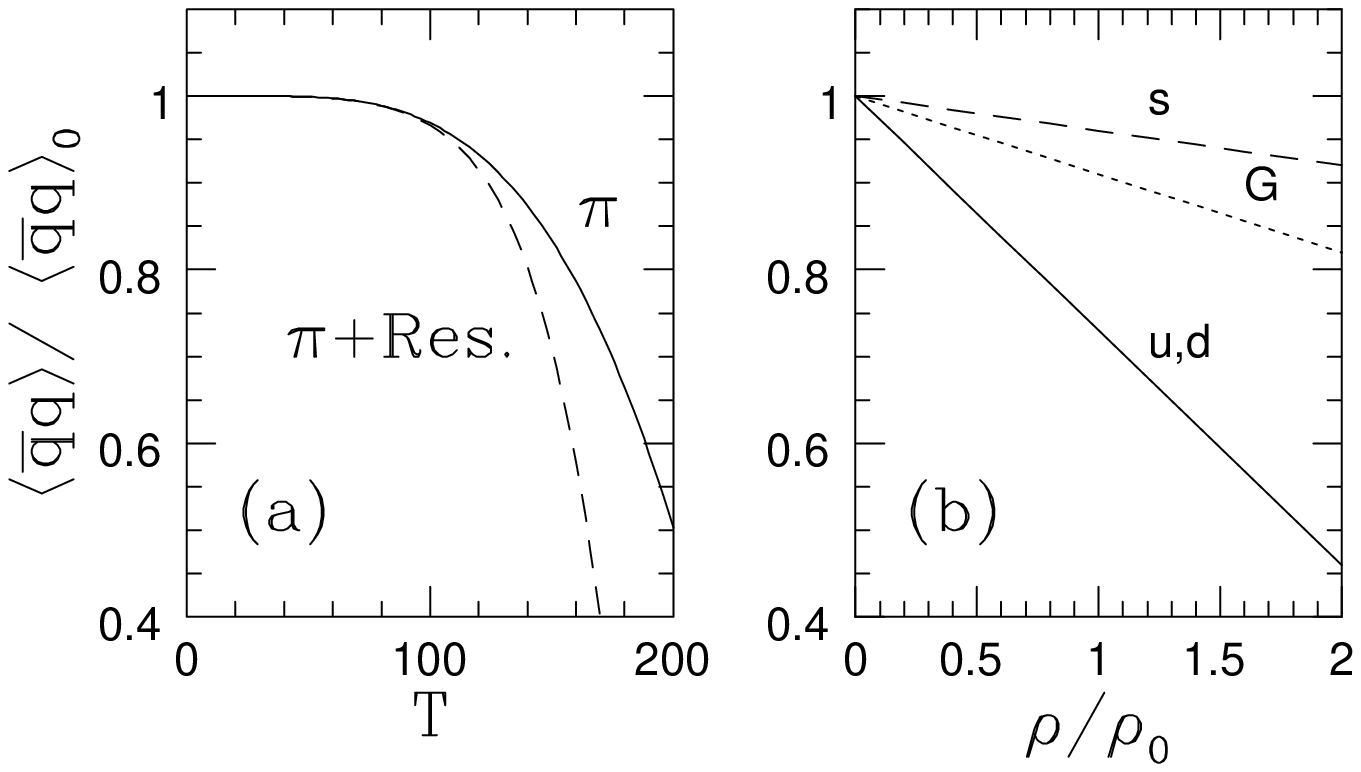}}
\noindent {\eightrm Figure 5. Medium modification of QCD condensates.
(a) Change of the  quark condensate in a pion gas as function of
temperature$^{12}$. (b) Dependence of the condensates of light quarks,
gluons and strange quarks as functions of the baryon density$^{31}$.}
\bigskip

Overall, the agreement between the predictions of the different
approaches agree remarkably well, reinforcing faith in their
accuracy.  In general, one finds that hadron structure is much less
disturbed by a baryon symmetric medium of finite temperature than by a
medium with finite net baryon density.  The right place to look for
medium effects on hadron structure is therefore in relativistic
nuclear collisions involving a large degree of baryon stopping, such
as the Brookhaven AGS --- especially with the ${}^{197}$Au beam --- or
possibly the CERN-SPS in Pb+Pb collisions.  In the case of matter
depleted of baryons the medium effects set in so late that it will
be very difficult to disentangle them from the quark-gluon plasma
phase transition.
\bigskip\medskip

\noindent {\bf QUARK-GLUON PLASMA }
\bigskip

The theory of the equation of state of quark matter is conceptually
much simpler, because it is directly based on the fundamental QCD
Lagrangian
$${\cal L}_{\hbox{\sevenrm QCD}} = -{\textstyle{1\over 4}} \sum_a
F_{\mu\nu}^a F^{a\mu\nu} + \sum_{f=1}^{N_f} \ov{\psi}
(i\gamma^{\mu}\partial_{\mu} - g\gamma^{\mu}A_{\mu}^a
\textstyle{{\lambda^a\over 2}}-m_f)\psi \eqno(13)$$
where the subscript $f$ denotes the various quark flavors $u,d,s,c$,
etc., and the nonlinear glue field strength is given by
$$F_{\mu\nu}^a = \partial_{\mu} A_{\nu}^a - \partial_{\nu}A_{\mu}^a +
gf_{abc} A_{\mu}^b A_{\nu}^c. \eqno(14)$$
QCD predicts a weakening of the quark-quark interaction at short
distances (or high momenta $Q^2$), because the one-loop series for
the gluon propagator yields a running coupling constant
$$g^2(Q^2) = {16\pi^2\over (11-{2\over 3}N_f)\ln (Q^2/\Lambda^2)}
\;\buildrel{Q^2\to\infty}\over\longrightarrow \; 0. \eqno(15)$$
The QCD scale parameter is now quite well determined$^{32}$ to be
$\Lambda \simeq$ 150 MeV.  The vanishing of the QCD coupling
constant at short distances, called ``asymptotic freedom'', has often
been taken to imply that interactions among quarks and gluons are
negligible in the limit of high temperature or high density.  As we
shall see below this is not the whole truth, because long-wavelength modes
continue to have an important influence on the properties of the
quark-gluon plasma.
\bigskip\medskip

\noindent{\bf Equation of State of Quark Matter}
\bigskip

So let us first suppose that interactions among quarks and gluons are
negligible at high energy density and see what we get.  At temperature
$T$ and quark chemical potential $\mu$ (one-third of the baryochemical
potential $\mu_{\hbox{\sevenrm B}} = 3\mu$), the energy density of
free gluons, quarks and antiquarks is (a detailed derivation can be
found in ref. 1):
$$\eqalignno{\varepsilon_g &= 16 {\pi^2\over 30} T^4. &(16a)\cr
\varepsilon_q + \varepsilon_{\ov{q}} &= 6N_f\left( {7\pi^2\over 120}
T^4 + {1\over 4}\mu^2T^2 + {1\over 8\pi^2} \mu^4\right). &(16b)\cr}$$
Since we have neglected the quark mass, we inserted a factor $N_f$
counting the number of quark flavors active at a given temperature
(essentially those with $m_f \le T$).  Clearly $N_f\ge 2$ at all
relevant temperatures, and $N_f=3$ in the range 200 MeV $\ll T<$ 1 GeV
where strange quarks can be considered as light, as well.  The other
thermodynamical quantities of interest, i.e. pressure $P$, entropy
density $s$, and baryon number density $\rho_{\hbox{\sevenrm B}}$, are
easily obtained from eqs. (16):
$$P = {1\over 3}\varepsilon, \quad s = \left( {\partial P\over \partial
T}\right)_{\mu}, \quad \rho_{\hbox{\sevenrm B}} = {1\over 3} \left(
{\partial P\over \partial \mu}\right)_T. \eqno(17)$$
Note that the simple relation between $\varepsilon$ and $P$ holds only
as long as all particles are considered massless.  The general relation
is
$$\varepsilon = T\left( {\partial P\over \partial T}\right)_{\mu} +
\mu \left( {\partial P\over \partial \mu}\right)_T - P \equiv Ts +
\mu_{\hbox{\sevenrm B}}\rho_{\hbox{\sevenrm B}}-P. \eqno(18)$$
In order to find the location of the phase transition toward hadronic
matter we have to incorporate the breaking of scale invariance
provided by QCD interactions.  The simplest way of achieving this is
by invoking the argument associated with the MIT-bag model:  Free
quarks and gluons can only propagate where the complex structure
of the real QCD vacuum has been destroyed.  The vacuum realignment
costs a certain amount of energy per unit volume, expressed by the
MIT-bag constant $\varepsilon_0= B \simeq$ (150-200 MeV)$^4$, and
the ``wrong'' vacuum is endowed with a negative pressure \break $P_0=-B$.
The relation $\varepsilon_0 = -P_0$ is mandated by the Lorentz
invariance of the vacuum state.  The negative sign is easily
understood as signal of the instability of the wrong vacuum state
which collapses if not supported by the pressure provided by free
partons propagating in the volume filled with it.  The equation of
state of the quark-gluon plasma then takes the simple form
$$\eqalignno{\varepsilon &= \varepsilon_g(T,\mu) +
\varepsilon_q(T,\mu) + \varepsilon_{\ov{q}}(T,\mu) + B; &(19a) \cr
P &= P_g(T,\mu) + P_q(T,\mu) + P_{\ov{q}}(T,\mu) - B. &(19b) \cr}$$
A lower limit of stability of the plasma state is obtained by setting
$P=0$.  This yields the stability line in the $T$-$\mu$ plane shown in
Figure 6.  Of course, the plasma phase becomes unstable against
formation of a gas of color-singlet hadrons even earlier, when its pressure
equals that of a hadron gas at the same temperature $T$ and chemical
potential $\mu$.
\bigskip

\centerline{\epsfbox{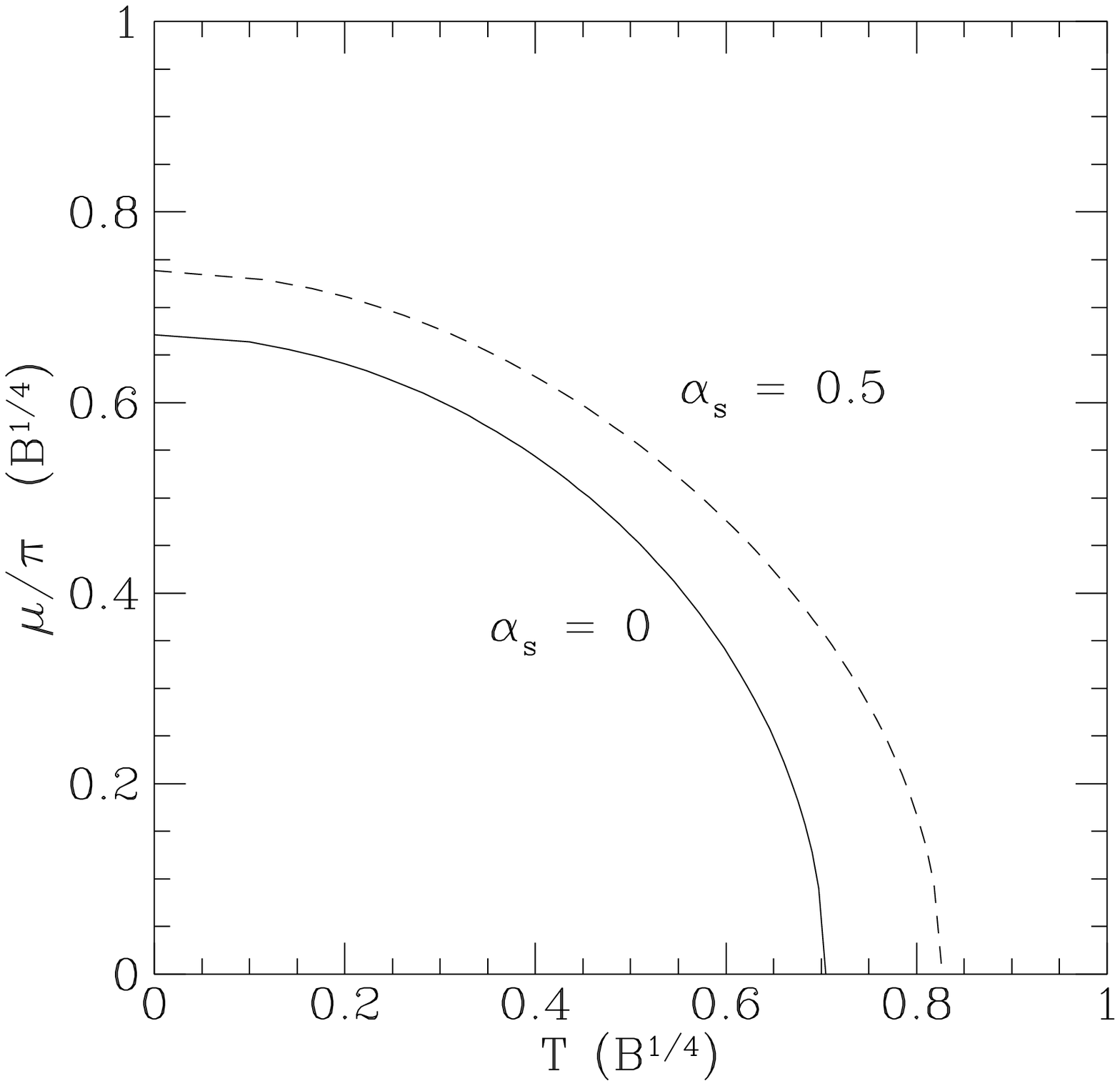}}
\noindent {\eightrm Figure 6. Stability line of the quark-gluon plasma
in the $\scriptstyle{T}$-$\scriptstyle{\mu}$ plane for two values of the
strong coupling constant.  The lines indicate where the pressure of a
gas of quarks and gluons vanishes. The quark-gluon plasma is unstable
in the lower left region of the figure.}
\bigskip

In order to explore this aspect further, let us restrict ourselves to the
baryon-free case $\mu=0$, and explore the coexistence between the
quark-gluon plasma phase given by eq. (19a) and hadronic matter,
represented by a gas of massless, noninteracting pions:
$$\varepsilon_{\pi} = 3{\pi^2\over 30}\; T^4, \quad P_{\pi} = {1\over
3} \varepsilon_{\pi}. \eqno(20)$$
$\varepsilon, P, \varepsilon_{\pi}$ and $P_{\pi}$ are shown as
functions of $T$ in Figure 7.
\bigskip

\centerline{\epsfbox{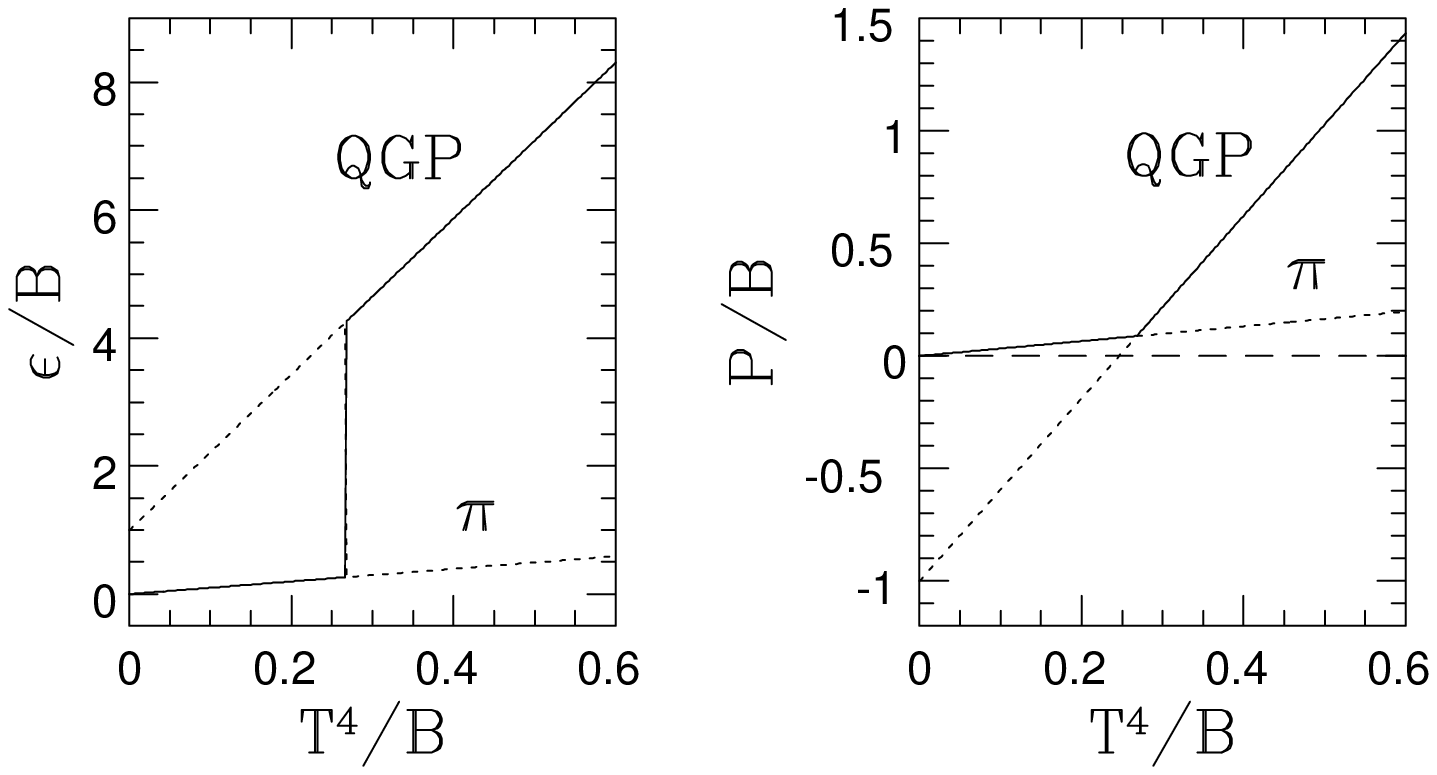}}
\noindent {\eightrm Figure 7. Energy density and pressure of a pion gas
and a free quark-gluon gas as function of temperature.  A first-order
phase transition occurs where the pressure of the two phases is
equal.  The solid line indicates the stable phase.}
\bigskip

Thermodynamical phase stability requires that the phase with the
larger pressure dominates, and phase equilibrium is achieved when
$P(T_c) = P_{\pi}(T_c)$.  As Figure 7b shows, one finds $T_c\simeq 0.75
B^{{1\over 4}}\simeq$ 150 MeV with $B\simeq$ (200 MeV)$^4$) in this
model.  Due to the vacuum rearrangement energy $B$, the energy density
between the two phases differs greatly at this point, by the amount
$\Delta\varepsilon_c \simeq 4B \simeq 0.8$ GeV/fm$^3$.  This simple
model obviously predicts a first-order phase transition between the
pion gas and quark-gluon plasma with a large latent heat
$\Delta\varepsilon_c$.

Of course, our model is grossly oversimplified because, as we saw
earlier, other hadron masses begin to decrease substantially around
$T\simeq$ 150 MeV, leading to an increase in the energy density of the
hadronic phase.  In parallel, interaction between quarks and gluons
cause a reduction in the energy density and pressure of the plasma phase.
In first-order perturbation theory, the modification of the plasma
equation of state is (for $N_f$ = 2):
$$\eqalignno{\varepsilon &= \left( 1 - {15\over 4\pi}\alpha_s\right) \;
{8\pi^2\over 15}\; T^4 + \left( 1 - {50\over 21\pi} \alpha_s\right)\;
{7\pi^2\over 10}\; T^4 +\cr
&\quad + \left( 1 - {2\over\pi}\;\alpha_s\right) \; {3\over \pi^2}\;
\mu^2 \left( \pi^2T^2 + {1\over 2}\mu^2\right) + B &(21) \cr}$$
yielding a reduction by about a factor 2 when  $\alpha_s$ = 0.5.

More reliable predictions concerning this phase transition can at
present only be obtained by numerical simulations of the QCD equation
of state on a discretized volume of space-time, usually referred to as
{\it lattice gauge theory}.  In this approach$^{33}$ one approximately
calculates the partition function for a discretized version of the QCD
Lagrangian (13) by Monte-Carlo methods.  In principle, this
technique should accurately describe the quark-gluon plasma as well as
the hadronic phase but, in practice, its accuracy especially at low
temperature is severely limited by finite size effects and other
technical difficulties.  Where the numerical results are most
reliable, i.e. for the pure gluon theory without dynamical quarks, the
calculations predict a sudden jump in the energy density at a certain
temperature while the pressure rises more gradually, as shown in
Figure 8.
\bigskip

\centerline{\epsfbox{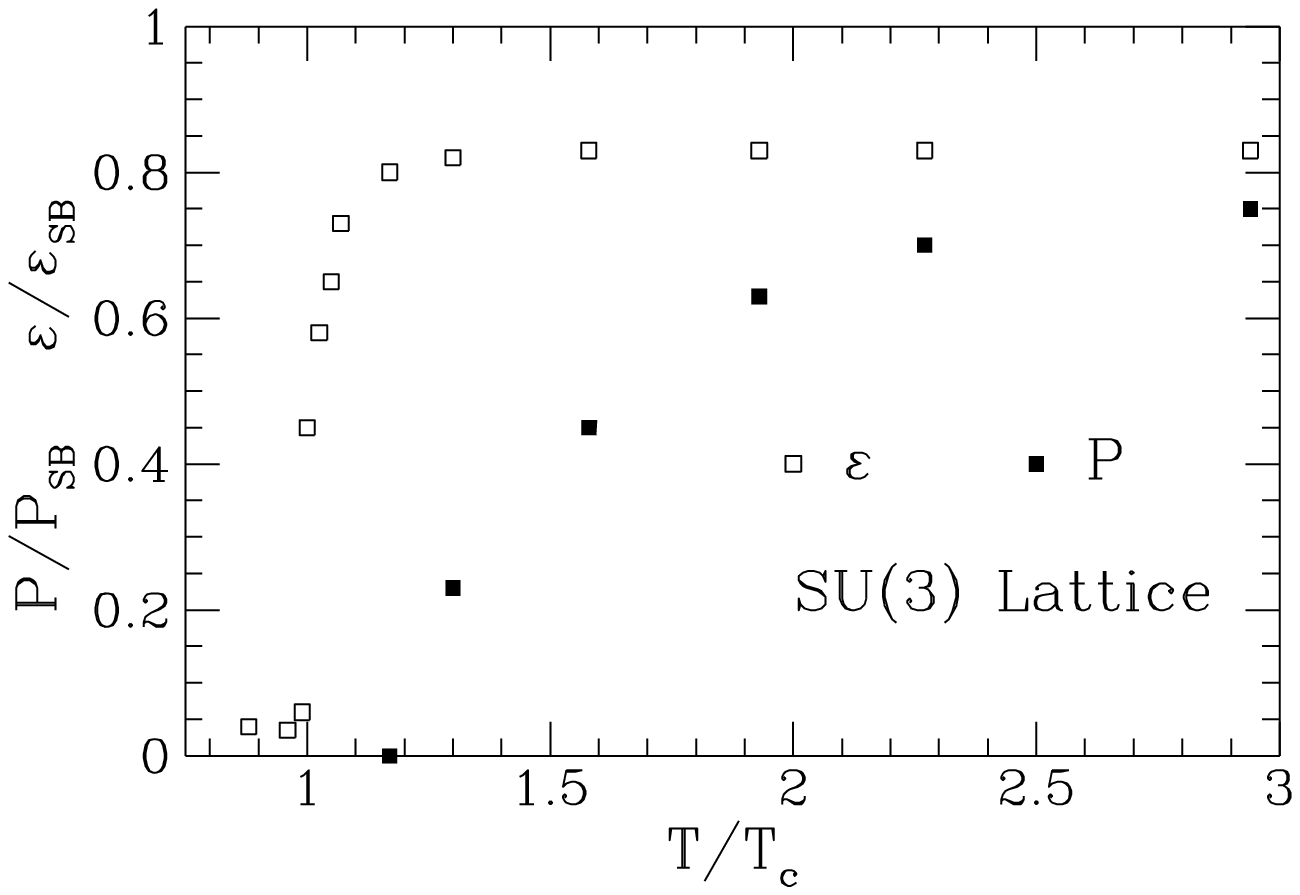}}
\noindent {\eightrm Figure 8. Energy density $\scriptstyle{\varepsilon}$
and pressure $\scriptstyle{P}$ of pure glue matter as calculated by
simulations of SU(3) lattice gauge theory$^{56}$.
$\scriptstyle{\varepsilon}$ and $\scriptstyle{P}$ are
plotted relative to the Stefan-Boltzmann limit, eq. (16a).
\bigskip}

When dynamical quarks are added, the picture becomes less clear for
two reasons.  One is that the calculations involving fermion fields on
the lattice are much more time consuming, and hence the numerical
results are less statistically meaningful and reliable.  Moreover, the
definition of quark confinement becomes rather fuzzy in the presence
of light quarks, because the color flux tube between two heavy quarks
can break by creation of a light quark pair:  $Q\ov{Q}\to
(Q\ov{q})(q\ov{Q})$.  E.g., highly excited states of charmonium can
break up into a pair of D-mesons.  Thus, in the calculations the
$Q\ov{Q}$ potential does not rise linearly with distance, but is
effectively screened.

For massless dynamical quarks there exists a new order parameter, the
quark-antiquark condensate $\langle 0 \vert\ov{q}q\vert 0\rangle$ in the
vacuum.  When it assumes a nonzero value, chiral symmetry is
spontaneously broken, as can be seen as follows:  The scalar quark
density has the chiral decomposition $\ov{q}q =
\ov{q}_Lq_R+\ov{q}_Rq_L$, hence the broken vacuum state contains pairs
of quarks of opposite chirality.  A left-handed quark, say, can
therefore annihilate on a left-handed antiquark in the vacuum
condensate, liberating its right-handed partner.  This process is
perceived as change of chirality of a free quark, which is exactly the
same result as that of a nonvanishing quark mass.  However, in reality
the mass of the light quarks $u,d$ is nonzero, and the chirality of
a light quark is never exactly conserved.

All one can do, therefore, is to look for sudden changes in the
distance at which color forces are screened, or in the quark
condensate.  If these are discontinuous, one deals with a phase
transition, otherwise with a possibly rapid, but continuous change of
internal structure as it occurs, e.g., in the transformation of an
atomic gas into an electromagnetic plasma.  The identification of the
nature of the phase change is complicated by finite size effects.  The
best published results, by the Columbia group$^{34}$, for a 16$^3
\times 4$ lattice indicate a surprisingly strong dependence of the
phase diagram on the magnitude of the strange quark mass.  For the
physical mass $m_{\hbox{\sevenrm s}} \simeq$ 150 MeV there seems to be
no discontinuity, but only a rapid change in the energy density over a
small temperature range (about 10 MeV).  However, it is probably
premature to consider this as the final word.
\bigskip

\noindent {\bf Intermezzo:  Astrophysical Implications}
\medskip

We will return to the results of lattice gauge theory in a moment, but
let us pause briefly to study some astrophysical implications of a
hadron-quark-gluon phase transition.  If it is indeed of first order,
as the results discussed so far suggest, both phases can coexist over
a certain range of temperatures.  In particular, the quark-gluon phase
can exist in a supercooled state between the critical temperature
$T_c$, where its pressure falls below that of an isothermal hadron
gas, and the temperature $T$, where its pressure becomes negative
signaling intrinsic instability of this phase (see Figure 7b).  The
conversion of the metastable quark-gluon phase into hadrons then
proceeds by the formation and subsequent growth of bubbles of hadronic
gas imbedded in the plasma.  The dynamics of bubble formation is
governed by the free energy difference as a function of bubble radius
$R$:
$$\Delta E(R) = -\Delta P {4\pi\over 3} R^3 + \sigma 4\pi R^2 + aR + \ldots ,
\eqno(22a)$$
where $\Delta P > 0$ is the difference in pressure between the two phases,
$\sigma$ is the surface tension of the interface, and $a$ is a
coefficient related to the Casimir energy of the bubble
configuration.  The function $\Delta E(R)$ is schematically depicted
in Figure 9a. The notable feature here is the fact that bubble
formation is thermodynamically discouraged for small bubbles.  Only
when a bubble grows to a radius $R>R_c$, where $R_c$ denotes the
location of the maximum $\Delta E_c \simeq 2\sigma/\Delta P$, due to
some thermal fluctuation will it continue to grow on its own.  As the
probability of forming such a {\it critical droplet}$^{35}$ is
obviously hindered by a factor $\exp (-\Delta E_c/T)$, the suppression
depends sensitively on the size of the interface energy $\sigma$.  Its
value has recently been calculated in the framework of QCD-related
models$^{36}$ as well as by lattice simulations$^{37}$.  All results
indicate that the surface tension should be small:
$$\sigma \simeq \hbox{20-50 MeV/fm$^2$}, \eqno(23)$$
implying that it is probably difficult to supercool a quark-gluon
plasma for a considerable time (see ref. 38 for a recent discussion).
\bigskip

\centerline{\epsfbox{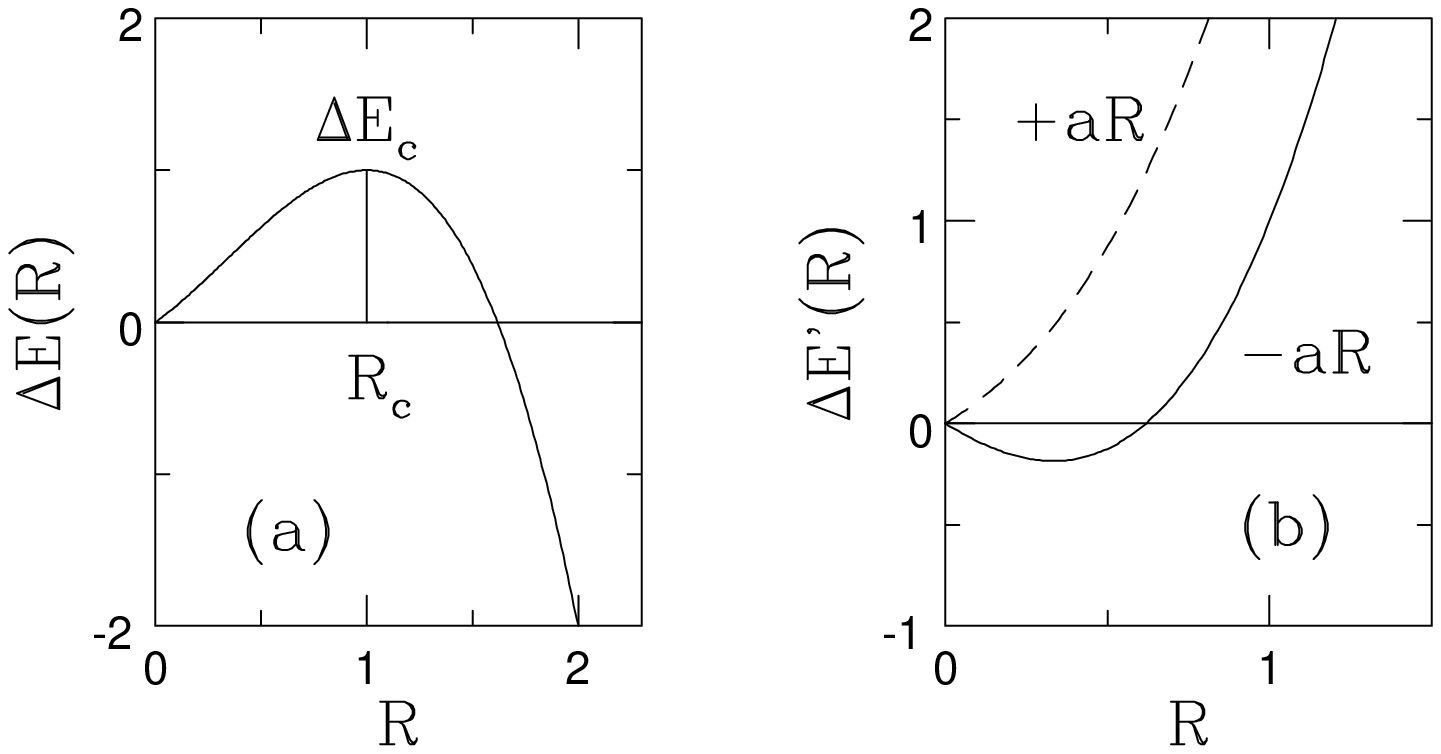}}
\noindent {\eightrm Figure 9. Energy of a hadronic bubble in the
quark-gluon plasma versus its radius:
(a) for $\scriptstyle{T<T_c}$, (b) for $\scriptstyle{T>T_c}$.
$\scriptstyle{R_c}$ in (a) indicates the critical droplet.
The solid curve in (b) predicts
the ``Swiss cheese'' instability of a quark-gluon plasma against
formation of small inhomogeneities.
\bigskip}

On the other hand the small surface tension raises the importance of
the next term in the series (22a), which is linear in the bubble
radius $R$.  That may have a dramatic effect when we consider the
opposite case, namely, superheated hadronic matter droplets forming
inside a large quark-gluon plasma bubble.  This is precisely the
situation one hopes to find right after a highly energetic nuclear
collision.  When we now ask about the stability of the plasma, we must
consider the reverse situation described by a droplet energy obtained
from (22a) by altering the signs of all terms with an odd power of $R$
(see solid curve in Figure 9b):
$$\Delta E' (R) = \Delta P {4\pi\over 3} R^3 + \sigma 4\pi R^2 - aR +
\ldots . \eqno(22b)$$
The change in sign of the linear (as well as the cubic) term was
pointed out by Lana and Svetitsky$^{39}$ who went on to study its
possible consequences.  Since the linear term dominates $\Delta
E'(R)<0$ for small $R$, and it is favorable to break up the plasma by
forming small droplets of superheated hadronic matter inside.  This
has been termed the ``Swiss cheese instability'' of the quark-gluon
plasma.  More detailed studies show the hadronic droplets to be
unstable against deformations, pointing to the possibility that the
plasma phase may be characterized by a complex structure of
inhomogeneities in the region of phase coexistence.  There is
preliminary support for this exotic picture from lattice
simulations$^{40}$.

Perhaps the most important effect caused by coexistence of the two
phases is due to the difference between the baryon densities in them
at equilibrium$^{41}$.  As discussed above, the baryon density in the
quark-gluon phase at given temperature $T$ and baryochemical potential
$\mu_{\hbox{\sevenrm B}}$ is (using eqs. 21, 17):
$$\rho_{\hbox{\sevenrm B}}^{(Q)}\simeq {2\over 3} \mu_{\hbox{\sevenrm
B}} \left( T^2 + {1\over 9\pi^2} \mu_{\hbox{\sevenrm B}}^2\right),
\eqno(24)$$
whereas the net baryon density in the hadronic phase (counting only
nucleons, not excited baryon resonances):
$$\rho_{\hbox{\sevenrm B}}^{(H)} \simeq 8 {\mu_{\hbox{\sevenrm
B}}\over T} \left( {m_NT\over 2\pi}\right)^{3/2} e^{-m_N/T},
\eqno(25)$$
where $m_N$ is the nucleon mass.  The ratio $\rho_{\hbox{\sevenrm B}}^{(Q)}
\big/\rho_{\hbox{\sevenrm B}}^{(H)}$ is very sensitive to the precise value of
the critical temperature $T_c$; but generally one finds
$\rho_{\hbox{\sevenrm B}}^{(Q)}\gg \rho_{\hbox{\sevenrm B}}^{(H)}$,
as illustrated in Figure 10.

\noindent Consequently, if the transition proceeds at or near chemical
equilibrium, baryons become enriched in the quark-gluon plasma phase.
For some time it was speculated that this effect could result in local
fluctuations of the proton-neutron ratio in the early universe that
were large enough to influence the course of nucleosynthesis.  The
argument goes roughly as follows:  Neutrons will quickly diffuse away
from a local surge in the baryon density, leaving behind a region of
abnormally large proton-neutron ratio.  Nucleosynthesis there will
result in a reduced number of neutron-rich isotopes, such as
${}^{4}$He and ${}^7$Li.  However, if the size of the region of
anomalous baryon concentration is too small (less than about 20 m)
neutrons can diffuse back during the era of nucleosynthesis and even
things out.  For the small value of the interface tension $\sigma$,
eq. (23), the effect of local inhomogeneities in the baryon density
is probably negligible$^{43,44}$.
\bigskip

\centerline{\epsfbox{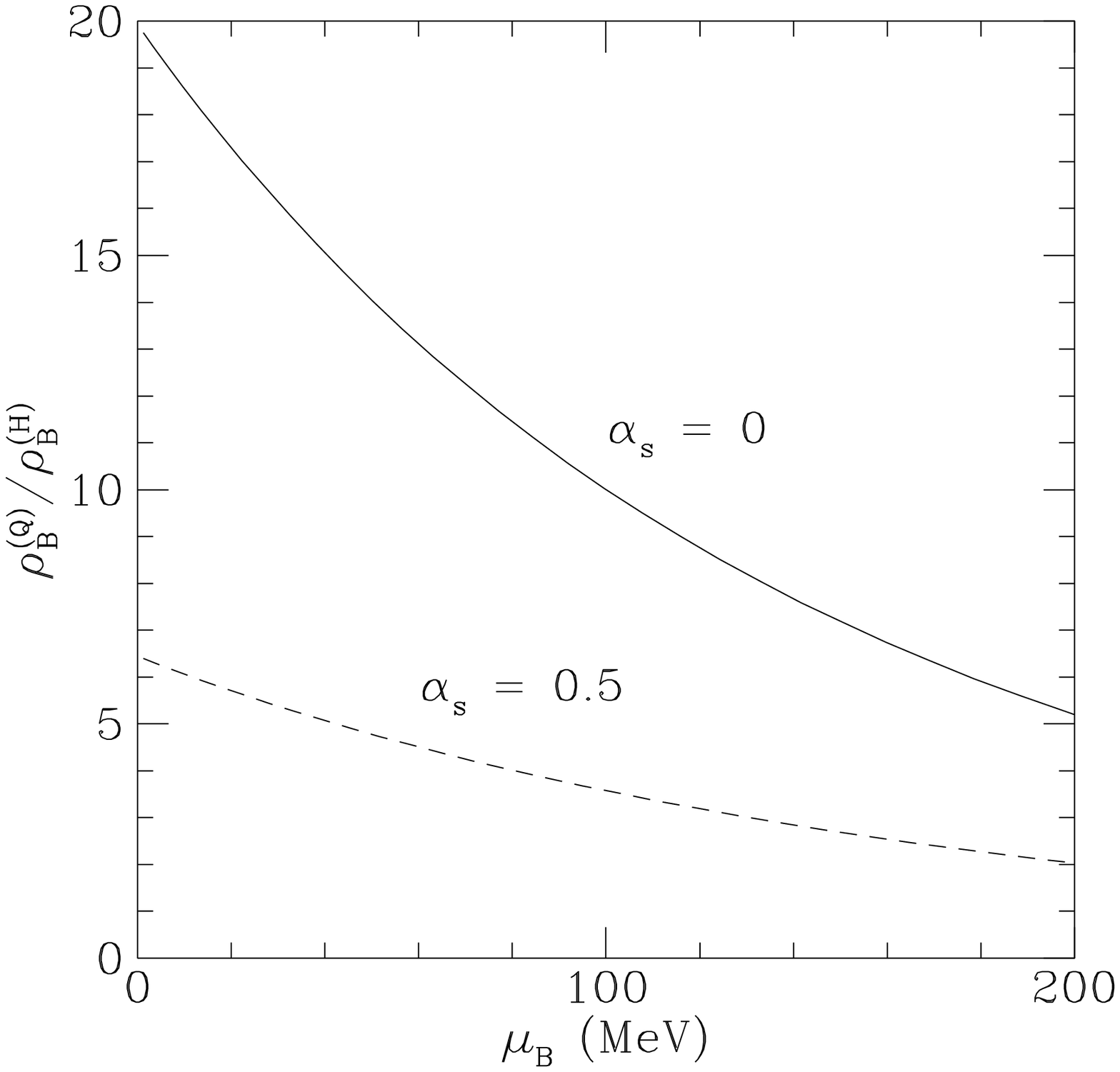}}
\noindent {\eightrm Figure 10. Ratio of the baryon densities of the
quark-gluon plasma and a hadronic gas at the coexistence line as
function of the baryochemical potential $\scriptstyle{\mu_B}$.
Conditions in the early universe correspond to the line
$\scriptstyle{\mu_B = 0}$.
\bigskip}

In relativistic heavy ion collisions, the most important result of the
concentration of baryons in the quark matter phase may be that it is
accompanied by an enrichment with {\it strange quarks} (as opposed to
strange antiquarks).  The interest in such a mechanism derives from
speculations that quark matter droplets with a high strangeness
content (so-called {\it strangelets}) might be stable or
metastable$^{45,41}$, because a large fraction of strange quarks
allows to construct electrically almost neutral nuclei with large
baryon number without raising the proton-neutron asymmetry.  As Liu
and Shaw$^{46}$, as well as Greiner, Koch and St\"ocker$^{47}$ showed,
the quark matter---hadron gas phase transition effectively acts as a
distillation process for strange quarks, if there is a net baryon
surplus$^{48}$.  Several experiments at Brookhaven (AGS) and CERN
(SPS) are searching, or preparing to search, for strangelets produced
in nuclear collisions$^{49}$.  Their existence would have far-reaching
technological implications$^{50}$.
\bigskip\medskip

\noindent{\bf INTERACTING QUARK-GLUON PLASMA}
\bigskip

Let us return now to our main theme, i.e. the physical properties of a
quark-gluon plasma.  So far, we have neglected interactions among
quarks and gluons in the deconfined phase, except in eq. (21), where
we included the contributions of one-loop diagrams to the energy
density and pressure of a quark gluon gas:
$$\eqalignno{ \varepsilon^{(1)} = 3P^{(1)} &= \left( 1- {15\over
4\pi}\alpha_s\right) {8\pi^2\over 15} T^4 + \left( 1-{50\over 21\pi}
\alpha_s\right) {7\pi^2\over 10} T^4 +\cr
&\quad +\left( 1 -{2\over \pi^2} \alpha_s\right) {3\over \pi^2} \mu^2
\left( \pi^2T^2 +{1\over 2}\mu^2\right), &(26)\cr}$$
where we have omitted the vacuum energy B, because it is a
non-perturbative contribution.  Graphically, eq. (26) corresponds to
the diagrams
\bigskip

\centerline{\epsfbox{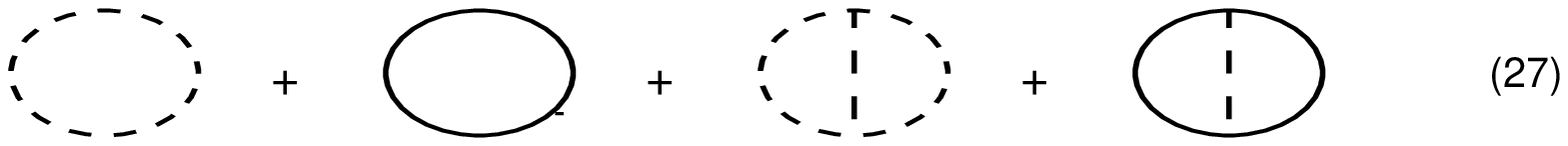}}
\bigskip

\noindent where dashed lines denote gluons, and straight lines denote quarks.
As (26) shows, we still have $P={1\over 3}\varepsilon$.  This changes
in the next order, because the two-loop contribution to the gluon
energy density is found to diverge!  The physical reason for this
divergence is that gluon and quark degrees of freedom develop an
effective mass, which leads to screening of long-range color-electric
forces.  Technically, the screening mass is obtained by summing an
infinite chain of one-loop insertions in the gluon propagator
\bigskip

\centerline{\epsfbox{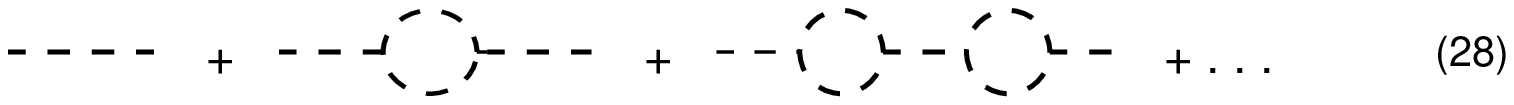}}
\bigskip

The contribution of all diagrams except the first two (which are
already included in (26)) can be summed analytically and yields a
contribution to the gluon energy of order $\alpha^{3/2}$ with a rather
large coefficient$^{51}$.
\bigskip\medskip

\noindent {\bf The QCD Plasmon}
\bigskip

We can obtain more insight into the properties of the interacting
quark-gluon plasma by looking at the gluon propagator, represented
graphically in (28), itself.  Because of gauge invariance, $k^{\mu}
D_{\mu\nu}(k)=0$, it can be decomposed into a longitudinal and a
transverse part, which are scalar functions of the variables $\omega=k^0$
and $k=\vert \hbox{\bf k}\vert$.  These are most conveniently
written in a form borrowed from electrodynamics of continuous media:
$$\eqalignno{D_L (\omega,k) &= {1\over \varepsilon_L(\omega,k)k^2},
&(29a)\cr
D_T(\omega,k) &= {1\over\varepsilon_T(\omega,k)\omega^2-k^2}, &(29b)
\cr}$$
where the color-dielectric functions are given by$^{52}$:
$$\eqalignno{\varepsilon_L(\omega,k) &= 1 + {g^2T^2\over k^2} \left[ 1
- {\omega\over 2k} \ln \left({\omega+k\over \omega-k}\right)\right];
&(30a) \cr
\varepsilon_T (\omega,k) &= 1 - {g^2T^2\over 2k^2} \left[ 1- \left(
1-{k^2\over \omega^2}\right) {\omega\over 2k} \ln \left(
{\omega+k\over \omega-k}\right)\right]. &(30b) \cr}$$
Several things are noteworthy about eqs. (29, 30).  First they imply
that static longitudinal color fields are screened:
$$D_L(0,k) = {1\over \varepsilon_L(0,k)k^2} = {1\over k^2 + g^2T^2} =
{1\over k^2+\lambda_D^{-2}} . \eqno(31a)$$
The Debye length obviously is $\lambda_D = (gT)^{-1}$.  On the other
hand, eqs. (29b,30b) show that static transverse (magnetic) color fields
remain unscreened at this level of approximation:
$$D_T(0,k) = -{1\over k^2}. \eqno(31b)$$
The static magnetic screeening length is of higher order in the
coupling constant; lattice gauge calculations$^{52}$ have shown that
$\lambda_M^{-1} = Cg^2T$.
\bigskip

\centerline{\epsfbox{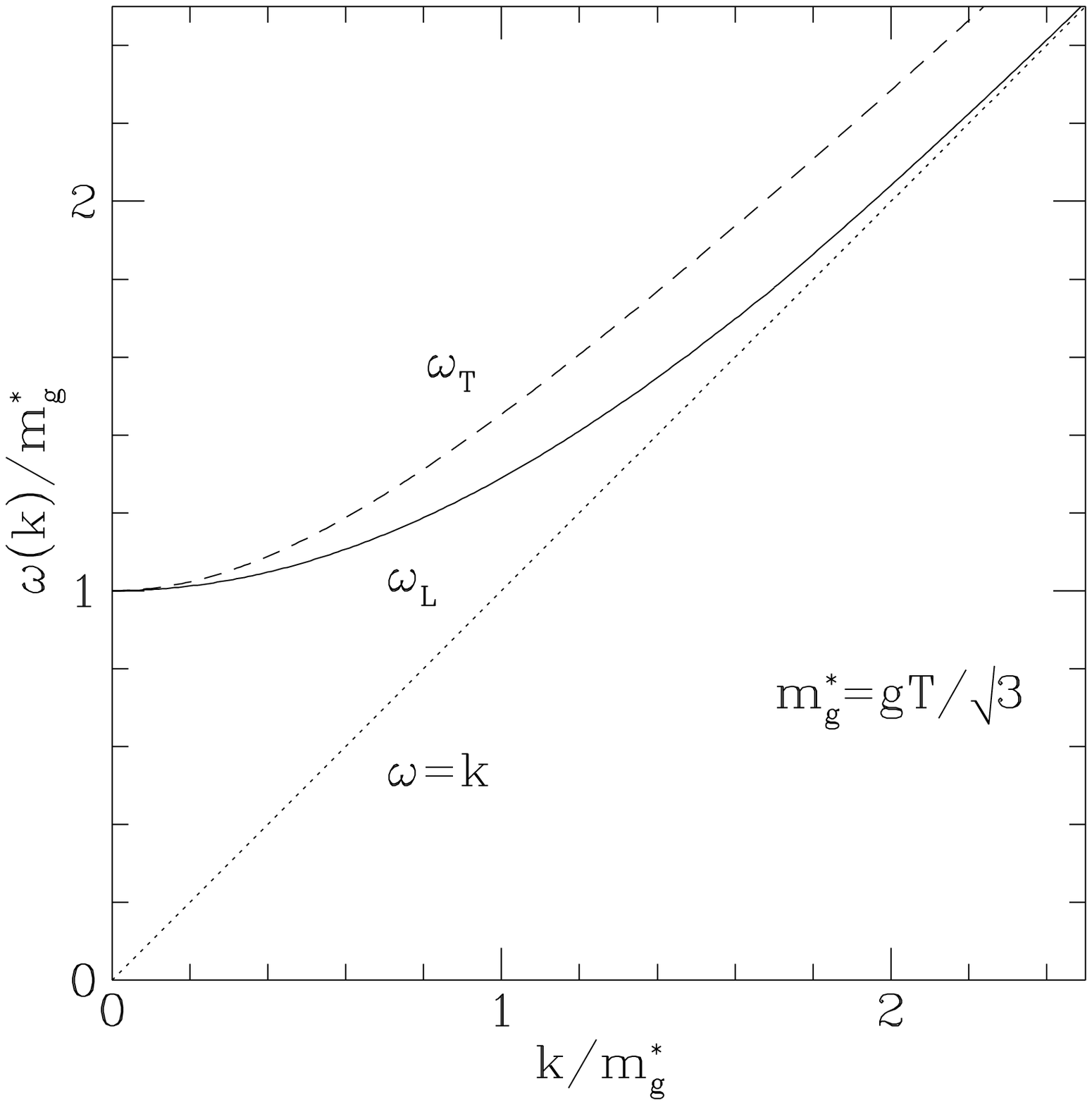}}
\noindent {\eightrm Figure 11. Gluon dispersion relation in the
perturbative quark-gluon plasma.  All quantities are measured in units
of the effective gluon mass $\scriptstyle{m_g^* = gT/\sqrt{3}}$.
Solid line: longitudinal plasmon mode;
dashed line: transverse collective gluon mode.
\bigskip}

For a finite frequency $\omega$ the in-medium propagators (29) have
poles corresponding to propagating, collective modes of the glue
field.  The dispersion relation for the longitudinal mode:
$$\varepsilon_L(\omega,k) = 0, \eqno(32a)$$
called the {\it plasmon}, has no counterpart outside the medium.  The
analogous relation for the transverse mode:
$$\varepsilon_T(\omega,k) = k^2/\omega^2 \eqno(32b)$$
describes the effects of the medium on the free gluon.  The behavior
of both modes is remarkably similar.  For $k\to 0$ they yield an
effective gluon/plasmon mass
$$\omega_L,\omega_T \buildrel{k\to 0}\over\longrightarrow \; m_g^* =
{1\over\sqrt{3}}gT, \eqno(33)$$
whereas for large momenta $(k\to\infty)$ one finds
$$\omega_L(k) \to k,\quad \omega_T(k) \to \sqrt{k^2+{1\over 2}g^2T^2}.
\eqno(34)$$
The full dispersion relations are shown in Figure 11.

For plasma
conditions realistically attainable in nuclear collisions ($T\simeq
250$ MeV, $g=\sqrt{4\pi\alpha_s}\simeq 2$) the effective
gluon mass is $m_g^* \simeq$ 300 MeV.  We must conclude, therefore,
that the notion of almost free gluons (and quarks) in the
high-temperature phase of QCD is quite far from the truth.  Certainly,
one has $m_g^*\ll T$ when $g\ll 1$, but this condition is {\it never}
really satisfied in QCD, because $g\simeq {1\over 2}$ even at the
Planck scale (10$^{19}$GeV), and $g<1$ only at energies above 100 GeV.
Let us discuss some consequences of these results:
\bigskip

\centerline{\epsfbox{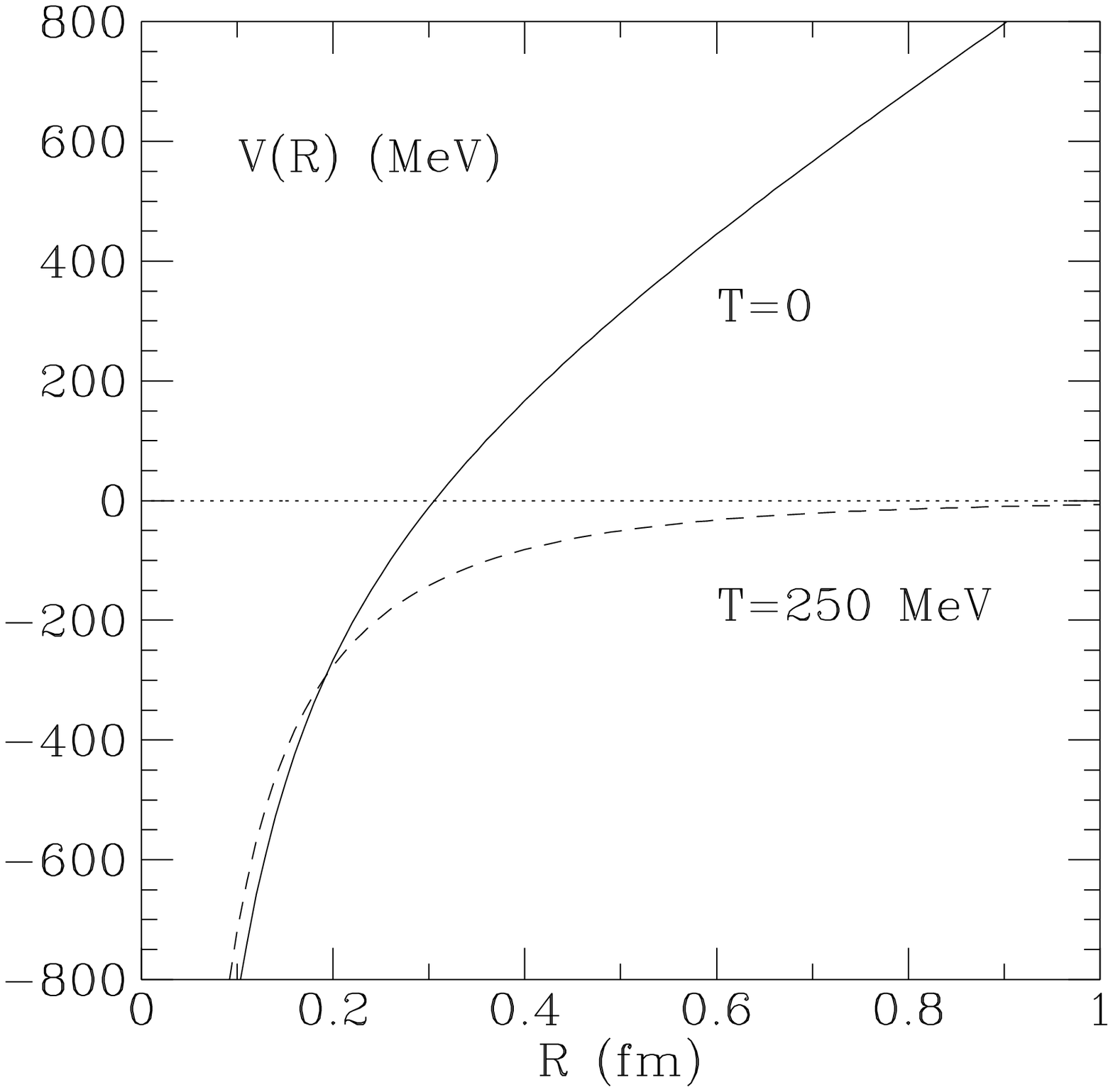}}
\noindent {\eightrm Figure 12. Effective quark-antiquark potential in QCD.
Solid line:  confining potential of a free $\scriptstyle{Q\ov{Q}}$
pair; dashed line: screened potential, eq. (35), of a
$\scriptstyle{Q\ov{Q}}$ pair imbedded in the quark-gluon plasma.
\bigskip}

\item{(1)} The potential between two static color charges, such as two heavy
quarks, is screened in the quark-gluon plasma phase.  The Fourier
transform of eq. (31a) yields the potential
$$V_{Q\ov{Q}}(r) \simeq {1\over r}e^{-r/\lambda_D} \eqno(35)$$
with screening length $\lambda_D \simeq 0.4$ fm at $T$ = 250 MeV.  This
screened potential is compared in Figure 12 with the confining potential
between a free quark-antiquark pair.  This screening of long-range
color forces is, of course, the origin of quark deconfinement in the
high-temperature phase.  An important consequence, to be discussed
later in the section on plasma signatures, is the disappearance of the
bound states of a charmed quark pair $(c\ov{c})$ in the quark-gluon
plasma$^{54}$.

\item{(2)} The color screening at large distances cures most infrared
divergences in scattering processes between quarks and gluons.  A
self-consistent scheme implementing this mechanism has been devised
by Braaten and Pisarski$^{55}$.  It involves the resummation of gluon
loops involving gluons with momenta of order $gT$ and has been shown
to be gauge invariant when also vertex corrections are taken into
account.

\item{(3)} The finite effective gluon mass $m_g^*$ leads to the
suppression of long-wavelength gluon modes with $k\le gT$ in the
quark-gluon plasma.  As a result, the relation $P= {1\over
3}\varepsilon$ is violated and the pressure is reduced.  We now have
two mechanisms that can be responsible for $P<{1\over 3}\varepsilon$:
an effective gluon mass $m_g^*$ and a nonvanishing vacuum energy $B$.
A fit to recent SU(3) lattice gauge theory results$^{56}$ with $m_g^*$
and $B$ taken as free parameters shows that probably both mechanisms
are at work$^{57}$ (see Figure 13).
\bigskip

\centerline{\epsfbox{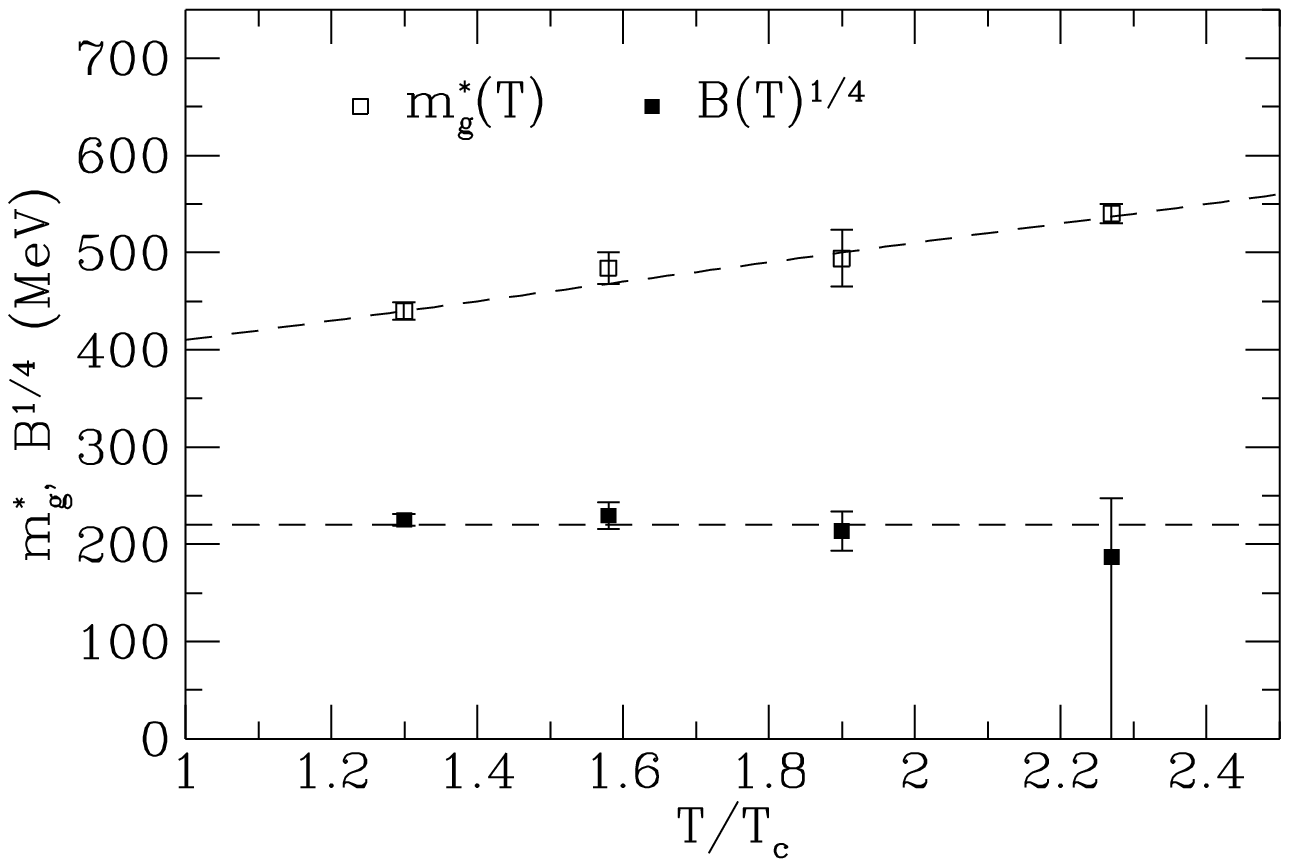}}
\noindent {\eightrm Figure 13. Effective mass $\scriptstyle{m_g^*}$ and
vacuum energy density $\scriptstyle{B}$ as function of temperature
(lower part) as deduced$^{57}$ from results of lattice gauge theory
simulations$^{56}$ for the energy density $\scriptstyle{\varepsilon}$
and pressure $\scriptstyle{P}$ of a gluon gas (see Figure 8).
\bigskip\medskip}

\noindent {\bf Properties of the Quark-Gluon Plasma}
\bigskip

\noindent {\it (a) Thermalization:}
\medskip

For those of us interested in the detection of a quark-gluon plasma
in nuclear collisions it is imperative to know something about its
rate of thermalization.  Does it thermalize sufficiently fast, so that
a thermodynamical description makes sense?  In the picture based on
quasi-free quarks and gluons moving through the plasma, thermalization
proceeds mainly via two-body collisions, where the color force between
the colliding particles is screened, as illustrated in Figure 14.  The
technically easiest way of looking at this is to consider the quark
(gluon) {\it damping rate $\gamma$}, i.e. the imaginary part of the
quark (gluon) self energy.  For quarks and gluons of typical thermal
momenta $(p\simeq T)$ one finds with the help fo the techniques
discussed in the previous section$^{55}$
$$\eqalignno{\gamma_q &= -\textstyle{{1\over 4p}} \hbox{\rm Im}
[\hbox{\rm Tr} (\gamma\cdot p\;\Sigma(p))]
\simeq \textstyle{{2\over 3}}\alpha_s T \left( 1+ \ln
\textstyle{{1\over\alpha_s}}\right), &(36a) \cr
\gamma_g &= -\hbox{\rm Im} [\omega_T(p)] \simeq
\textstyle{{9\over 4}}\gamma_q,  &(36b)\cr}$$
where the factor ${9\over 4}$ reflects the ratio of the quadratic
Casimir invariants $C_2$ of the octet and triplet representations of
color-SU(3).  Gluons simply have a higher color charge than quarks
$(C_2=3$ versus $C_2 = {4\over 3})$ and therefore scatter more often.
\bigskip

\centerline{\epsfbox{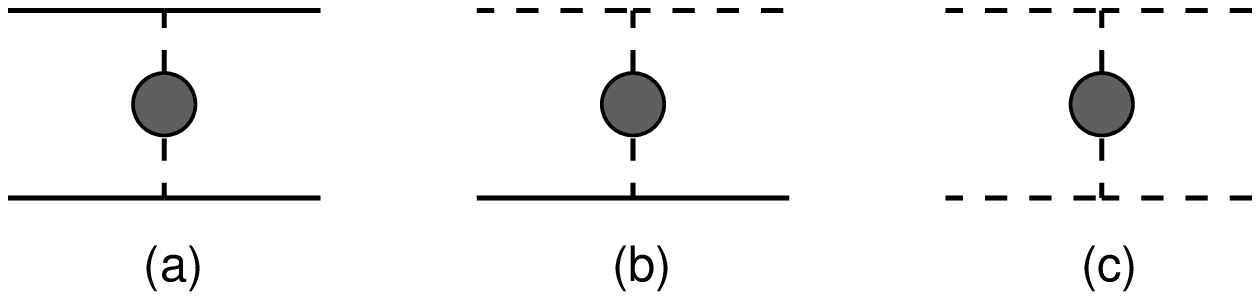}}
\noindent {\eightrm Figure 14. QCD diagrams describing scattering
processes contributing to thermalization of the quark-gluon plasma.
\bigskip}

One can argue that a better way to look at thermalization is to consider
the rate of momentum transfer between particles, i.e. weighting the
scattering diagrams of Figure 14 by $\sin^2\theta$, where $\theta$ is the
scattering angle.  Here one finds$^{58}$:
$$\Gamma_q^{(\hbox{\sevenrm tr})} \simeq 2.3\alpha_s^2 T \ln
\textstyle{{1\over \alpha_s}}; \quad \Gamma_g^{(\hbox{\sevenrm tr})}
\simeq 3\Gamma_q^{(\hbox{\sevenrm tr})}. \eqno(37)$$

\noindent The {\it transport rate}
$\Gamma^{(\hbox{\sevenrm tr})}$ is closely related to the shear
viscosity of the quark gluon plasma.  Both approaches yield quite
similar numbers for values of the strong coupling constant in the
range $\alpha_s$ = 0.2-0.5.  The characteristic equilibration time, as
defined as the inverse of the rates $\gamma_i$ or $\Gamma_i$ are:
$$\tau_g \simeq 1\; \hbox{fm/c}, \quad \tau_q \simeq 3\; \hbox{fm/c},
\eqno(38)$$
i.e. gluons thermalize about several times faster than quarks$^{59}$.
Initially, therefore, probably a rather pure glue plasma is formed in
heavy ion collisions, which then gradually evolves into a chemically
equilibrated quark-gluon plasma.  We shall later return to this point in
the context of our discussion of microscopic models of relativistic
nuclear collisions, but it is worthwhile to point out one consequence:
During its hottest phase the QCD plasma is mainly composed of gluons,
which are not accessible to electromagnetic probes, such as lepton
pairs and photons.  The gluon plasma can only be probed by strongly
interacting signals, such as charmed quarks or jets.
\medskip

\noindent {\it (b) Energy loss of a fast parton:}
\medskip

One possible way of probing the color structure of QCD matter is by
the energy loss of a fast parton (quark or gluon).  The mechanisms are
similar to those responsible for the electromagnetic energy loss of a
fast charged particle in matter, i.e. energy may be lost either by
excitation of the penetrated medium or by radiation.  Although
radiation is a very efficient energy loss mechanism for relativistic
particles, it is strongly suppressed in a dense medium by the
Landau-Pomeranchuk effect$^{60}$.  In the case of QCD this effect has
recently been analyzed comprehensively$^{61}$, and the suppression of
soft radiation is now firmly established.  This limits the radiative
energy loss to about 1 GeV/fm.  Excitational energy loss proceeds via
collisions with quarks and gluons from the plasma.  Here again color
screening plays an essential role, reducing the rate of energy loss to
about 0.3 GeV/fm for a fast quark$^{62,63}$.
\vfill\eject

\noindent {\bf Entropy production}
\medskip

The correspondence principle asserts that highly excited states of a
quantum system usually exhibit quasi-classical behavior.  Similarly,
the high-temperature limit of a quantum system is quasi-classical,
since thermal fluctuations dominate over quantum
fluctuations.\footnote{*}{As is well-known from the case of
electromagnetic black-body radiation, the classical limit is only
reached for long-wavelength modes, but not for modes with wavelength
$\lambda\ll \hbar /T$.  The quasi-classical description therefore requires
an ultraviolet cut-off, eliminating these modes.} Since the
quark-gluon plasms is the high-temperature phase of QCD, one may ask
whether classical, thermal QCD can yield useful results.  In view of
the remark made in the footnote, most predictions of classical thermal
QCD will be cut-off dependent and therefore not completely reliable,
except possibly for those quantities that do not involve Planck's
constant $\hbar$.  A simple dimensional analysis reveals that the
combination $g^2T$  does not contain factors of $\hbar$,
and therefore one may suspect that the thermalization rate
$\gamma_g$, eq. (36b), can be obtained by classical considerations.
This permits an entirely different, nonperturbative determination of
the thermalization rate of gluonic matter by simulation of the
dynamics of SU(3) gauge theory in real time on a lattice.

The approach to thermal equilibrium in a classical dynamical system is
governed by the rate of entropy production.  More precisely,
thermodynamic (ensemble) averages can be applied to an isolated
classical system, if they coincide with the long-time averages, e.g.
of a quantity $A(t)$:
$${1\over T} \int_0^T dt \;\;A(t) \buildrel {T\to \infty}\over
\longrightarrow \langle A\rangle_{\hbox{\sevenrm ensemble}};
\eqno(39)$$
such systems are called {\it ergodic}.  It was shown in the important
work of Krylov, Kolmogorov, and others$^{64}$ that ergodicity is
really of practical use only when the system exhibits the property of
{\it mixing}, meaning that the equality (39)
 is approached uniformly
throughout almost the entire phase space at an exponential rate.  This
condition, in turn, is satisfied if classical trajectories
$\hbox{\bf x}(t)$ are {\it unstable} against small fluctuations
almost everywhere in phase space.  Such systems are also called {\it
strongly chaotic}.
\bigskip

\centerline{\epsfbox{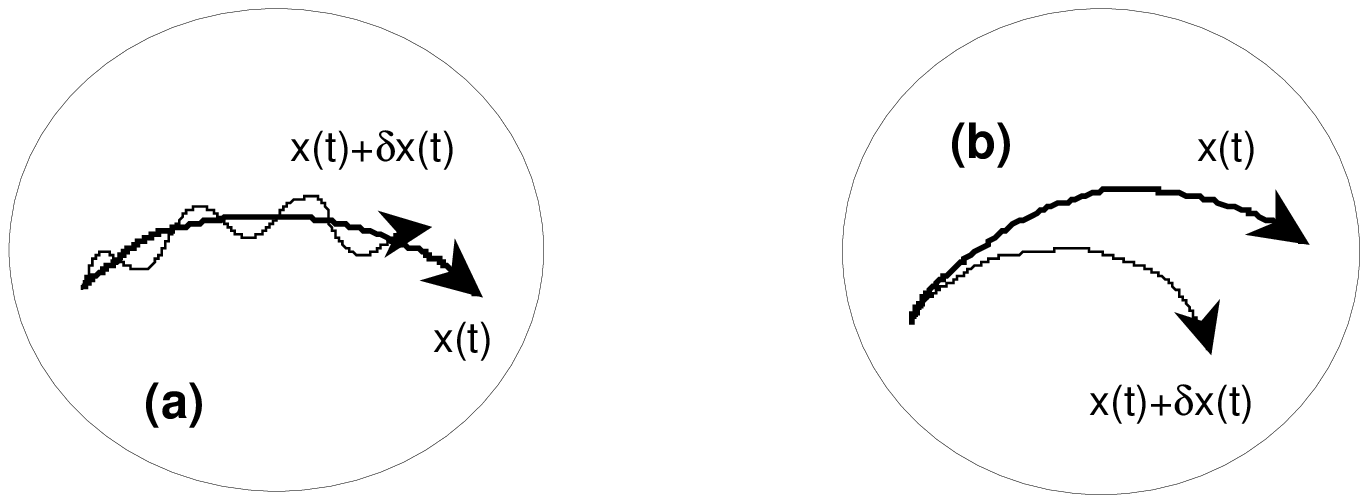}}
\noindent {\eightrm Figure 15. (a) Stable phase space trajectory;
(b) unstable trajectory.
\bigskip}

Consider two neighboring trajectories, i.e. solutions of the classical
equations of motion, {\bf x}$(t)$ and {\bf x}$^{\prime}(t)=$ {\bf
x}$(t)+\delta\hbox{\bf x}(t)$.  The trajectory is {\it unstable}, if
the norm of an infinitesimal derivation grows exponentially with time
$$\vert\delta\hbox{\bf x}(t)\vert\;\;
\buildrel {t\to\infty}\over \longrightarrow\;\;
D_0 \exp(\lambda t), \quad \lambda > 0, \eqno(40)$$
as illustrated in Figure 15.  $\lambda$ is called a {\it Lyapunov exponent}.
In general, chaotic systems possess several positive Lyapunov exponents,
depending on the direction of the initial fluctuation $\delta\hbox{\bf
x}(0)$, but usually the {\it maximal} Lyapunov exponent $\lambda_0$
dominates in eq. (40) for any arbitrarily chosen $\delta\hbox{\bf
x}(0)$.

We are now ready to discuss entropy growth.  For a classical system,
entropy is defined in terms of the volume in phase space covered by an
ensemble of identical systems:\footnote{*}{The constant is undefined in
classical physics; its value is fixed when the volume $\Delta\Gamma$ is
measured in units of $\hbar^N$, where $N$ is the number of degrees of
freedom.}
$$S = \ln (\Delta\Gamma) + \hbox{const.} \eqno(41)$$
At first one might think
that $\Delta\Gamma$ remains constant in time for a Hamiltonian system
on account of Liouville's theorem.  However, for a strongly chaotic
system this notion conflicts with the finite resolution of any
measurement, ultimately with the quantum uncertainty limit $\hbar^N$.
The occupied phase space volume $\Delta\Gamma$ must, therefore, be
smeared out with the finite resolution:
$$\Delta\Gamma \to \ov{\Delta\Gamma}, \quad S = \ln
(\ov{\Delta\Gamma}) + \hbox{const.}, \eqno(42)$$
a process usually referred to as {\it coarse-graining}.  For a
strongly chaotic, or mixing system the instability of trajectories
leads to a growing filamentation of the phase space volume occupied by
an ensemble, as illustrated in Figure 16.
\vskip2.2917truein

\noindent {\eightrm Figure 16. Filamentation of the occupied phase
space volume permits a chaotic system to ``fill'' the available phase
space homogeneously without violating Liouville's theorem (from
Zaslavsky$^{64}$).
\bigskip}

\noindent A careful analysis$^{64}$ of this process shows that the
coarse-grained entropy of a strongly chaotic system grows linearly
with time:
$$ds/dt = \langle h\rangle_{\Gamma}, \eqno(43)$$
where the right-hand side denotes the phase-space average of the sum
of all positive Lyapunov exponents:
$$h = \sum_{\alpha} \lambda_{\alpha}\theta(\lambda_{\alpha}).
\eqno(44)$$
In this way entropy growth, and ultimately thermalization, in
classical dynamical systems is intimately connected to the instability
of its trajectories in phase space.

One may ask whether this has any relevance for quantum systems?
Intuitively, one would suppose that it must, because highly excited
states of a quantum system usually exhibit many aspects of classical
dynamics.  However, the notion of a Lyapunov exponent has no
immediately counterpart in quantum mechanics, because one cannot
define a trajectory in phase space for quantum mechanics.  Recently,
however, Gun He has pointed out that there exists a closely related
concept in quantum mechanics, namely the growth of local
inhomogeneities in the Wigner function of an ensemble of quantum
systems$^{65}$.  In simple, analytically tractable examples one finds
that the exponential growth of the variance of this generalized phase
space distribution is governed by the same exponent as the divergence
of trajectories in the classical system.  This can be utilized to
introduce a meaningful concept of entropy for quantum systems that
differs from von Neumann's information entropy, but provides a useful
description of the approach to equilibrium$^{65}$.

Now back to QCD!  It has been known for a long time that non-abelian
gauge theories exhibit elements of chaos in the classical
limit$^{66}$.  Recently, Trayanov and myself showed$^{67}$ that randomly
chosen field configurations in SU(2) lattice gauge theory are
characterized by a universal Lyapunov exponent, which scales with the
average energy density.  This analysis has been extended to SU(3) by
C. Gong$^{68}$, who found the result
$$h \simeq \textstyle{{1\over 10}} g^2 \langle E_p\rangle,
\eqno(45)$$
where $\langle E_p\rangle$ denotes the average energy per lattice
plaquette.  For a thermalized system the relation $\langle E_p\rangle
= {16\over 3}T$ holds in SU(3); hence one obtains an entropy growth rate for
SU(3) gauge theory of
$$dS/dt = \langle h\rangle \simeq 0.54\; g^2T. \eqno(46)$$
This value coincides numerically with (twice) the thermal damping rate
of long-wavelength gluons
$$\gamma_g^{(0)} = {6.635\over 4\pi} g^2T \simeq 0.264 g^2T.
\eqno(47)$$
The factor two between (46) and (47) may be understood by the remark
that (47) gives the damping rate of the gluon {\it amplitude}; the
decay rate of the probability is $2\gamma_g$.  Although it is easy to
see similarities between $\langle h\rangle$ and $2\gamma_g$, their
identity has not been established, and their numerical equality
may be accidental.  Independent of this, however, eq. (46) provides us
with a precise value of the gluon thermalization time in the
high-temperature, classical limit, defined as the characteristic
entropy growth time near equilibrium:
$$\tau_s = \langle h\rangle^{-1} \simeq 1.85/g^2T. \eqno(48)$$
Using the one-loop expression for the running thermal coupling constant
$$g^2(T) = {16\pi^2\over 11\ln (\pi T/\Lambda)^2} \eqno(49)$$
one finds thermalization times of the order of 0.4 fm/c, as shown in
Figure 17.  This value is comfortably short on the time-scale of
relativistic heavy ion collisions, where the high-density phase is
predicted to last for several fm/c.
\bigskip

\centerline{\epsfbox{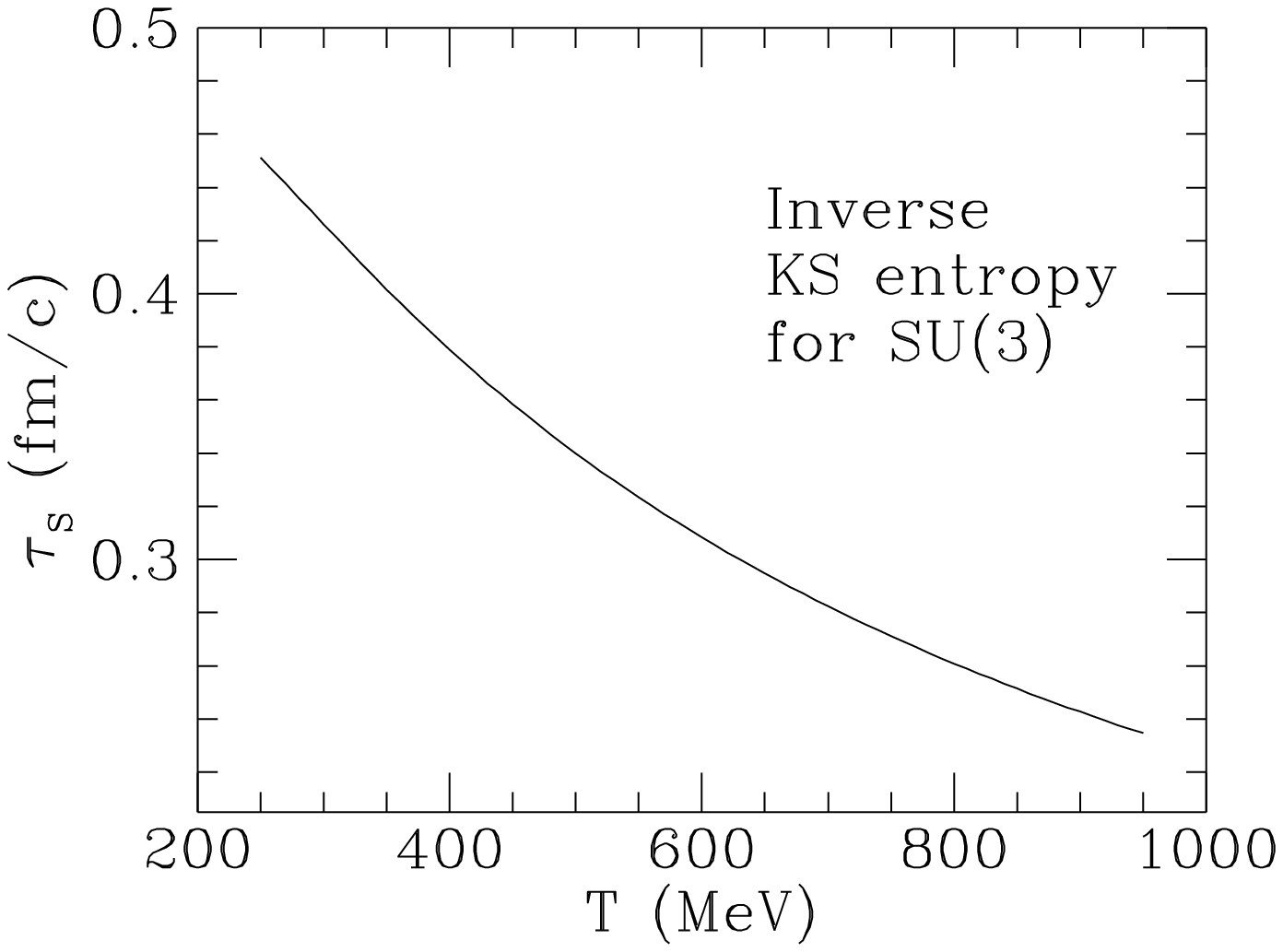}}
\noindent {\eightrm Figure 17. Thermalization time of gluonic matter
close to thermal equilibrium, as obtained from the Lyapunov exponent
of thermal SU(3) lattice gauge theory$^{68}$.
\bigskip}

The advantage of the definition (48) of the thermalization is that it
is also meaningful for field configurations that are far from the
average thermal configuration.  Numerical studies of the rate of
instability of coherent field configurations in SU(2) gauge theory
have yielded even higher rates of entropy growth for the same energy
density$^{67}$.  More general investigations of this phenomenon would
be helpful.
\bigskip\medskip

\noindent {\bf FORMATION OF THE QUARK-GLUON PLASMA}
\bigskip

\noindent {\bf Approaches to the Thermalization Problem}
\bigskip

In the previous section we discussed how the quark-gluon plasma
equilibrates when it is already close to thermal equilibrium.
Relativistic heavy ion collisions pose a very different problem:  how
do the fully coherent parton wave functions of two nuclei in their
ground states evolve into locally quasi-thermal distributions of
partons as they are characteristic of the quark-gluon plasma state?
There are mainly two approaches to this problem that have been
extensively investigated:  (a) QCD string breaking and (b) the
partonic cascade.
\bigskip

\centerline{\epsfbox{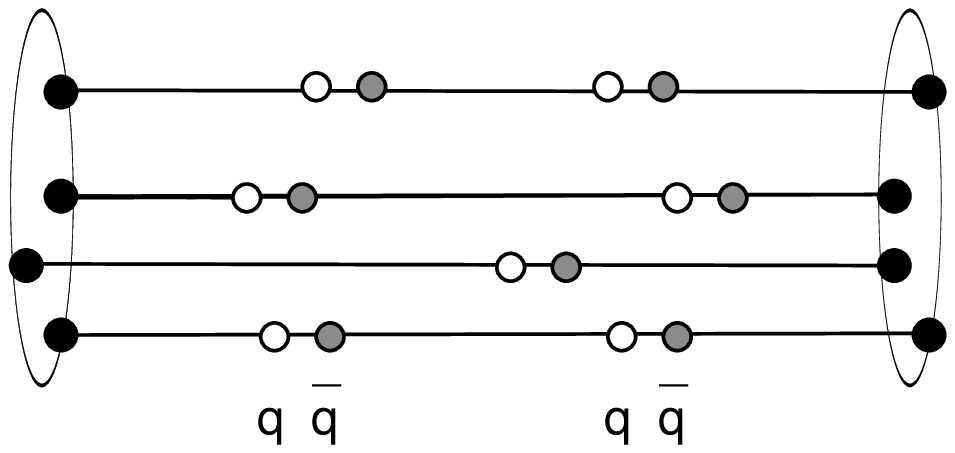}}
\noindent {\eightrm Figure 18. String-based picture of the formation
of a quark-gluon plasma.  Primary interactions lead to color flux
tubes which break by quark pair production.
\bigskip}

In the string picture, developed from models of soft hadron-hadron
interactions, one assumes that nuclei pass through each other at
collider energies with only a small rapidity loss (on average about one
unit), drawing color flux tubes, or strings, between the ``wounded''
nucleons.  If the area density of strings is low (not much greater
than 1 fm$^{-2}$) they are supposed to fragment independently by quark
pair production on a proper time scale of order 1 fm/c.  Most
realizations of this picture are based on the Lund string
model$^{69}$, e.g. Fritiof$^{70}$, Attila$^{71}$, Spacer$^{72}$,
Venus$^{73}$, QGSM$^{74}$ and RQMD$^{75}$.  When the density of strings
grows further, at very high energy and for heavy nuclei, the formation of
``color ropes'' instead of elementary flux tubes has been postulated$^{76}$.

Alternatively, a continuum description based on the Schwinger model of
(1+1)-dimensional QED with heuristic
back-reaction---``chromohydrodynamics''---has been invoked to describe
the formation of a locally equilibrated quark-gluon plasma$^{77,78}$.
One general aspect of these models is that initially part of the
kinetic energy of the colliding nuclei is stored in coherent glue
field configurations, which subsequently decay into quark pairs.  The
flux tubes carry no identifiable entropy.  The entropy associated with
a thermal state is produced in the course of pair creation.  In
particular, there is no distinction between gluon and quark
thermalization.
\vskip2.083truein

\centerline{ {\eightrm Figure 19. Schematic view of a parton cascade
(from ref.~85).}}
\bigskip

The parton cascade approach, whose basic concepts were developed by
Boal$^{79}$, Hwa and Kajantie$^{80}$ and Blaizot and A. Mueller$^{81}$,
is founded on the parton picture and renormalization-group improved
perturbative QCD.  Whereas the string picture runs into conceptual
difficulties at very high energy, when the string density becomes large,
the parton cascade becomes invalid at lower energies, where most partonic
scatterings occur at energies that are too low to be described by
perturbative QCD.  Nevertheless, the two descriptions may well be just
two different formulations of the same physical processes, since there
exists a remarkable similarity between the states produced by QCD
bremsstrahlung and QCD flux tubes$^{82}$.

Let us look at the ``big picture'', illustrated in Figure 20, where we
distinguish three regimes in the evolution of an ultra-relativistic
heavy-ion collision.  Immediately after the Lorentz contracted nuclear
``pancakes'' have collided, scattered partons develop an incoherent
identity and evolve into a quasithermal phase space distribution by
free streaming separation of the longitudinal spectrum (a).  Rescattering
of these partons finally leads to the thermalization after a time of
the order of 1 fm/c.  The thermalized quark-gluon plasma then evolves
according to the laws of relativistic hydrodynamics (b), until it has
cooled to the critical temperature $T_c \simeq $ 150-200 MeV, where it
begins to hadronize (c).  The physics governing the evolution during the
three stages is very different.  In this section we will concentrate
mainly on the first, i.e. the preequilibrium state, because it is here
where much recent progress has been made in our understanding.
\bigskip

\centerline{\epsfbox{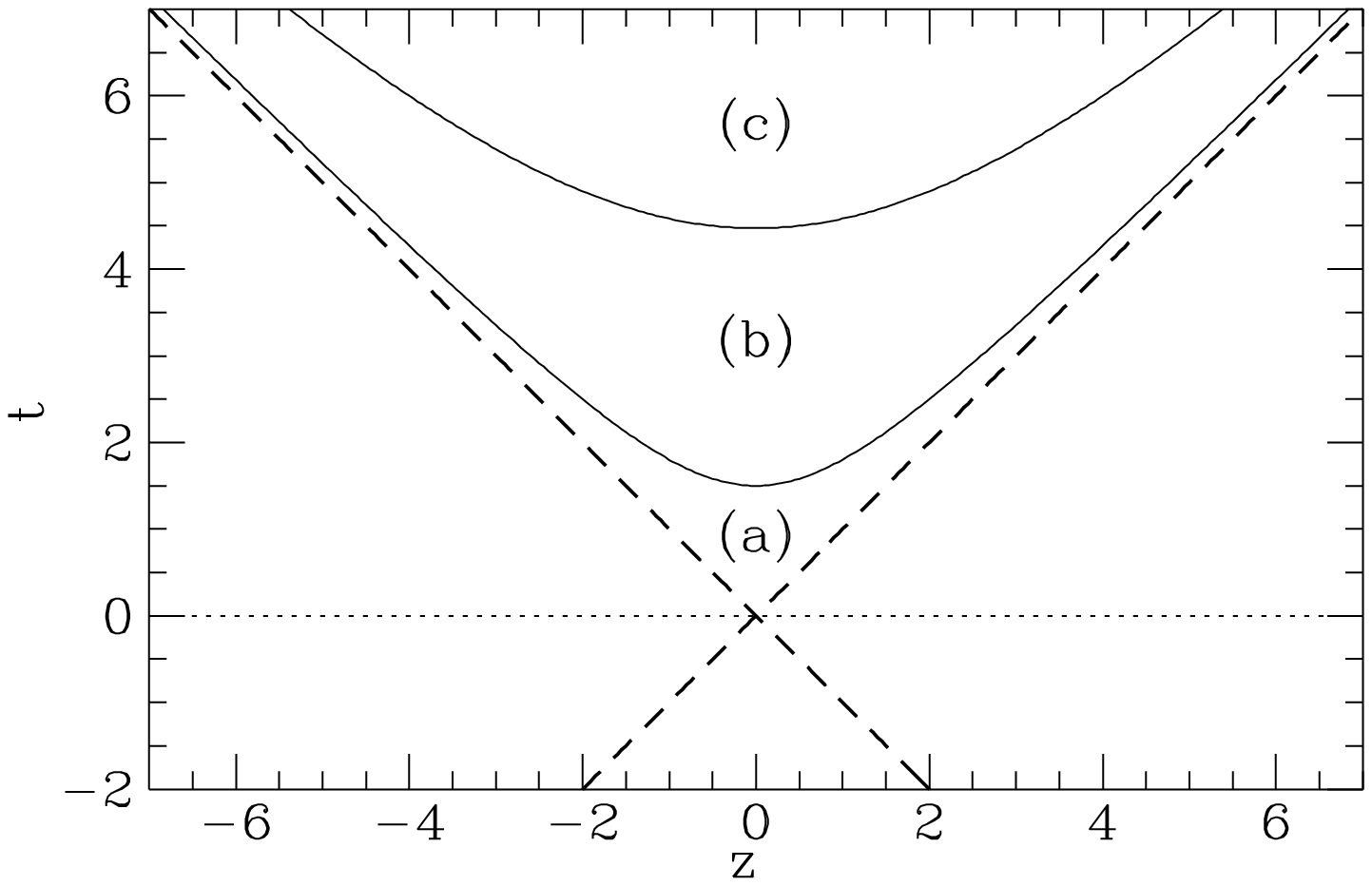}}
\noindent {\eightrm Figure 20. One-dimensional space-time picture of
the evolution of an ultrarelativistic nuclear collision,
distinguishing three regimes:  (a) Pre-equilibrium state, (b)
thermalized quark-gluon plasma, (c) hadronization phase.
$\scriptstyle{z}$ denotes the beam axis. The dashed lines indicate
the colliding nuclei.}
\bigskip\medskip

\noindent {\bf Parton Cascades}
\bigskip

As mentioned above, a fast parton of energy $E_p$ penetrating a
quark-gluon plasma loses about 1 GeV of energy in every 1 fm traversed.
Its thermalization is therefore of the order
$$\tau_{\hbox{\sevenrm th}} = \left( {E_p\over \hbox{GeV}}\right)\;
\hbox{fm/c}, \eqno(50)$$
which can be many tens or hundreds of fm/c for a past parton.
How, then, is it possible that two nuclei colliding at 100 GeV/u
(RHIC) or 3000 GeV/u (LHC) can deposit enough energy while penetrating
each other to thermalize within 1 fm/c proper time?  The resolution of
this almost paradoxical situation has two aspects:

\item{(a)} The emphasis is on {\it proper time} $\tau$.  $\tau$ = 1
fm/c  corresponds to a very long time in the laboratory frame for matter
produced at high rapidity.  To wit, thermalization measured in the
rest frame of the fast parton takes only
$$\tau_{\hbox{\sevenrm th}} = \int {dt\over\gamma_p} = \int {m_p^*\over
E_p} {dE_p\over \vert dE/dx\vert} = {m_p^*\over \vert dE/dx\vert} \ln
(E_p/3T) \eqno(51)$$
where $m_p^*$ is the effective parton mass, and $3T$ is the average
thermal energy of a parton.

\item{(b)} Nucleons contain partons of all possible energies.  Soft
partons, i.e. partons with $x\ll 1$ mainly contribute to matter formed
at central rapidity in the c.m. frame, while hard partons, i.e. those
with $x\simeq 0.3-1$, are the source of matter at large rapidity in
the c.m. frame.  It is quite illuminating to consider what would
happen , if it were possible to instantaneously ``shatter'' the
complete parton wave function of a fast nucleon.  Indeed, one may
ideally think of the collision of two nuclei at ultrarelativistic
energies as an event during which only the phase coherence of the
initial parton distributions is almost totally destroyed, but their
longitudinal momentum distribution remains rather unchanged.  Let us
see, therefore, what would happen if the nuclear collision acts
as such a ``phase filter''$^{83}$.
\vfill\eject

\noindent {\bf Intermezzo:  Nuclear Collisions as ``Phase Filter''}
\bigskip

We calculate the rapidity distributions of those quantities of the
parton distribution which are relevant to experimental data
regarding the final state.  One candidate is the entropy distribution,
$dS/dy$, the other is the rapidity distribution of transverse energy,
$dE_T/dy$.  For the calculation of $dS/dy$ and $dE_T/dy$ let us
consider the phase space density of partons of flavor $i$ with
degeneracy $d_i$:
$$f_{i,A}(\vec P,\vec r) = {(2\pi)^3\over d_i} {dN_A\over d^3pd^3r} =
{(2\pi)^3\over d_i} {1\over V_A^*} {dN_A\over d^3p}. \eqno(52)$$
Here $A$ denotes the mass number of the colliding nuclei and $V_A^*$
is the effective nuclear volume to be specified below.  The momentum
dependence of phase-space density can be derived from the measured
parton structure functions:
$${dN_a\over d^3p} = {1\over P}\; \tilde{F}_{i,A}(x,p_T^2) \approx
{1\over P}\; AF_i(x) {1\over 2\pi p_0^2}\;
\exp (-p_T^2\big/ 2p_0^2) \eqno(53)$$
where we assumed the factorization of the longitudinal and transverse
momenta:  $x$ is the Bjorken variable, the fraction of the nucleon's
longitudinal momentum $P$ carried by a parton.  $F_i(x)$ is
the measured parton distribution in the nucleon, where shadowing
effects are neglected here.  One normally assumes a Gaussian
distribution of width $p_0 \simeq 0.3$ GeV/c for the transverse parton
momenta.

To determine the spatial dependence of the phase space density, we use
the concept of {\it distributed contraction} of partons$^{80}$.  Let us
call $Y$ the beam rapidity and $y$ the parton rapidity in the c.m.
system.  Then only those partons occupy the Lorentz contracted region
of fast moving nucleons with radius $R_N$ which satisfy the
uncertainty relation
$$p_L \cdot {2R_N\over \cosh Y} \ge {1\over 2}. \eqno(54)$$
Here $p_L$ is the longitudinal momentum of the parton and $R_N/\cosh Y
= R_N/\gamma$ is the longitudinal size of the contracted nucleons.
Partons with smaller longitudinal momenta than (54) are more spread
out in the longitudinal direction to satisfy the Heisenberg
uncertainty relation, they form a cloud, containing mostly gluons and
sea-quarks, surrounding the contracted nucleons (see Figure 21).
\bigskip

\centerline{\epsfbox{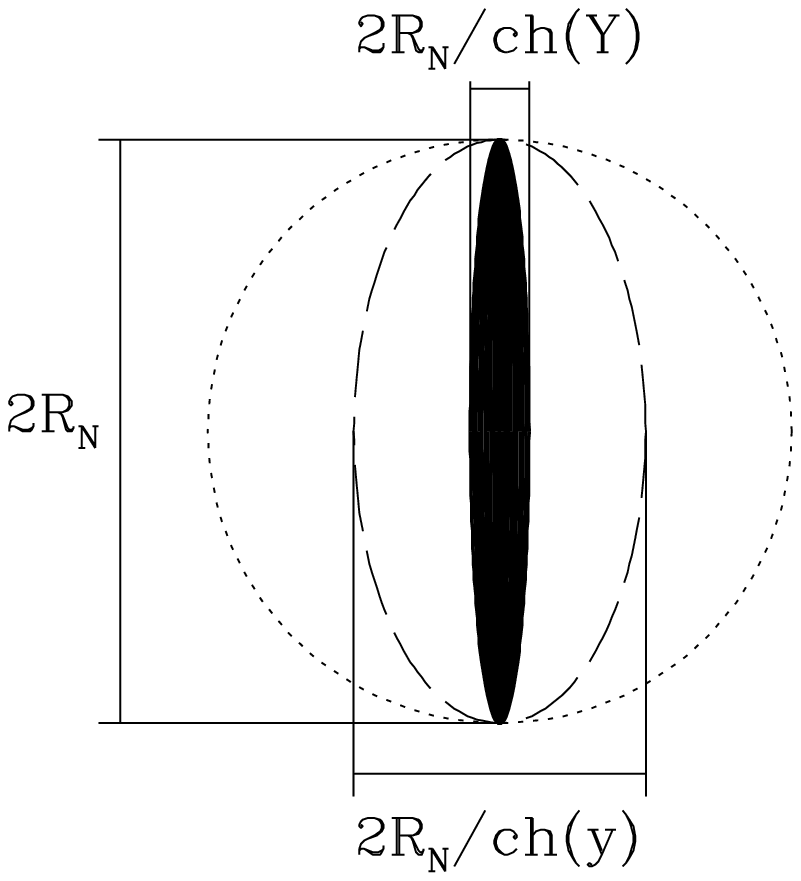}}
\noindent {\eightrm Figure 21. The longitudinal spread of the
distributions of partons with rapidity $\scriptstyle{y}$ around a
Lorentz contracted proton moving with rapidity $\scriptstyle{Y}$.
The spread is determined by the uncertainty relation.}
\bigskip

The effective spatial volume is then:
$$V_A^* (y) = {4\pi R_A^2\over 3} {R_N\over \cosh \tilde y}, \eqno(55)$$
where $\tilde y = \hbox{min}(y,Y)$.  One obtains a critical value $x_c
= \cosh Y/(4R_NP) \approx 0.1$.  For $x\le x_c$ distributed contraction
sets in. Considering the phase-space element
$${d^3pd^3r\over (2\pi)^3} = {V_A^*(y)\over (2\pi)^2}\;
p_T^2\;dp_T\;\cosh\; y, \eqno(56)$$
we can now calculate the rapidity distributions:
$$\eqalignno{ {dS\over dy} &= \int_0^{\infty} {dp_T\over (2\pi)^2}\;
V_A^*\; p_T^2\;\cosh\; y \sum_i d_i \left[ f_i \ln f_i - (1\pm f_i)
\ln (1\pm f_i)\right], &(57) \cr
{dE_T\over dy} &= \int_0^{\infty} {dp_T\over (2\pi)^2}\; V_A^*\; p_T^3 \;
\cosh\; y \sum_i d_if_i, &(58) \cr}$$
where
$$f_i = {1\over d_i} {(2\pi)^3\over V_A^*P} AF_i(x) {1\over 2\pi
p_0^2} \exp (-p_T^2\big/ 2p_0^2). \eqno(59)$$

Using the approximation, $\Sigma_iF_i(x) \approx D/x$ for sea partons,
one obtains the following expression for the integral (58):
$${dE_T\over dy} = {ADp_0\sqrt{\pi/2} \over \tanh y}. \eqno(60)$$

The evaluation of $dS/dy$ is more complicated and it is not expected
to be linear in $A$, but it only depends on $\tanh y$, as well.  These
results mean that the concept of ``{\it phase filtering}'' leads to
the prediction of a boost invariant phase-space distribution of
partons in the rapidity region $1<y<Y$, where $\tanh y\simeq 1$, as it
was postulated by Bjorken$^{84}$.  In the region $0 < y < 1$ we have a
strong rapidity dependence in expressions (57,58), because the function
$(\tanh y)^{-1}$ diverges in the limit $y\to 0$.  However, a detailed
analysis$^{83}$ shows that this is balanced by the need to introduce a
lower cut-off in the $p_T$ integrations, allowing for the continuation
of the constant rapidity distributions down to $y=0$.
\bigskip

\centerline{\epsfbox{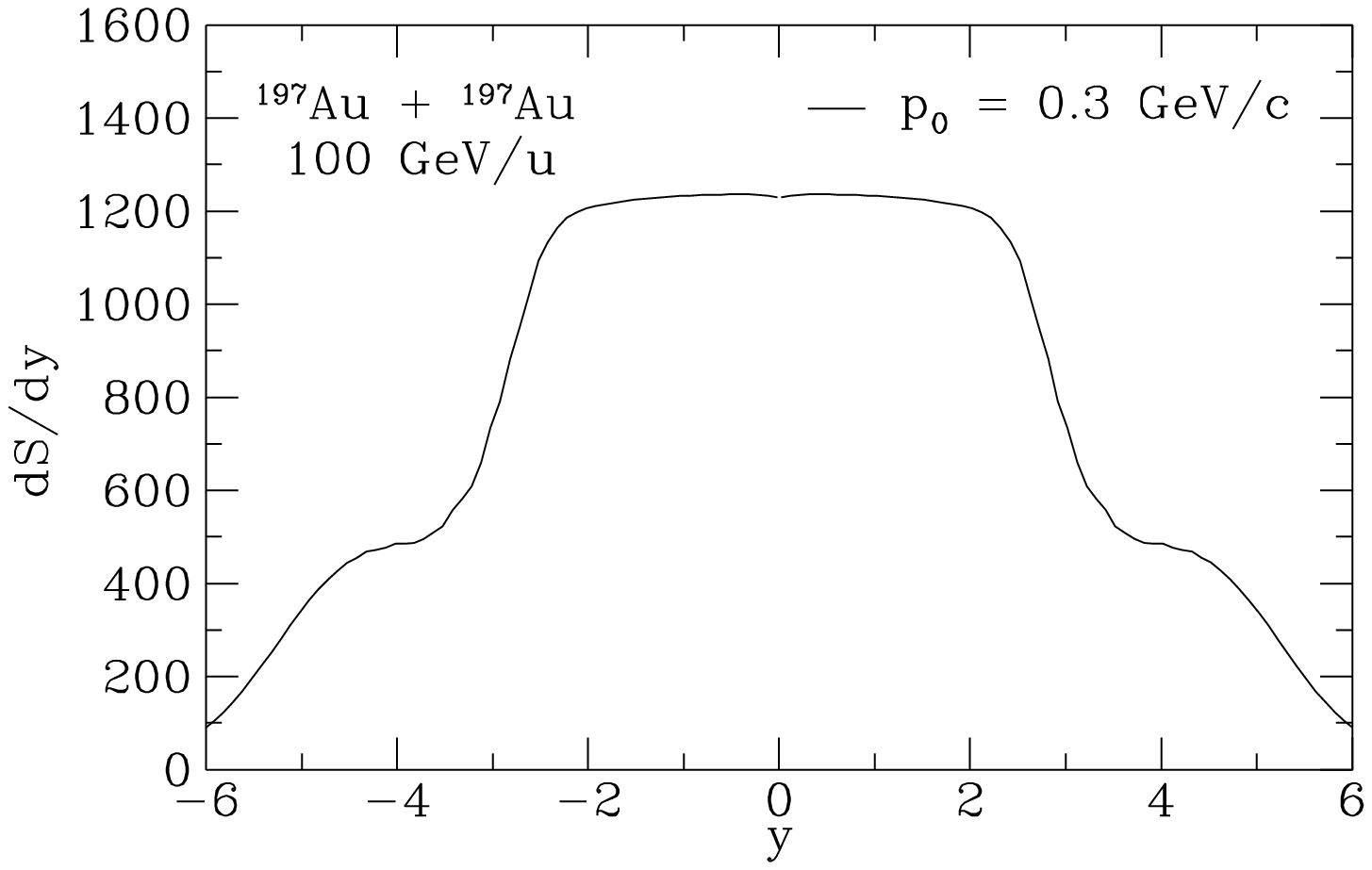}}
\noindent {\eightrm Figure 22.  Rapidity distributions of partons in
collisions at RHIC (Au+Au at $\scriptstyle{E_{cm}}$ (A=100 GeV),
assuming complete ``phase filtering'' of the initial distributions.
$\scriptstyle{p_0}$ parametrizes the initial transverse momentum
distribution of partons in the nucleon.}
\bigskip

Let us consider a Au+Au collision at the characteristic energy
of the Brookhaven RHIC machine (100 GeV/u), for which our
considerations are likely to be applicable.  As seen in Figure 22 an
extended central plateau develops, where we predict a hadron
multiplicity $dN/dy > C^{-1}(dS/dy) = 400$, with $C\simeq 3.6$ for a
thermal ultra-relativistic Bose gas.  This multiplicity may rise further
due to rescattering and fragmentation.  These effects can be studied
systematically in a complete parton cascade model$^{85}$, as we will
discuss further below.
\bigskip

\noindent {\bf Parton Cascades (continued)}
\bigskip

Having discussed the initial parton structure of the colliding nuclei,
let us examine the scattering event more closely.  When the two
Lorentz-contracted nuclei collide, some of the partons will scatter
and then continue to evolve incoherently from the remaining partons.
Three aspects of these individual parton scatterings are worth
discussing:

\item{(a)} A parton-parton scattering can be described by perturbative
QCD, if the momentum transfer involved is sufficiently large.
Nobody knows for sure where perturbative QCD becomes invalid, but
typical choices$^{85-87}$ for the momentum cut-off are
$p_T^{\hbox{\sevenrm min}} \simeq$ 1.7-2 GeV/c.

\item{(b)} The elementary scattering has a finite space-time duration,
which is given by the amount of off-shell propagation of exchanged
virtual gluons or quarks.  For the typical momentum transfer
$p_T=p_T^{\hbox{\sevenrm min}}$ this range is
$$\Delta t, \Delta x \simeq 1/p_T^{\hbox{\sevenrm min}} \simeq 0.1\;
\hbox{fm(/c)}. \eqno(61)$$
More precisely, this estimate applies to the QCD diagrams involving
$t$-channel exchange, which dominate the total parton cross-section.
For $s$-channel diagrams, e.g. $\ov{q}q$ annihilation into a lepton
pair, the elementary time scale is of order $\hat s^{-1/2}$, the
inverse scattering energy in the parton c.m. frame.

\item{(c)} The time it takes for the scattered parton wave functions to
decohere from the initial parton cloud depends on the transverse
momenta of the scattered partons.  Usually, one argues that the
partons must have evolved at least one-half transverse wavelength
$\lambda_T=\pi/p_T$ away from their original position before they can
be considered as independent quanta.  The considerations underlying
this argument are similar to those for the Landau-Pomeranchuk
effect$^{60,61}$.  The critical issue here is that parton-parton cross
sections are defined as squares of S-matrix elements which involve
integration over all space and time from the infinite past to the
infinite future.  In the course of the nuclear interaction, however,
that integration cannot be extended beyond their previous interaction
point or their point of ``formation''.  If the plane wave factors
representing the partons have not performed at least one-half complete
oscillation the limited integral is not reasonably well represented by
an S-matrix element.  This is probably the weakest point in the line
of arguments for a parton cascade and clearly would warrant more
careful study.

\noindent In any event, one concludes that a minimal time of the order of
0.1-0.3 fm/c must pass before scattered partons can be considered as
incoherent field quanta, which fully contribute to the entropy and
can rescatter as independent particles.

What happens after the initial scattering?  The longitudinal parton
momentum is little changed on average, i.e. $p_z' \simeq p_z = xP$,
hence the scattered parton appears at rapidity
$$y \simeq \ln (2p_z/p_T'), \eqno(62)$$
where $p_T'$ is its final transverse momentum.  Let us recall that the
partons with momentum $p_z$ were originally localized in a
longitudinal interval (see Figure 20):
$$\Delta z \simeq {1\over 2p_z} \simeq {1\over p_T' e^y}. \eqno(63)$$
Hence, immediately after the nuclei have collided, the fast partons
are highly localized, whereas the distribution of soft partons is more
fuzzy.  Now let us make the outrageous assumption that the cloud of
once scattered partons expands without any further interaction!
Obviously, this implies the gradual separation of partons according
to their velocity, because fast partons will leave the original
interaction site quickly, while slower partons stay behind.  Because
of the approximate boost invariance already present in the initial
state, we restrict the following discussion to the central rapidity
interval, say, $-1<y<1$.  In particular all partons with momenta
greater than $\vert p_z\vert$ will have moved away from the central
plane $z=0$ after a time
$$\tau(p_z) = {\Delta z\over v_z} \simeq {1/2p_z\over p_z/E} =
{\sqrt{p_z^2+p_T^{\prime\,2}}\over 2p_z^2}. \eqno(64)$$
Hence the longitudinal momentum spread of partons remaining at $z=0$
drops rapidly.  It becomes equal to the average transverse momentum
after:
$$\tau(p_z\simeq \langle p'_T\rangle) = {1\over \sqrt{2} \langle
p'_T\rangle} \simeq {1\over 1.4 \hbox{GeV/c}} = 0.15\; \hbox{fm/c},
\eqno(65)$$
where we have assumed an average transverse momentum of 1 GeV/c, in
line with results of numerical simulations $^{85,86}$.

At that moment the distribution of scattered partons in the most
central slab ($z \simeq 0$) is approximately isotropic.  It is also
approximately {\it thermal} ?  That depends on the actual {\it phase
space density} of partons in the vicinity of $z=0$.  Only if every
phase space cell up to $\vert\vec p\vert \simeq \langle p'_T\rangle$
is occupied with probability close to one, can we speak of a
quasi-thermal distribution.  So let us estimate the phase space
density. According to eq. (62), counting only the gluon distribution
$G(x)$, the number of partons in the original
nuclear distributions, per unit of rapidity, is:
$${dN\over dy} \simeq {dN\over d(\ln x)} \simeq 2A x G(x)
\buildrel x\to 0 \over \longrightarrow = 6A. \eqno(66)$$
Here we have neglected nuclear shadowing, since we are aiming
at a crude estimate.  Let us further assume that the nuclei are
sufficiently ``thick'' and the collision energy is high enough that
most available partons actually are scattered.  The gluon phase space
volume to be occupied is:
$${dV_p\over dy} = 16 \pi R_A^2\; \Delta z\; {\Delta p_z\over dy}\;
{\pi\langle p_T^{\prime\,2}\rangle\over (2\pi)^3}, \eqno(67)$$
where the factor 16 counts the color-spin degeneracy and $R_A = r_0
A^{1/3}$ stands for the nuclear radius.  From the uncertainty relation
(63) we have $\Delta z\cdot \Delta p_z \simeq 1$; and since we have
waited until $\Delta p_z$ has become equal to $\langle p'_T\rangle$,
the rapidity spread of partons is $\Delta y\simeq 1$.  Combining
everything, we find
$$dV_p/dy \simeq {2\over \pi} R_A^2 \langle p_T^{\prime\,2}\rangle.
\eqno(68)$$
Asking for full phase space occupation means requiring $dV_p/dy \le
dN/dy$, or
$$A^{1/6} \ge {r_0\langle p'_T\rangle\over \pi} \simeq 2, \eqno(69)$$
i.e. $A\ge 50$, if we use $\langle p'_T\rangle \simeq $ 1 GeV/c.  Of
course, not every available gluon from the initial structure functions
will scatter when the nuclei collide. On the other hand, the fragmentation
of scattered partons due to gluon bremsstrahlung effectively increases
the number of scattered partons. Quantitative predictions can be made using
a complete simulation of perturbative parton interactions, such as the
parton cascade$^{85}$ or {\cs Hijing}$^{87}$.  A calculation with
{\cs Hijing} for the system Au + Au at LHC energy was reported in ref. 88.
The result was $dN/dy \simeq 800$, $\langle p'_T\rangle \simeq 1.75$ GeV/c.
The available phase space, according to eq. (68) is:
$${dV_p\over dy} = {2\over \pi} \left( {6\hbox{fm} \cdot 1.75
\hbox{GeV} \over 0.2 \hbox{GeV fm}}\right)^2 \simeq 1500, \eqno(70)$$
i.e. phase space is about half occupied on average.  The situation is
less favorable with smaller nuclei or at lower energies.  Figure 23
shows the number of primary parton-parton collisions for
nucleon-nucleon collisions as well as for collisions of medium-sized
\bigskip

\centerline{\epsfbox{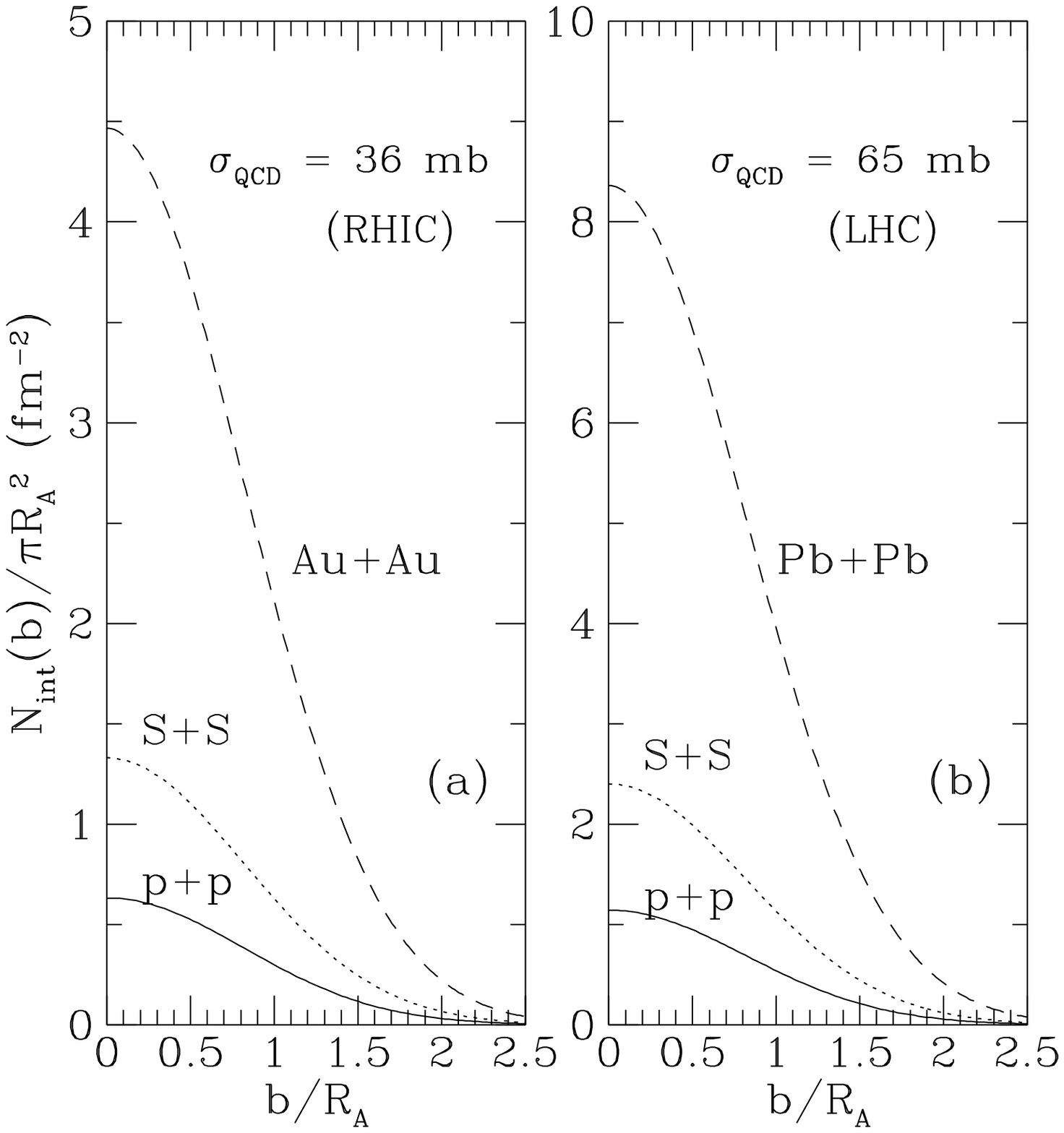}}
\noindent {\eightrm Figure 23. Area density of primary parton
scatterings for various projectiles at RHIC and LHC energies, as a
function of the reduced impact parameter b.  $\scriptstyle{R_A}$
denotes the projectile radius.}
\bigskip

\noindent and heavy nuclei, respectively, in the RHIC and LHC energy
ranges$^{86}$.  The advantage of heavy nuclei, such as Au or Pb, for
achieving a high density of scattered partons is obvious.  Sulfur
nuclei do not bring a great improvement over central nucleon-nucleon
interactions.  The increase in c.m. energy between RHIC and LHC by a
factor 30 brings about a considerable increase in the parton density,
as shown in Figure 24.  The rise is not only due to the
increased number of primary scatterings, but also partially caused by
the growing amount of initial- and final-state gluon radiation.  The
calculations also show that only a small fraction of scattered partons
are quarks.  The assumption of fully thermalized quark distributions
at very early times, $\tau \ll 1$ fm/c, is therefore most likely
unwarranted$^{89}$.

Complete calculations following the evolution of the parton
distributions microscopically until the attainment of thermal
equilibrium have been carried out recently by K. Geiger$^{85,91}$.  A
detailed account of these is given in Geiger's lecture at this School.
Suffice it to mention here that he finds almost fully thermalized
phase space distributions of gluons in Au + Au collisions at RHIC
energy ($E_{cm}$ = 100 GeV/u) with $T\simeq$ 325 MeV after proper time
$\tau \simeq$ 1.8 fm/c (see Figure 25).  If this captures the truth,
there can be no doubt that a quark-gluon plasma will be observed
in experiments at RHIC.
\bigskip

\centerline{\epsfbox{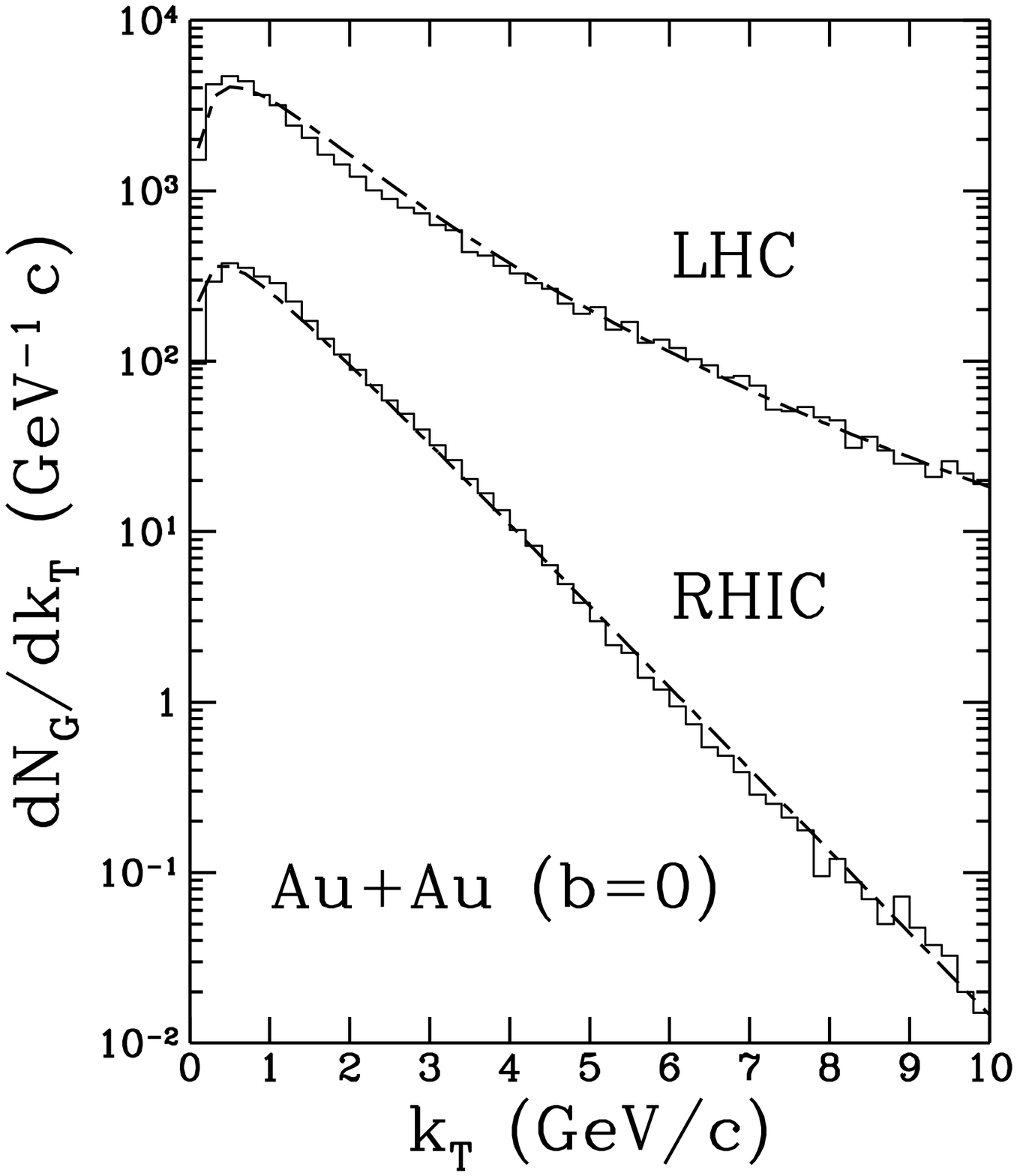}}
\noindent Figure 24. {\eightrm Distributions of scattered gluons
predicted by {\cs Hijing} for Au+Au collisions at RHIC and LHC.  The
predictions contains elementary QCD scattering cross sections and the
radiative cascades of final state partons in the leading logarithmic
approximation$^{86}$.}
\bigskip
\vskip2.5truein

\noindent{\eightrm Figure 25. Rapidity and transverse momentum
distribution of final-state partons as predicted by the parton
cascade$^{91}$.  The values of $\scriptstyle{T}$ are obtained by a fit
of the analytical isotropic fireball formula to the numerical results.}
\bigskip\medskip

\noindent {\bf Pre-Equilibrium Parton Physics}
\bigskip

The complete spectrum of phenomena occurring during the approach toward
local thermal equilibrium still awaits exploration.  Here I will
discuss just two aspects:  color screening$^{88}$ and charm
production$^{92}$, because they are dominated by the evolution of the
gluon distribution, which is better understood at this moment.  The
experimentally very interesting contribution to lepton-pair
production, which has also been studied$^{93}$, is more sensitive to
the quark distribution.
\medskip

\noindent {\it Color screening}$^{88}$:  The screening length
$\lambda_0$ of longitudinal color fields is important, because it
defines the range over which coherent color-electric fields can
extend.  If it is larger than the characteristic confinement radius
$\Lambda^{-1} \simeq$ 1 fm, long-range color-electric fields arrange
themselves as flux tubes.  In the one-loop approximation, $\lambda_D$
is determined by the momentum distribution $f$({\bf k}) of gluons
according to:
$$\lambda_D^{-2} = - {3\alpha_s\over \pi^2} \lim_{\vert \hbox{\bf q}
\vert\to 0} \int d^3k {\vert\hbox{\bf k}\vert\over \hbox{\bf q}
\cdot \hbox{\bf k}}\; \hbox{\bf q}\cdot \nabla_{\hbox{\bf k}}\;
f(\hbox{\bf k}). \eqno(71)$$

A simple result can be obtained when the gluon distribution is
exponential in $k_T$ and flat in rapidity $y$:
$$\lambda_D^{-2} \simeq {3\pi \alpha_s\over R_A^2\tau \langle
k_T\rangle}\; {dN\over dy}, \eqno(72)$$
where $dN/dy$ is the rapidity density of gluons.  At the earliest time
$t_i \simeq \langle k_T\rangle^{-1}$, when the scattered gluons become
incoherent, the calculations of ref. 88 based on the {\cs Hijing} code
predict
$$\eqalignno{ \lambda_D &\simeq 0.4\; \hbox{fm \quad (RHIC)}, \cr
\lambda_D &\approx 0.15\; \hbox{fm\quad (LHC)} . &(73)\cr}$$
We conclude that the color screening length at LHC energies will be so
short, even after the first sequence of parton interactions, that
coherent flux tubes cannot develop.  The Lund model with its formation
of independent strings that break up by creation of quark pairs is
simply not applicable under these conditions.  At RHIC, on the other
hand, the screening length appears to be marginally favoring a parton
cascade description, but models with partially fusing flux tubes
(``color ropes'') may also be able to describe certain aspects of the
pre-equilibrium phase of the nuclear collision.
\medskip

\noindent {\it Charm production}$^{92}$:
Charmed quark pairs are predominantly produced in collisions between
two gluons.  Since the charm production threshold is rather high,
about 3 GeV, the rate of production in a thermalized quark-gluon
plasma with temperature $T\simeq$ 300 MeV is negligible.  Conventional
wisdom until recently was, therefore, that most charmed quarks are
produced in primary parton interactions$^{94}$.  In view of the high
density of scattered partons with transverse momenta well above 1 GeV
at collider energies, one may suspect that there is a sizable contribution to
charm production from secondary parton collisions.  This is borne out
by a calculation of secondary charm production based on the initial
scattered gluon distribution predicted by {\cs Hijing}$^{92}$.  The
additional charmed quarks populate predominantly the central rapidity
plateau, where initial charm production is reduced by gluon shadowing
effects (see Figure 26).  At LHC energies the total yield of secondary
charmed quarks may be twice as large as that of primary charmed quarks.

The amount of secondary charm production is sensitive to the
thermalization time of the parton distribution.  It depends on the
ratio $\langle\sigma_c\rangle\big/\langle\sigma_{\hbox{\sevenrm
tot}}\rangle$, where $\langle\sigma_{\hbox{\sevenrm tot}}\rangle$ is the
average total parton-parton cross section that governs thermalization,
while $\langle\sigma_c\rangle$ denotes the averaged cross section for
charm production.  $\langle\sigma_{\hbox{\sevenrm tot}}\rangle^{-1}$
is proportional to the thermalization time $\tau_{\hbox{\sevenrm
th}}$.  A measurement of the total yield of charmed particles (mostly
D-mesons) in the central rapidity region would, therefore, provide
valuable information on the time-scale of thermalization$^{92}$.
\bigskip

\centerline{\epsfbox{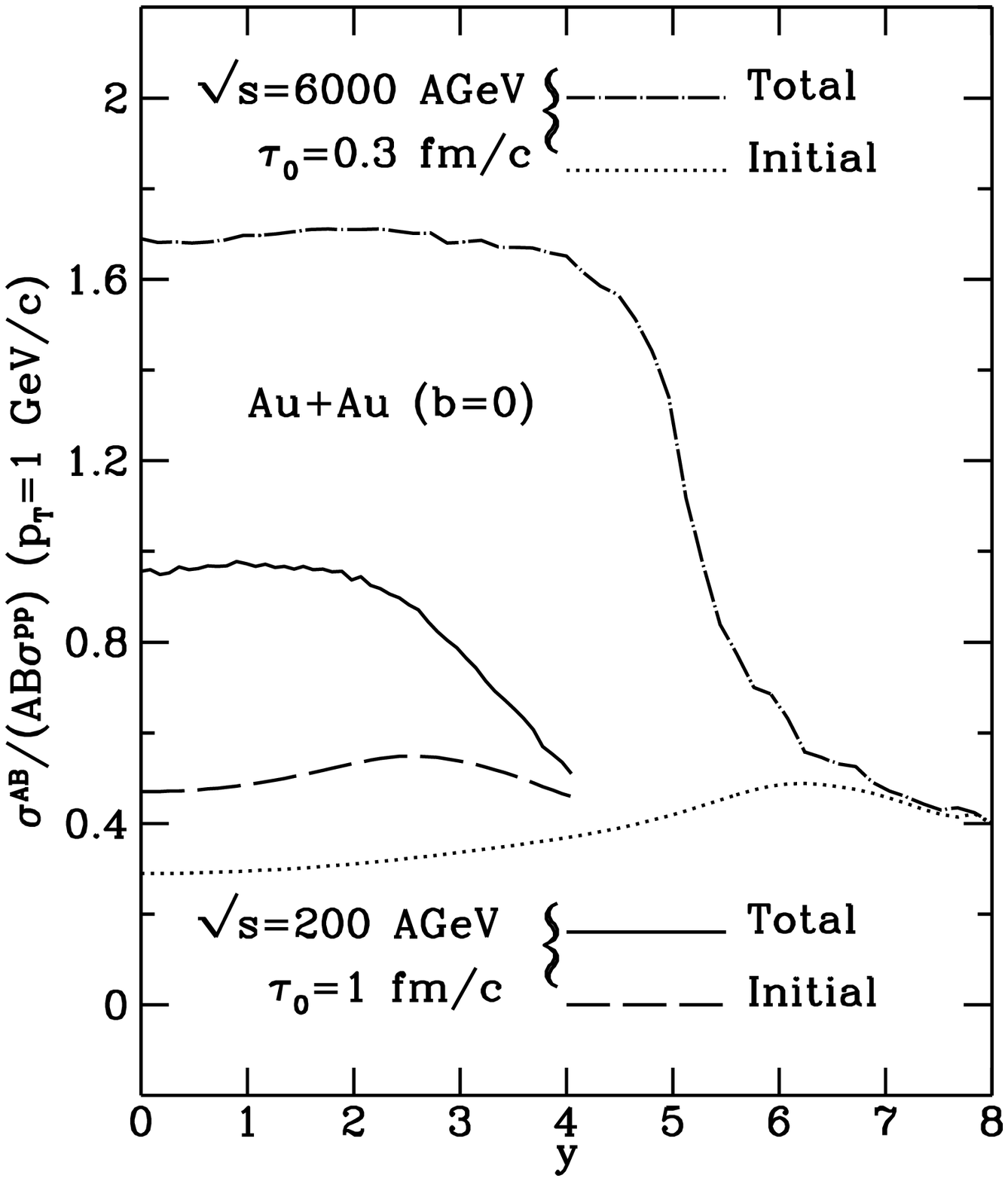}}
\noindent {\eightrm Figure 26. Rapidity distributions of charmed quarks
produced in secondary gluon interactions, in comparison to the primary
parton model predictions.  The calculations are based on Duke-Owens
structure functions with nuclear gluon shadowing.}
\bigskip\medskip

\noindent {\bf Plasma Evolution and Hadronization}
\bigskip

Once the quark-gluon plasma has reached local thermal equilibrium,
its further evolution can be described without reference to the parton
reactions at the microscopic level. This concept was first quantitatively
developed by Bjorken$^{84}$. The hydrodynamic equations for
an ultrarelativistic plasma with $P={1\over 3}\varepsilon$
admit a boost-invariant solution describing a longitudinally
expanding fireball with constant rapidity density. When transverse
expansion effects are taken into account, longitudinal boost
invariance is partially destroyed, but the overall picture
remains intact. This scenario has been thoroughly studied by
a large number of theorists. Because the results are accessible
in several fine reviews$^{95}$, I will refrain from discussing
it here in detail.

An important aspect of the late evolution of the quark-gluon plasma
is its hadronization. Mostly it is assumed that the plasma expands
and cools until it reaches the critical temperature $T_c\simeq
200 \hbox{MeV}$ and then converts into a hadronic gas while
maintaining thermal and chemical equilibrium. More detailed descriptions
of the dynamics of hadronization have been developed in connection
with the problem of strangeness production, which is reviewed in
the next chapter.

A totally different approach, explored only very recently, consists
in following the partonic reactions at a microscopic level, until
the parton density has become sufficiently low to permit the formation
of individual hadrons$^{91}$. A great deal is known about the mechanism
of final state hadron production in e$^+$e$^-$- and NN-scattering but,
unfortunately, we do not know whether this knowledge applies to the
hadronization of a quark-gluon plasma in bulk. Therefore, the treatment
of hadron formation at the end of a partonic cascade is frought with
a great deal of uncertainty.

At present, it is hard to judge the merits
of either approach on a theoretical basis. Their validity and usefulness
simply depends on whether the microscopic processes during hadronization
proceed approximately at thermodynamical equilibrium or not. The lesson
from nuclear collision processes at much lower energies (below 1 GeV/u)
has been that both scenarios are possible: neutron evaporation from
highly excited nuclear fragments is well described by thermodynamics,
but nuclear multi-fragmentation, where Coulomb forces play a dominant
role, requires a more detailed theoretical treatment.
\bigskip\medskip

\noindent {\bf QUARK-GLUON PLASMA SIGNATURES}
\bigskip

All theory of the quark-gluon plasma would be largely academic if there
were no reliable signatures to observe its formation and to study its
properties experimentally.  It is impossible to present a complete
review of quark-gluon plasms signatures here.  I will, therefore, only
try to capture the essential ideas and the current status of the
theoretical studies on the most promising quark-gluon plasma signals.  Anyone
interested in more details is referred to the review of Kajantie
and McLerran$^{96}$  and to the proceedings of the Strasbourg
workshop$^{97}$.

In order to shed some light on the connections between the many
proposed quark-gluon plasma signatures I will group them in five categories,
according to the physical properties of superdense hadronic matter to
which they are sensitive.  These are:

\item{1.} thermodynamic variables measuring the equation of state;

\item{2.}  probes for chiral symmetry restoration;

\item {3.} probes of the color response function;

\item{4.} probes of the electromagnetic response function;

\item{5.}  ``exotic'' signatures of the quark-gluon plasma.
\bigskip\medskip

\noindent {\bf Thermodynamic Variables}
\bigskip

The basic idea behind this class of signatures is to measure the
equation of state of superdense hadronic matter, i.e. the dependence
of energy density $\epsilon$, pressure $P$, and entropy density $s$ on
temperature $T$ and baryochemical potential $\mu_B$.  Here one wants
to search for a rapid rise in the effective number of degrees of
freedom, as expressed by the ratios $\epsilon/T^4$ or $s/T^3$, over a
small temperature region.  These quantities would exhibit a
discontinuity, if there were a first-order phase transition, and if we
were dealing with systems of infinite extent.  More realistically, we
can expect a steep, step-like rise.  According to recent lattice
simulations this rise should occur over a temperature range of less
than 10 MeV.

Of course, one requires measurable observables that are related to the
variables $T, s,$ or $\epsilon$.  It is customary to identify those
with the average transverse momentum $\langle p_T\rangle$, and with the
rapidity distribution of hadron multiplicity $dN/dy$, or transverse
energy $dE_T/dy$, respectively$^{98}$.  One can then, in principle,
invert the $\epsilon-T$ diagram and plot $\langle p_T\rangle$ as
function of $dN/dy$ or $dE_T/dy$.  If there occurs a rapid change in
the effective number of degrees of freedom, one expects an S-shaped
curve, as shown in Figure 27, whose essential characteristic feature is
the saturation of $\langle p_T\rangle$ during the persistence of a
mixed phase, later giving way to a second rise when the structural
change from color-singlet to colored constituents has been completed.
Detailed numerical studies in the context of the hydrodynamical model
have shown that this characteristic feature is rather weak in
realistic models, unless rehadronization occurs like an explosive
process$^{99}$.

In order to trace this curve in nuclear collisions one probably has to
vary the beam energy in rather small steps.  This has not been done,
so far, but it will be possible at RHIC.  In nucleon-antinucleon
collisions, however, one may make use of the existence of large
fluctuations in the total multiplicity even from central $N-N$
collisions.  Using this tool, the E-735 collaboration at
Fermilab$^{100}$ found a continued rise of $\langle p_T\rangle$ for
antiprotons and hyperons with multiplicity, reaching 1 GeV/c for the
most violent events $(dN/dy > 20)$.  When these data are analyzed in
terms of a simple model, where one assumes that all hadrons are emitted
from a longitudinally and transversely expanding fireball$^{101}$, one
finds that the surface velocity at high $dN/dy$ must take on quite
large values for the hadrons, reaching up to $v/c$ = 0.8.  Studying
the hydrodynamical evolution that might lead to this final state, it
is hard to believe that such a ``flow'' pattern can be produced at the
level of hadrons, because the drag exerted by the dominant pions on
the nucleons is far too weak to accelerate these to such speed.
\bigskip

\centerline{\epsfbox{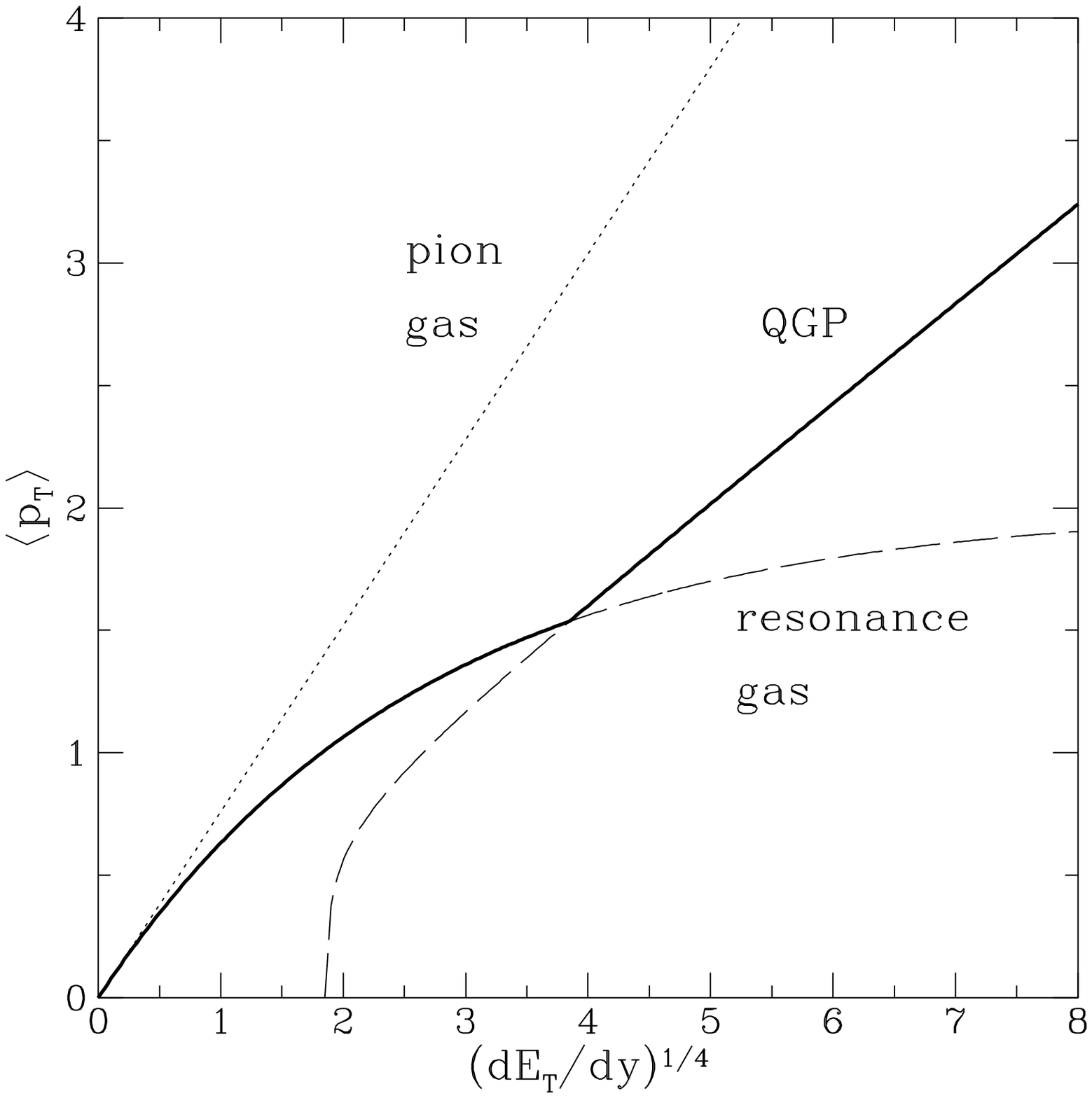}}
\noindent {\eightrm Figure 27. Average transverse momentum of emitted
hadrons as function of transverse energy $\scriptstyle{dE_T/dy}$,
representing the maximal energy density reached in a collision.  The
different curves correspond to:  (a) pion gas, (b) Hagedorn resonance
gas, (c) quark-gluon plasma.}
\bigskip

The question then is:  What produces the apparent transverse flow?
Clearly, it must be established at the quark-parton level.  It could
be a consequence of expansion of a quark-gluon plasma or mixed phase.
This could be tested by inspection of the full $p_T$ spectra at high
multiplicity$^{102}$.  Alternatively, the transverse ``flow'' might be
generated by the superposition of several extended minijets, as argued
by Gyulassy and Wang$^{103}$.  It is not entirely clear that the two
pictures are substantially different.  Minijets might be the
microscopic mechanism by which the transverse expansion of a quark-gluon plasma
is
produced.

Models of the space-time dynamics of nuclear collisions need
independent confirmation, especially concerning the correctness of
their geometrical assumptions.  Such a check is provided by identical
particle interferometry, e.g. of $\pi\pi, KK$, or $NN$
correlations$^{104}$, which yield information on the reaction
geometry.  By studying the two-particle correlation function
in different directions of phase space, it is possible to
obtain measurements of the transverse and longitudinal size, of the
lifetime, and of flow patterns of the hadronic fireball at the moment
where it breaks up into separate hadrons.  The transverse sizes found
in heavy ion collisions$^{105}$, as well as in $N-\ov{N}$ collisions
at high multiplicity$^{100}$ are larger than the radius of the incident
particle, clearly exposing the fact that produced hadrons rescatter
before they are finally emitted.  If interferometric size
determinations would be possible on an event-by-event basis when Pb or
Au beams become available, the correlation of global parameters like
$\langle p_T\rangle$ and $dN/dy$ with the fireball geometry could
be performed for each individual collision event.  This will allow
for much more precise study of the thermodynamic properties of
superdense hadronic matter and may prove to be a sharp tool in the
experimental search for a phase transition.
\bigskip\medskip

\noindent {\bf Chiral Symmetry Restoration}
\bigskip

The two most often proposed signatures for a (partial) restoration of
chiral symmetry in dense hadronic matter are enhancements in
strangeness and antibaryon production.  The basic argument in both
cases is the reduction in the threshold for production of strange
hadrons (from about 700 MeV to 300 MeV) and baryon-antibaryon pairs
(from about 2 GeV to almost zero).  As Rafelski pointed out over a
decade ago, the optimal signal is obtained by considering strange
antibaryons, which combine both signatures$^{106}$.  The enhanced
strange quark production in a chirally restored, deconfined
quark-gluon plasma$^{107}$ leads to chemical equilibrium abundances
for all strange hadrons, which would be difficult to understand on the
basis of hadronic reactions alone$^{108}$.

It has also been pointed out that strange particles, and especially
antibaryons, would be produced more abundantly, if their masses would
be modified even in the hadronic phase due to medium effects$^{109}$.
As discussed in an earlier section, the mass of K-mesons can be
substantially lowered at finite baryon number density and the
effective mass of antibaryons might be substantially reduced.

An enhancement of strange particle production in nuclear collisions
has been observed by many experiments$^{110}$.  However, we have also
learned that such an enhancement alone does not make a reliable
signature for the quark-gluon plasma.  Strange particles, especially $K$ and
$\Lambda$ can be copiously produced in hadronic reactions before the
nuclear fireball reaches equilibrium.  This mechanism seems to work
very efficiently at AGS---as well as SPS---energies$^{111,112}$.  The
cascade leading to enhanced $K$ and $\Lambda$ production has been
studied in detail in the framework of the RQMD model, where it was
found that most of the enhancement comes from $\pi N$ reactions
initiated by the pions produced in first $NN$ collisions, which have a
strong nonthermal spectrum$^{111}$.

Maybe the most spectacular data in this respect are those obtained by
the WA85 collaboration$^{113}$ at CERN, who find the following abundance
ratios at mid-rapidity and for momenta $4p_T > 1\; \hbox{GeV/c}$:
$$\eqalign{ \ov{\Lambda}\big/\Lambda &= 0.13 \pm 0.03, \cr
\Xi\big/\ov{\Lambda} &= 0.6 \pm 0.2, \cr} \qquad
\eqalign{ \ov{\Xi}\big/\Xi &= 0.39\pm 0.07; \cr
\Xi\big/ \Lambda &= 0.2 \pm 0.04. \cr}$$
It is difficult to understand why the production of the doubly strange
$\ov{\Xi}$ should be particularly enhanced in a hadronic scenario,
because the medium effects are expected to act mainly on the light
quark content of baryons.  Lattice calculations and sum rule
estimates indicate that the light $q\ov{q}$ condensate is more rapidly
depleted than the $s\ov{s}$ condensate in the medium.
Rafelski$^{114}$ has argued that the observed ratios correspond to
those found in a quark-gluon plasma that is about half equilibrated in
its strangeness content.  More recently, Rafelski and Tounsi$^{115}$
have argued that strange baryon ratios seen by WA85 and other CERN heavy
ion experiments can be consistently explained {\it either} by a
quark-gluon plasma or hadronic gas with the parameters $T$ = 220 MeV
and $\mu_B$ = 340 MeV.  Davidson et al.$^{116}$ have also pointed out
that the ratios found by WA85 are close to those of a hadronic gas with
effective volume correction in complete chemical equilibrium.  However,
even if this is so, the fundamental question raised by the data is:  how
chemical equilibrium be attained during the short life of a hadronic
fireball in any other way than through an intermediate quark-gluon
plasma phase?

Recently, attempts have been made to explain the enhanced
$\ov{\Lambda}$ production seen by NA35 at midrapidity in terms of new
mechanisms in the framework of collision models based on the string
picture.  Aichelin and Werner$^{117}$ invoke the formation of ``double
strings'' connected to the same leading quark to enhance the
production of baryons containing strange quarks in the VENUS code.  H.
Sorge et al$^{118}$ introduced a mechanism for string fusion into
``color ropes''$^{76}$, which break faster and more often produce
strange quarks and diquarks, into the RQMD model.  This leads not only
to strongly enhanced $\ov{\Lambda}$ production but also to a
significant increase in the prediction for the number of produced
antiprotons.  Unfortunately no data on $\ov{p}$ production at the
CERN-SPS are presently available to test this prediction.

Strangeness enhancement has also been seen in the $\phi$-meson channel
by the NA38 experiment$^{119}$.  Koch and Heinz$^{120}$ have argued that
this effect can be understood as addition $\phi$-production due to
rescattering of secondaries, in combination with the small absorption
cross section of the $\phi$-meson.  The required density of scatterers
agrees well with that invoked for explanation of the observed
$J/\psi$-suppression (see below).
\bigskip\medskip

\noindent {\bf Color Response Function}
\bigskip

The basic aim in the detection of a color deconfinement phase
transition is to measure changes in the color response function
$$\Pi_{\mu\nu}^{ab} (q^2) = \int d^4x\; d^4y\; e^{iq(x-y)} \langle
j_{\mu}^a(x) j_{\nu}^b(y)\rangle. \eqno(74)$$
Although this correlator is not gauge invariant (except in the limit
$q\to 0$), its structure can  be probed in two ways (see Figure 28):
\bigskip

\centerline{\epsfbox{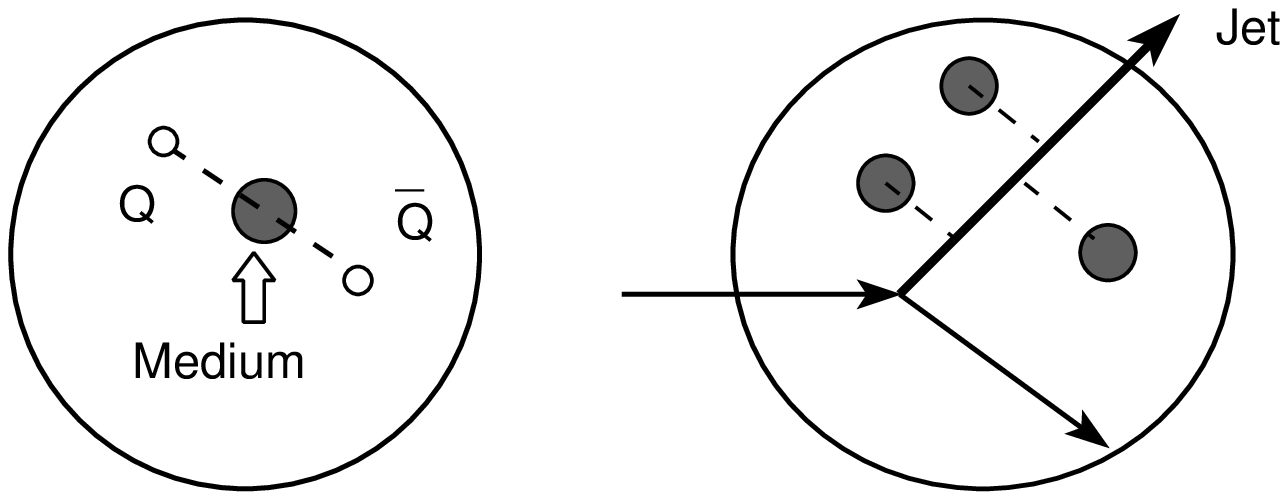}}
\noindent {\eightrm Figure 28.  Medium corrections to (a) the heavy
quark-antiquark potential and (b) the energy loss of a penetrating
parton (QCD jet) are sensitive to the color structure of dense
hadronic matter.}
\bigskip

\item{1.} The screening length $\lambda_D\delta^{ab} =
\Pi_{00}^{ab}(0)^{-1/2}$ leads to dissociation of bound states of a
heavy quark pair, such as ($c\ov{c}$).

\item{2.} The energy loss $dE/dx$ of a quark jet in a dense medium is
sensitive to an average of $\Pi_{\nu\mu}^{ab}(q^2)$ over a wide range
of $q$.
\medskip

Let me begin with the energy loss of a fast quark in the quark-gluon plasma,
which was first studied by Bjorken$^{121}$ in perturbative QCD.  The
connection between energy loss of a quark and the color-dielectric
polarizability of the medium was recently investigated by several
authors$^{62,63}$ in analogy to the theory of electromagnetic energy
loss.  The basic formula is:
$${dE\over dx} = - {C\alpha_s\over 2\pi^2v} \int d^3k\omega d\omega
\left[ {1\over k^2} \hbox{Im} {1\over \epsilon_L} + \left( v^2 -
{\omega^2\over k^2}\right) \hbox{Im} {1\over \omega^2\epsilon_T-
k^2}\right] \delta(\omega-\hbox{\bf v} \cdot \hbox{\bf k}), \eqno(75)$$
where $\epsilon_{L/T}$ denote the longitudinal and transverse
components of the color-dielectric given explicitly in eqs. (30).
$C$ is the Casimir operator for the color representation of the
penetrating particle ($C = {4\over 3}$ for quarks, $C$= 3 for gluons).  Using
eqs. (30) the expression can be evaluated analytically, yielding the
energy loss of a fast quark:
$${dE\over dx} = -{8\pi\over 3} \alpha_s^2T^2 (1+ N_f/6) \left[
{1\over v} - {1- v^2\over 2v^2}\; \ln {1+v\over 1-v} \right] \;
\ln (q_+/q_-). \eqno(76)$$
This result includes both single-particle collisional energy loss and
energy loss through excitation of plasmons.  All the medium dependence
resides in the cut-off momenta $q_{\pm}$ in the logarithm:  $q_+
\approx 2 \sqrt{TE}$, while $q_-$ is a function of the Debye screening
mass $m_D = \lambda_D^{-1}$ in the quark-gluon plasma phase.  The magnitude of
the
energy loss is critically influenced by the strong coupling constant
$\alpha_s$, whose value unfortunately is not well known.  The choices
0.2 and 0.3 preferred by various authors$^{61,62}$ leads to a stopping
power between 0.4 and 1 GeV/fm for a fast quark.  This may be a little
smaller than the energy loss of a fast quark in nuclear matter.

In addition, a fast quark loses energy by radiating gluons.  Although
this mechanism is strongly suppressed a high energy by the
Landau-Pomeranchuk effect, it still contributes of the order of 1
GeV/fm to the energy loss$^{61}$.  Adding the two contributions it
thus appears that the stopping power of a fully established quark-gluon plasma
is
probably slightly higher than that of hadronic matter.  However, in
the vicinity of the deconfinement phase transition (if it exists!)
there might be a region where the stopping power of strongly
interacting matter {\it decreases} with growing energy density.  One
would expect this effect to be particularly pronounced if the phase
transition is of second order.  The critical opalescence would in this
case strongly suppress the emission of gluon radiation from the fast
parton.  If such an effect could be observed, e.g. by varying the
transverse energy produced, it would clearly point toward a strong
structural rearrangement in dense hadronic matter, possibly toward a
quark-gluon plasma which is dominated by the propagation of collective
color modes.

The suppression of $J/\psi$ production, originally proposed by Matsui
and Satz$^{54}$, is based on a simple, yet elegant idea:  The ground
state of $(c\ov{c})$ pair does not exist when the color screening
length $\lambda_D=1/gT$ is less than the bound state radius $\langle
r^2_{J/\psi}\rangle^{1/2}$ (see also Figure 12).  Lattice simulations of
SU(3) gauge theory$^{123,124}$ show that this condition should be
satisfied slightly above the deconfinement temperature $(T/T_c >
1.2)$.  The screening length appears to be even shorter, when
dynamical fermions are included in the lattice simulations$^{125}$.
In addition, the D-meson is expected to dissociate in the deconfined
phase, lowering the energy threshold $\Delta E^*$ for thermal break-up
of the $J/\psi$.  Blaschke$^{126}$ has estimated, using the kinetic
relation
$$\sigma_{\hbox{\sevenrm diss}}^{J/\psi} \approx \sigma_0\; e^{-\Delta
E^*/T}, \eqno(77)$$
that the dissociation probability jumps significantly already at
$T_c$, and reaches unity at $T/T_c \simeq 1.2$.

The $J/\psi$ may still survive, if it escapes from the ``dangerous''
region before the $c\ov{c}$ pair has been spatially separated by more
than the size of the bound state, i.e. more than about 0.5 fm.  This
may happen, if the quark-gluon plasma cools very fast, or if the $J/\psi$
has sufficiently high transverse momentum$^{127}$:  $p_T \ge 3$ GeV/c.
The details of $J/\psi$ suppression near $T_c$ are quite complicated
and could require a rather long lifetime of the quark-gluon plasma state
before becoming clearly visible$^{128}$.

On the other hand, the $J/\psi$ may also be destroyed in a hadronic
scenario without phase transition by sufficiently energetic collisions
with comoving hadrons$^{129}$, leading to dissociation into a pair of
D-mesons.  This mechanism has recently been analyzed carefully by
Gavin$^{130}$ and by Vogt et al.$^{131}$.  In addition, the dependence
of the suppression factor $S$ on transverse momentum of the $J/\psi$
is explained by a broadening of the transverse momentum distribution of
projectile gluons due to prescattering$^{132}$.   Drell-Yan data$^{133}$
of Fermilab experiment E772 indicate that this proceeds like a random
walk leading to a broadening $\Delta\langle p_T^2\rangle$ which grows
like $A^{-1/3}$. These effects combine to
explain most of the NA38 data.  The result of these studies$^{130,131}$
is that the pattern of $J/\psi$ suppression observed in experiment
NA38 at CERN$^{134}$ can be understood on the basis of ``standard''
hadronic interactions, if one assumes comoving hadronic matter at
density of at least 1/fm$^3$ and an absorption cross section of the
order of 2 mb.

Karsch and Satz have recently analyzed whether this ambiguity persists
at RHIC and LHC energies$^{135}$.  Assuming that a hadronic phase can be
formed at energy densities of $4 - 7.5$ GeV/fm$^3$, they find little
difference in $S(p_T)$ in the accessible $p_T$ range.  However, they
predict a substantial difference in the suppression of the $\Upsilon$
resonance at LHC energies$^{136}$.
\bigskip

\noindent {\bf Electromagnetic Response Function}
\bigskip

Electromagnetic signals for the quark-gluon plasma are in many respects
ideal because they probe the earliest and hottest phase of the
evolution of the fireball, and are not affected by final state interactions.
Their drawbacks are (a) the rather small count rates and (b) the relatively
large backgrounds from hadronic decay processes, especially $\pi^0$
and $\eta$ decays. Electromagnetic signals probe the structure of the
electromagnetic current response function:
$$\Pi_{\mu\nu} (q^2) = \int d^4 xd^4 y\; e^{iq(x-y)} \langle j_{\mu} (x)
j_{\nu} (y)\rangle. \eqno(78)$$
In the hadronic phase, $\Pi_{\mu\nu}(q^2)$ is dominated by the $\rho^0$
resonance at 770 MeV, whereas perturbative QCD predicts a broad
continuous spectrum above twice the thermal quark mass $m_q =
gT/\sqrt{6}$.  At low $q^2\ll$ 100 MeV collective modes are predicted
to exist in both phases.  In first approximation the collective
quark-gluon plasma excitation, the {\it plasmino}$^{137}$, has a somewhat
higher effective mass than the collective $\pi^+\pi^-$ mode$^{138}$,
but its influence is hidden under strong nonresonant effects of
soft QCD interactions in the plasma that cause a strong increase
at low $q^2$ (see Figure 29).
Unfortunately, these interesting modifications below the mass of
the vector mesons will probably be overwhelmed by background
from Dalitz pairs$^{139}$.

On the other hand, the production of lepton pairs with large
invariant mass in the quark-gluon plasma phase
may be sensitive to pre-equilibrium phenomena,
e.g. collective plasma oscillations of large amplitude$^{78}$.
Such oscillations are known to occur in the framework of the
chromo-hydrodynamic model where the collision energy is first stored
in a coherent color field which later breaks up into $q\ov{q}$ pairs.
The ensuing collective flow could enhance the production of lepton
pairs of high invariant mass.
\bigskip

\centerline{\epsfbox{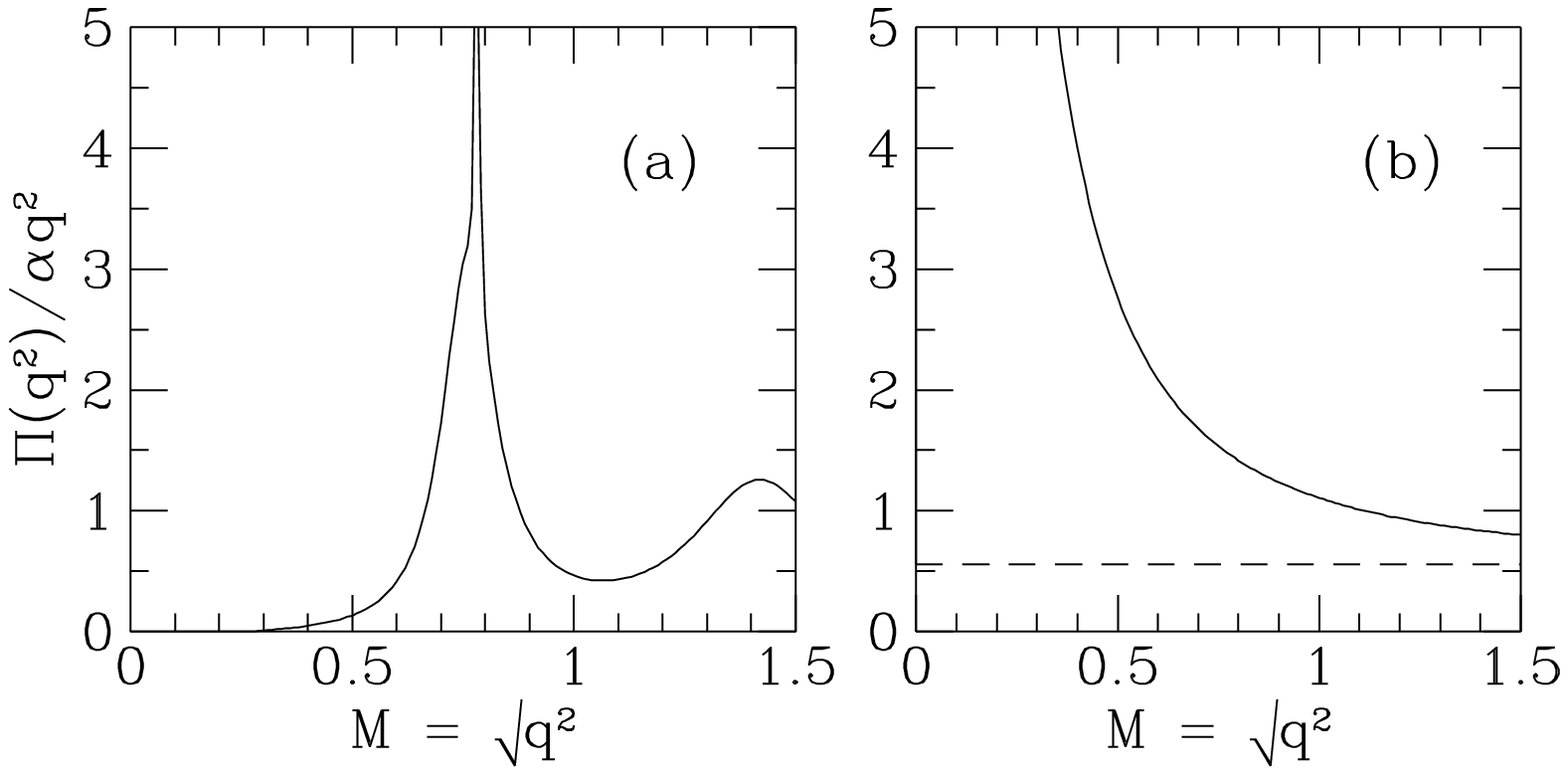}}
\noindent {\eightrm Figure 29. Electromagnetic response function of
dense hadronic matter (a) and quark-gluon plasma (b), as function of
the invariant mass $\scriptstyle{\sqrt{q^2}}$ of a virtual photon.
The hadronic
response function is dominated by the neutral vector meson resonances
$\scriptstyle{\rho^0, \omega}$, and $\scriptstyle{\phi}$.
The dashed line in (b) shows the contribution from free quarks,
while the solid line includes QCD interactions
($\scriptstyle{\alpha_s=0.3}$). Both cases exhibit collective modes
at low $\scriptstyle{q^2}$, which are not shown here.}
\bigskip

The suggestion by Siemens and Chin$^{140}$ that the disappearance of
the $\rho$-meson peak in the lepton pair mass spectrum would signal
the deconfinement transition has recently been revived$^{141}$.  The
basic idea is to utilize the fact that the quark-gluon plasma phase should
exhibit
the higher temperature than the hadronic phase, and therefore lepton
pairs from the quark-gluon plasma should dominate at high $p_T$ over those
originating from hadronic processes.  Unfortunately, the reasoning probably
breaks down when one allows for collective transverse flow.  Because
of its larger mass, the $\rho$-meson spectrum is much more sensitive
to the presence of flow than the quark spectrum in the quark-gluon plasma
phase$^{142}$.

Nonetheless, the lepton pairs from $\rho$-meson decay can be a very
useful tool for probing the hadronic phase of the fireball.  Heinz and
Lee$^{143}$ have pointed out that the $\rho$-peak is expected to grow
strongly relative to the $\omega$- and $\phi$- peaks in the electron
pair mass spectrum, if the fireball lives substantially longer than 2
fm/c.  This occurs because of the short average lifetime of the $\rho$
(1.3 fm/c), so that several generations of thermal $\rho^0$-mesons
would contribute to the spectrum.  In the limit of a very long-lived
fireball the ratio of lepton pairs from $\rho^0$- and $\omega$- decays
would approach the ratio of their leptonic decay widths (11:1).  The
$\rho/\omega$- ration can therefore serve as a fast ``clock'' for the
fireball lifetime.

The widths and positions of the $\rho$, $\omega$, and $\phi$ peaks
should also be sensitive to medium induced changes of the hadronic
mass spectrum, especially to precursor phenomena associated with
chiral symmetry restoration.  This has been  studied extensively$^{144}$.
The general conclusion, however, is that
these modifications are probably small except in the immediate
vicinity of the phase transition.  Changes are predicted to occur
sooner, if the hadronic phase contains an appreciable net baryon
density.  E.g. a change in the K-meson mass could be detected via the
induced change in the width of the $\phi$-meson$^{145}$.

Direct photons, the second electromagnetic probe of dense matter, must
face the formidable background from $\pi^0$ and $\eta$- decays.
Whether these decays can be reconstructed and subtracted with
sufficient reliability remains questionable, despite the remarkable
achievements in this respect by experiment WA80 at CERN$^{146}$.  But
even if it were possible, it is not clear what direct photons would
tell.  A new calculation by photon emission from hadronic matter and
quark-gluon plasma at the same temperature ($T$=200 MeV) by Kapusta,
Lichard and Seibert$^{147}$ has yielded virtually idential results for
the two scenarios.  This is by no means trivial, because the fireball is
optically thin.  The result is a consequence of the presence of
thermal $\rho$-mesons and only occurs when the process $\pi\rho \to
\pi\gamma$ is taken into account.

Altogether, the prospects for an unambiguous quark-gluon plasma signal
from the electromagnetic sector are doubtful.  Moreover, the count rates
predicted at the future heavy ion colliders (RHIC and LHC) are quite
small.  The situation is reviewed in more detail by V.
Ruuskanen$^{148}$.  However, one should recall that electromagnetic
probes are most sensitive to the earliest phase of a nuclear
collision$^{90}$, and thus might prove to be a valuable tool for the
detection of pre-equilibrium phenomena$^{93}$, even if they should
eventually turn out not to be good probes of the thermal quark-gluon plasma.
\bigskip\medskip

\noindent {\bf Exotica}
\bigskip

It would be nice if the formation of quark-gluon plasma would be associated
with the
appearance of completely novel phenomena:  there would be no ambiguity
in such signatures.  Indeed one should remember that the
proposal to look for quark-gluon plasma in nuclear collisions$^{149}$ was
originally
derived from the apparent existence of unexplained phenomena observed
in cosmic ray interaction, such as the famous ``Centauro'' events.
The most probable exotic objects that might be formed from quark-gluon plasma
are
{\it strangelets}$^{46-48}$.  As explained earlier, this name
describes metastable objects with baryon number $A\ge 2$ that contain
several strange quarks.  The simplest such object is the strangeness
$S=-2$ dibaryon, the $H$-particle, which is predicted to be metastable
in the original MIT bag model$^{151}$ and might be produced in relativistic
nuclear collisions$^{152}$.  Experiments$^{49}$ searching for
strangelets produced in relativistic heavy ion reactions are in
progress at BNL, and in preparation at CERN.

Recently there has been speculation about the possible formation of
locally ``disoriented'' chiral vacua in relativistic nuclear
collisions$^{153}$.  Such states would decay into a large number of
pions, possibly with a strong isospin imbalance as observed in the
Centauro events.  They might be produced by some not well understood
collective emission process, or by spontaneous symmetry breaking when
the dense hadronic system returns from the chirally restored phase.
\bigskip\medskip

\noindent {\bf CONCLUSIONS}
\bigskip

It is appropriate to conclude this review of quark-gluon plasma
physics by emphasizing the positive aspects.  Many of the proposed
quark-gluon plasma signals have actually been observed already in the
present experiments:  $J/\psi$ suppression; enhanced production of
strange hadrons, most notably of strange antibaryons; increase in
transverse momenta of emitted particles.  None of these results has been
demonstrated to be an unambiguous signal of the quark-gluon plasma, so far.
However, one should bear in mind that the experiments were all performed
with systems that were too small (Si and S are hardly ``heavy'' nuclei)
and at energies too low to expect the formation of a full-fledged,
sufficiently long-lived quark-gluon plasma state.  In view of this,
the experimental results are encouraging.

On the theoretical side, a better understanding of what we really mean
by a ``quark-gluon plasma signature'' is required$^{154}$.  In
practical terms, we need a consistent formulation of what precisely is
measured by $J/\psi$-suppressing and antihyperon enhancement.  How do
these signatures depend on the color response function or the quark
correlation function $\langle q\ov{q}\rangle$ in the medium?  Finally,
we need to understand how pre-equilibrium processes influence
predictions for the proposed signatures.  These are difficult
questions but, as I have tried to show, our understanding of the
physics of the quark-gluon plasma has progressed to the point where
these problems can be seriously addressed.
\bigskip\medskip

\noindent {\bf Acknowledgements}
\bigskip

I want to thank the organizers of the NATO-Advanced Study Institute
at Il Ciocco, especially H. Gutbrod and J. Rafelski for providing
opportunities of intense discussion that helped clarify several points
contained in these lecture notes.  I am deeply indebted to T. Bir\'o,
K. Geiger, C. Gong, P. L\'evai, J. Rau, M. Thoma, A. Trayanov,
and X. N. Wang, who helped develop many of the ideas presented here.
I thank W. A. Zajc for pointing out several errors. This work was partially
supported by the U.S. DOE (Grant DE-FG05-90ER40592), NCSC and NATO.

\bigskip

\noindent {\b REFERENCES}
\bigskip

{\frenchspacing
\item{1.} B. M\"uller, {\sl The Physics of the Quark-Gluon Plasma},
Lecture Notes in Physics, Vol. 225 (Springer-

\itemitem{}Verlag, Berlin-Heidelberg 1985).

\item{2.} L. McLerran, {\sl Rev. Mod. Phys. {\bf 58}}, 1021 (1986).

\item{3.} {\sl Quark-Gluon Plasma}, edited by R. C. Hwa
(World Scientific, Singapore, 1991).

\item{4.} For a recent review of cosmological implications of the
quark-gluon phase transition, see:

\itemitem{} K. A. Olive, {\sl Science {\bf
251}}, 1194 (1991) and references therein.

\item{5.} See e.g.: E. W. Kolb and M. S. Turner, {\sl The Early
Universe} (Addison-Wesley, Redwood City,

\itemitem{} 1990), Chap. 3.5.

\item{6.} A. Guth, {\sl Phys. Rev. {\bf D23}}, 347 (1981).

\item{7.} L. F. Abbot and S. Y. Pi, {\sl Inflationary Cosmology},
Reprint volume (World Scientific, Singapore,

\itemitem{} 1986).

\item{8.} J. Ellis, J. I. Kapusta, and K. A. Olive, {\sl Nucl. Phys.
{\bf B348}}, 345 (1991).

\item{9.} N. K. Glendenning, {\sl Phys. Rev. Lett. {\bf 63}}, 2629
(1989).

\item{10.} R. Hagedorn, in: {\sl Quark Matter '84}, ed. by K.
Kajantie, {\sl Lecture Notes in Physics } Vol. 221, p. 53

\itemitem{} (Springer-Verlag, Berlin-Heidelberg, 1985).

\item{11.} J. D. Walecka, {\sl Phys. Lett. {\bf 59B}}, 109 (1975); J.
Theis, et al., {\sl Phys. Rev. {\bf D28}}, 2286 (1983).

\item{12.} J. Gasser and H. Leutwyler, {\sl Phys. Lett. {\bf 188B}},
477 (1987).

\item{13.} P. Carruthers, {\sl Collective Phenomena, {\bf 1}}, 147
(1973).

\item{14.} J. C. Collins and M. Perry, {\sl Phys. Rev. Lett. {\bf
34}}, 1353 (1975).

\item{15.} G. Baym and S. A. Chin, {\sl Phys. Lett. {\bf 62B}}, 241
(1976);  S. A. Chin, {\sl Phys. Lett. {\bf 78B}}, 552 (1978).

\item{16.} B. Friedman and L. McLerran, {\sl Phys. Rev. {\bf D17}},
1109 (1978); L. D. McLerran, {\sl Phys. Rev. {\bf D24}},

\itemitem{} 450 (1981).

\item{17.} E. V. Shuryak, {\sl Phys. Rep. {\bf 61}}, 71 (1980).

\item{18.} R. Hagedorn, {\sl Suppl Nuovo Cimento {\bf 3}}, 147 (1965).

\item{19.} See e.g.: M. B. Green, J. H. Schwarz, and E. Witten,
{\sl Superstring Theory}, Vol. 1, chap. 2.3.5.

\itemitem{}  (Cambridge University Press, Cambridge, 1987).

\item{20.} J. Kapusta, {\sl Phys. Rev. {\bf D23}}, 2444 (1981).

\item{21.} K. Sailer, B. M\"uller, and W. Greiner, {\sl Int. J. Mod.
Phys. {\bf A4}}, 437 (1989).

\item{22.} N. Cabibbo and G. Parisi, {\sl Phys. Lett. {\bf 59B}}, 67
(1975).

\item{23.} R. Hagedorn and J. Rafelski, {\sl Phys. Lett. {\bf 97B}},
136 (1980).

\item{24.} E. Shuryak, {\sl Nucl. Phys. {\bf A533}}, 761 (1991); E.
Shuryak and V. Thorsson, {\sl Nucl. Phys. {\bf A536}}, 739

\itemitem{} (1992).

\item{25.} C. Gong, {\sl J. Phys. {\bf B18}}, L123 (1992).

\item{26.} T. Abbott, et al. [E-802 Collaboration], {\sl Phys. Rev.
Lett. {\bf 64}}, 847 (1990).

\item{27.} G. E. Brown and M. Rho, {\sl Phys. Rev. Lett. {\bf 66}},
2720 (1991).

\item{28.} See e.g.: L. J. Reinders, H. R. Rubinstein, and S. Yazaki,
{\sl Phys. Rep. {\bf 127}}, 1 (1985).

\item{29.} R. J. Furnstahl, T. Hatsuda, and S. H. Lee, {\sl Phys. Rev.
{\bf D42}}, 1744 (1990).

\item{30.} G. Adami, T. Matsuda, and I. Zahed. {\sl Phys. Rev. {\bf
D43}}, 921 (1991).

\item{31.} T. D. Cohen, R. J. Furnstahl, and D. K. Griegel, {\sl Phys.
Rev. {\bf C45}}, 1881 (1992).

\item{32.} A. X. El-Khadra, G. Hockney, A. S. Kronfeld, and P. B.
Mackenzie, {\sl Phys. Rev. Lett. {\bf 69}}, 729

\itemitem{} (1992).

\item{33.} M. Creutz, {\sl Quarks, Gluons and Lattices} (Cambridge
University Press, Cambridge, 1983).

\item{34.} F. R. Brown, et al., {\sl Phys. Rev. Lett. {\bf 65}}, 2491
(1990);  see also the lecture by F. Karsch at this

\itemitem{} School.

\item{35.} See e.g. the lecture by L. P. Csernai at this School.

\item{36.} E. Farhi and R. L. Jaffe, {\sl Phys. Rev. {\bf D30}}, 2379
(1984); I. Mardor and B. Svetitsky, {\sl Phys. Rev.

\itemitem{} {\bf D44}}, 878 (1991).

\item{37.} S. Huang, J. Potvin, C. Rebbi, and S. Sanielevici, {\sl
Phys. Rev. {\bf D42}}, 2864 (1990); {\bf D43}, 2056E

\itemitem{} (1991);
K. Kajantie, L. Karkkainen, and K. Rummukainen, {\sl Nucl. Phys. {\bf
B357}}, 693 (1991).

\item{38.} L. P. Csernai and J. I. Kapusta, preprint TPI-MINN-92-10-T,
University of Minnesota, March

\itemitem{} 1991; {\sl Phys. Rev. Lett. {\bf 69}},
737 (1992).

\item{39.} G. Lana and B. Svetitsky, {\sl Phys. Lett. {\bf B285}}, 251
(1992).

\item{40.} J. Potvin, Talk at the Workshop on QCD Vacuum Structure,
Paris, June 1992 (to be published).

\item{41.} E. Witten, {\sl Phys. Rev. {\bf D30}}, 272 (1984).

\item{42.} J. H. Applegate and C. J. Hogan, {\sl Phys. Rev. {\bf
D31}}, 3037 (1985); {\bf D34}, 1938 (1986); C. Alcock,

\itemitem{} G. M. Fuller, and G. J. Mathews, {\sl Astrophys. J.
{\bf 320}}, 439 (1987).

\item{43.} M. Kurki-Suonio and R. Matzner, {\sl Phys. Rev. {\bf D39}},
1046 (1989); {\sl Phys. Rev. {\bf D42}}, 1047 (1990);

\itemitem{} H. Kurki-Suonio, et al.,
{\sl Astrophys. J. {\bf 353}}, 406 (1990).

\item{44.} G. J. Mathews, B. Meyer, C. R. Alcock, and G. M. F\"uller,
{\sl Astrophys. J. {\bf 358}}, 406 (1990).

\item{45.} A. Bodmer, {\sl Phys. Rev. {\bf D4}}, 1601 (1971); S. Chin
and A. Kerman, {\sl Phys. Rev. Lett. {\bf 43}}, 1291

\itemitem{} (1978); E. Farhi
and R. Jaffe, {\sl Phys. Rev. {\bf D30}}, 2379 (1984).

\item{46.} H. Liu and G. Shaw, {\sl Phys. Rev. {\bf D30}}, 1137 (1984).

\item{47.} C. Greiner, P. Koch, and H. St\"ocker, {\sl Phys. Rev.
Lett. {\bf 58}}, 1825 (1987);  {\sl Phys. Rev. {\bf D38}}, 2797

\itemitem{} (1988).

\item{48.} C. Greiner, D. Rischke, H. St\"ocker, and P. Koch, {\sl Z.
Phys. {\bf C38}}, 283 (1988); C. Greiner and

\itemitem{} H. St\"ocker, {\sl Phys. Rev. {\bf D44}}, 3517 (1991);
G. Shaw, G. Benford, and D. Silverman, {\sl Phys. Lett. {\bf 169B}},
275 (1986).

\item{49.} Brookhaven Experiments E814 (J. Barrette, et al., {\sl Phys.
Lett. {\bf B252}}, 550 (1990)), E864

\itemitem{} (J. Sandweis, et al.); E878 (H. J. Crawford, et al.);
CERN-Experimental Proposal SPSLC/ P-268 (K. Pretzl, et al.).

\item{50.} G. Shaw, M. Shin, R. Dalitz, and M. Desai, {\sl Nature {\bf
337}}, 436 (1989); M. S. Desai and G. Shaw,

\itemitem{} Technological Implications of Strange Quark Matter,
to appear in {\sl Nucl. Phys. B.}

\item{51.} J. Kapusta, {\sl Nucl. Phys. {\bf B148}}, 461 (1979).

\item{52.} V P. Silin, {\sl Sov. Phys. JETP {\bf 11}}, 1136 (1960); V.
V. Klimov, {\sl Sov. Phys. JETP {\bf 55}}, 199 (1982);

\itemitem{} H. A. Weldon, {\sl Phys. Rev. {\bf D26}}, 1394 (1982).

\item{53.} A. Billoire, G. Lazarides, and Q. Shafi, {\sl Phys. Lett.
{\bf 103B}}, 450 (1981); T. A. DeGrand and

\itemitem{} D. Toussaint, {\sl Phys. Rev. {\bf D25}}, 526 (1982).
These authors find a coefficient $C=0.27 \pm$ 0.03 for SU(2) gauge theory.

\item{54.} T. Matsui and H. Satz, {\sl Phys. Lett {\bf 178B}}, 416
(1986).

\item{55.} R. D. Pisarski, {\sl Phys. Rev. Lett. {\bf 63}}, 1129
(1989); E. Braaten and R. D. Pisarski {\sl; Phys. Rev. {\bf D42}},

\itemitem{} 2156 (1990).

\item{56.} A. Ukawa, {\sl Nucl. Phys. {\bf A498}}, 227c (1989).

\item{57.} T. Bir\'o, P. L\'evai, and B. M\"uller, {\sl Phys. Rev. {\bf
D42}}, 3078 (1990).

\item{58.} M. H. Thoma, {\sl Phys. Lett. {\bf B269}}, 144 (1991); G.
Baym, H. Monien, C. J. Pethick, and

\itemitem{} D. G. Ravenhall, {\sl Phys. Rev. Lett. {\bf 64}}, 1867 (1990).

\item{59.} E. Shuryak, {\sl Phys. Rev. Lett. {\bf 68}}, 3270 (1992).

\item{60.} A. B. Migdal, {\sl Sov. Phys. JETP {\bf 5}}, 527 (1957).

\item{61.} A. H. S\o rensen, {\sl Z. Phys. {\bf C53}}, 595 (1992); M.
Gyulassy and X. N. Wang, to be published.

\item{62.} M. H. Thoma and M. Gyulassy, {\sl Nucl. Phys. {\bf B351}},
491 (1991); E. Braaten and M. H. Thoma,

\itemitem{} {\sl Phys. Rev. {\bf D44}}, R2525 (1991).

\item{63.} S. Mr\'owczy\'nski, {\sl Phys. Lett. {\bf B269}}, 383 (1991).

\item{64.} N. S. Krylov, {\sl Works on the Foundation of Statistical
Physics} (Princeton University Press,

\itemitem{} Princeton, 1979); A. N. Kolmogorov, {\sl Dokl. Akad. Nauk
SSSR {\bf 119}}, 861 (1958) and {\bf 124}, 754 (1959); Ya. G. Sinai,
{\sl Dokl. Akad. Nauk SSSR {\bf 124}}, 768 (1959) and {\bf 125}, 1200
(1959).  See also: G. M. Zaslavsky, {\sl Chaos in Dynamic Systems}
(Harwood, Chur, 1985).

\item{65.} Gun He, {\sl Phys. Lett. {\bf A149}}, 95 (1990).

\item{66.} S. G. Matinyan, G. K. Savvidy, and N. G.
Ter-Arutyunyan-Savvidy, {\sl Sov. Phys. JETP {\bf 53}}, 421

\itemitem{} (1981); {\sl JETP Lett. {\bf 34}}, 590 (1981); See also:
C. Gong, B. M\"uller, and A. Trayanov, preprint DUKE-TH-92-34 and references
therein.

\item{67.} B. M\"uller and A. Trayanov, {\sl Phys. Rev. Lett. {\bf
68}}, 3387 (1992).

\item{68.} C. Gong, preprint DUKE-TH-92-41.

\item{69.} B. Anderson, G. Gustafson, G. Ingelman, and T. Sj\"ostrand,
{\sl Phys. Rep. {\bf 97}}, 31 (1983).

\item{70.} B. Nilsson-Almquist and E. Stenlund, {\sl Comp. Phys. Comm.
{\bf 43}}, 387 (1987).

\item{71.} M. Gyulassy, preprint CERN-TH-4784 (1987, unpublished).

\item{72.} T. Cs\H{o}rg\H{o}, J. Zim\'anyi, J. Bondorf, and H. Heiselberg,
{\sl Phys. Lett. {\bf B222}}, 115 (1989).

\item{73.} K. Werner, {\sl Z. Phys. {\bf C42}}, 85 (1989).

\item{74.} N. S. Amelin, K. K. Gudima, and V. D. Toneev, {\sl Yad.
Fiz. {\bf 51}}, 512 (1990).

\item{75.} M. Sorge, H. St\"ocker, and W. Greiner, {\sl Nucl. Phys.
{\bf A498}}, 567c (1989);  {\sl Ann. Phys. {\bf 192}}, 266

\itemitem{} (1989).

\item{76.} T. S. Bir\'o, H. B. Nielsen, and J. Knoll, {\sl Nucl. Phys.
{\bf B245}}, 449 (1984).

\item{77.} A. Bia\l as and W. Czy\'z, {\sl Phys. Rev. {\bf D31}}, 198
(1985); {\sl Nucl. Phys. {\bf B267}}, 242 (1986); S. Kagiyama,

\itemitem{} A. Nakamura, and A. Minaka, {\sl Prog. Theor. Phys. {\bf 75}},
319 (1986).

\item{78.} K. Kajantie and T. Matsui, {\sl Phys. Lett. {\bf B164}},
373 (1985); G. Gatoff, A. K. Kerman, and

\itemitem{} T. Matsui, {\sl Phys. Rev. {\bf D36}}, 114 (1986); M. Asakawa
and T. Matsui, {\sl Phys. Rev. {\bf D43}}, 2871 (1991); G. Gatoff,
preprint ORNL/CCIP/91/24, Oak Ridge, 1991.

\item{79.} D. Boal, {\sl Phys. Rev. {\bf C33}}, 2206 (1986).

\item{80.} R. C. Hwa and K. Kajantie, {\sl Phys. Rev. Lett. {\bf 56}},
696 (1986).

\item{81.} J. P. Blaizot and A. H. Mueller, {\sl Nucl. Phys. {\bf
B289}}, 847 (1987).

\item{82.} F. Niedermayer, {\sl Phys. Rev. {\bf D34}}, 3494 (1986).

\item{83.} P. L\'evai and B. M\"uller, preprint DUKE-TH-90-10.

\item{84.} J. D. Bjorken, {\sl Phys. Rev. {\bf D27}}, 140 (1983).

\item{85.} K. Geiger and B. M\"uller, {\sl Nucl. Phys. {\bf B369}},
600 (1992);  see also lecture by K. Geiger at this

\itemitem{} School.

\item{86.} (a) T. Sj\"ostrand and M. van Zijl, {\sl Phys. Rev. {\bf
D36}}, 2019 (1987);

\item{} (b) N. Abou-El-Naga, K. Geiger, and B. M\"uller,
{\sl J. Phys. {\bf G18}}, 797 (1992).

\item{87.} X. N. Wang and M. Gyulassy, {\sl Phys. Rev. {\bf D44}},
3501 (1991).

\item{88.} T. Bir\'o, B. M\"uller, and X. N. Wang, {\sl Phys. Lett. {\bf
B283}}, 171 (1992).

\item{89.} A recent calculation$\scriptstyle{^90}$ of lepton-pair
production from the quark-gluon plasma assumes that quarks

\itemitem{} come into thermal equilibrium with $T$ = 900 MeV at
LHC energy.  Our arguments indicate that, although this value of $T$ may
adequately describe the quark spectrum, the phase space density of
quarks will be far below thermal, strongly reducing the lepton-pair yield.

\item{90.} J. Kapusta, L. McLerran, and D. K. Srivastava, {\sl Phys.
Lett. {\bf B283}}, 145 (1992).

\item{91.} K. Geiger, preprints UMSI 92/113, 92/174 and 92/175,
University of Minnesota (1992).

\item{92.} B. M\"uller and X. N. Wang, {\sl Phys. Rev. Lett. {\bf
68}}, 2437 (1992).

\item{93.} I. Kawrakow and J. Ranft, preprint UL-HEP-92-08, Leipzig
(1992); B. K\"ampfer and

\itemitem{} O. P. Pavlenko, {\sl Phys. Lett. {\bf B289}}, 127 (1992).

\item{94.} J. Cleymans and R. Philippe, {\sl Z. Phys. {\bf C22}}, 271
(1984); J. Cleymans and C. Vanderzande,

\itemitem{} {\sl Phys. Lett. {\bf 147B}}, 186 (1984).

\item{95.} J.P. Blaizot and J.Y. Ollitrault, in: Ref. 3, p. 393;
see also: H. von Gersdorff, L. McLerran,

\itemitem{} M. Kataja, and P. V. Ruuskanen, {\sl Phys Rev. {\bf D34}},
794 (1986); M. Kataja, P. V. Ruuskanen,

\itemitem{} L. McLerran, and H. von Gersdorff, {\sl Phys Rev. {\bf D34}},
794 (1986).

\item{96.} K. Kajantie and L. McLerran, {\sl Ann. Rev. Nucl. Sci. {\bf
37}}, 293 (1987).

\item{97.} {\sl QGP Signatures}, edited by V. Bernard, et al. (Editions
Fronti\`eres, Paris, 1990).

\item{98.} L. van Hove, {\sl Phys. Lett. {\bf 118B}}, 138 (1982); {\sl
Z. Phys. {\bf C21}}, 93 (1983).

\item{99.} H. von Gersdorff, {\sl Nucl. Phys. {\bf A461}}, 251c (1987).

\item{100.} T. Alexopoulos. et al., {\sl Phys. Rev. Lett. {\bf 64}},
991 (1990);  see also L. Gutay's lecture at this School.

\item{101.} P. L\'evai and B. M\"uller, {\sl Phys. Rev. Lett. {\bf 67}},
1519 (1991).

\item{102.} Such data are now becoming available [A. Goshaw, Duke
University, private communication.]

\item{103.} X. N. Wang and M. Gyulassy, {\sl Phys. Lett. {\bf B282}},
466 (1992).

\item{104.} See e.g. the lecture by W. Zajc at this School.

\item{105.} M. Lahanas, et al. [NA35 collaboration], {\sl Nucl. Phys.
{\bf A525}}, 327c (1991).

\item{106.} J. Rafelski, {\sl Phys. Rep. {\bf 88}}, 331 (1982).

\item{107.} J. Rafelski and B. M\"uller, {\sl Phys. Rev. Lett. {\bf
48}}, 1066 (1982); {\bf 56}, 2334E (1986).

\item{108.} P. Koch, B. M\"uller, and J. Rafelski, {\sl Phys. Rep.
{\bf 142}}, 167 (1986).

\item{109.} C. M. Ko, et al., {\sl Phys. Rev. Lett. {\bf 66}}, 2577
(1991).

\item{110.} T. Abbott, et al. [E-802 collaboration], {\sl Phys. Lett.
{\bf B197}}, 285 (1987); {\sl Phys. Rev. Lett. {\bf 64}}, 847

\itemitem{} (1990); S. E. Eiseman, et al. [E-810 collaboration]. {\sl Phys.
Lett. {\bf B248}}, 254 (1990); J. Bartke, et al. [NA35 collaboration],
{\sl Z. Phys. {\bf C48}}, 191 (1990); H. van Hecke, et al.
[HELIOS collaboration], {\sl Nucl. Phys. {\bf A525}}, 227c (1991);
S. Abatzis, et al. [WA85 collaboration], {\sl Phys. Lett. {\bf B270}},
123 (1991); E. Andersen, et al. [NA36 collaboration], submitted to
{\sl Phys. Lett. B}; For reviews see:  O. Hansen, {\sl Comments Nucl. Part.
Phys. {\bf 20}}, 1 (1991); G. Odyniec, preprint LBL-29996, published in
ref. 97.

\item{111.} R. Mattiello, H. Sorge, H. St\"ocker, and W. Greiner, {\sl
Phys. Rev. Lett. {\bf 63}}, 1459 (1989).

\item{112.} N. N. Nikolaev, {\sl Z. Phys. {\bf C44}}, 645 (1989).

\item{113.} E. Quercigh, Lecture at this School.

\item{114.} J. Rafelski, {\sl Phys. Lett. {\bf 262B}}, 333 (1991), and
lecture at this School.

\item{115.} J. Letessier, A. Tounsi, and J. Rafelski, preprint
PAR/LPTHE/92-23, Paris (1992).

\item{116.} N. J. Davison, H. G. Miller, R. M. Quick, and J. Cleymans,
{\sl Phys. Lett. {\bf 255B}}, 105 (1991).

\item{117.} J. Aichelin and K. Werner, preprint HD-TVP-91-15 and
HD-TVP-91-18, Heidelberg (1991).

\item{118.} H. Sorge, M. Berenguer, H. St\"ocker, and W. Greiner,
preprint LA-UR-92-1078; see also the lecture

\itemitem{} by M. Sorge at this School.

\item{119.} J. P. Guillaud, et al. [NA38 collaboration], {\sl Nucl.
Phys. {\bf A525}}, 449c (1991).

\item{120.} P. Koch and U. Heinz, {\sl Nucl. Phys. {\bf A525}}, 293c
(1991);  see also lecture by P. Koch at this School.

\item{121.} J. D. Bjorken, Fermilab pub. 82/59, Batavia (unpublished).

\item{122.} Y. Koike and T. Matsui, preprint, U. of MD PP \#91-223,
Maryland (1991).

\item{123.} T. A. DeGrand and C. E. DeTar, {\sl Phys. Rev. {\bf D34}},
2469 (1986); K. Kanaya and H. Satz, {\sl Phys.

\itemitem{} Rev. {\bf D34}}, 3193 (1986).

\item{124.}  F. Karsch, {\sl Z. Phys. {\bf C38}}, 147 (1988).

\item{125.} F. Karsch and H. W. Wyld, {\sl Phys. Lett. {\bf 213B}},
505 (1988).

\item{126.} D. Blaschke, {\sl Nucl. Phys. {\bf A525}}, 269c (1991).

\item{127.} F. Karsch and R. Petronzio, {\sl Phys. Lett. {\bf 212B}},
255 (1988); J. P. Blaizot and J. Y. Ollitraut,

\itemitem{} {\sl Phys. Lett. {\bf 199B}}, 499 (1987).

\item{128.} S. Hioki, T. Kanki, and O. Miyamura, {\sl Prog. Theor.
Phys. {\bf 84}}, 317 (1990);  {\bf 85}, 603 (1991).

\item{129.} S. Gavin, M. Gyulassy, and A. Jackson, {\sl Phys. Lett.
{\bf 207B}}, 257 (1988).

\item{130.} S. Gavin, R. Vogt, {\sl Nucl. Phys. {\bf B345}}, 104
(1990);  S. Gavin, preprint HU-TFT-91-33, Helsinki

\itemitem{} (1991).

\item{131.} R. Vogt, S. J. Brodsky, and P. Hoyer, {\sl Nucl. Phys.
{\bf B360}}, 67 (1991).

\item{132.} J. Blaizot and J. Y. Ollitraut, {\sl Phys. Lett. {\bf
217B}}, 392 (1989).

\item{133.} J. M. Moss, et al. [E-772 collaboration], {\sl Nucl. Phys.
{\bf A525}}, 285c (1991).

\item{134.} A. Guichard, et al. [NA38 collaboration], {\sl Nucl. Phys.
{\bf A525}}, 467c (1991).

\item{135.} F. Karsch and H. Satz, preprint CERN-TH-5900/90, {\sl Z.
Phys. C} (in press).

\item{136.} Possibly because their quark-gluon plasma scenario is
oversimplified.  Also the analysis of the

\itemitem{}  hadronic scenario is based on unrealistically high
energy densities in a  pure pion gas.

\item{137.} H. A. Weldon, {\sl Phys. Rev. Lett. {\bf 66}}, 283 (1991).

\item{138.} C. Gale and J. Kapusta, {\sl Phys. Rev. {\bf D43}}, 3080
(1991).

\item{139.} E. Braaten, R. D. Pisarski, and T. C. Yuan, {\sl Phys.
Rev. Lett. {\bf 64}}, 2242 (1990).

\item{140.} P. J. Siemens and S. A. Chin, {\sl Phys. Rev. Lett. {\bf
55}} 1266 (1985).

\item{141.} D. Seibert, {\sl Phys. Rev. Lett. {\bf 68}}, 1476 (1992).

\item{142.} M. Kataja, P. V. Ruuskanen, J. Letessier, and A. Tounsi,
preprint, University of Jyv\"askyl\"a and

\itemitem{} LPTHE, Univ. Paris VII (1991).

\item{143.} U. Heinz and K. S. Lee, {\sl Phys. Lett. {\bf 259B}}, 162
(1991).

\item{144.} H. W. Barz, G. Bertsch, B. L. Friman, H. Schulz and S.
Boggs, {\sl Phys. Lett. {\bf 265B}}, 219 (1991);

\itemitem{} C. Chanfray and P. Schuck, preprint, Grenoble 1991;
Z. Aouissat, G. Chanfray, P. Schuck, and G. Welke, preprint,
Grenoble 1991; C. M. Ko, P. L\'evai and W. J. Qin, preprint,
Texas A\&M University 1991.

\item{145.} D. Lissauer and E. V. Shuryak, {\sl Phys. Lett. {\bf
253B}}, 15 (1991); P. Z. Bi and J. Rafelski,

\itemitem{} {\sl Phys. Lett. {\bf 262B}}, 485 (1991).

\item{146.} R. Albrecht, et al. [WA80 collaboration], {\sl Z. Phys. {\bf
C51}}, 1 (1991).

\item{147.} J. Kapusta, P. Lichard, and D. Seibert, {\sl Phys. Rev.
{\bf D44}}, 2774 (1991).

\item{148.} P. V. Ruuskanen, {\sl Nucl. Phys. {\bf A525}}, 255c
(1991);  see also the lecture by P. V. Ruuskanen at this

\itemitem{} School.

\item{149.} S. A. Chin and A. K. Kerman, {\sl Phys. Rev. Lett. {\bf
43}}, 1292 (1979).

\item{150.} M. Tamada, {\sl Nuovo Cim. {\bf 41B}}, 245 (1977).

\item{151.} R. L. Jaffe, {\sl Phys. Rev. Lett. {\bf 38}}, 195 (1977);
{\bf 38}, 1617E (1977).

\item{152.} C. B. Dover, P. Koch and M. May, {\sl Phys. Rev. {\bf
C40}}, 115 (1989).

\item{153.} A. A. Anselm and M. G. Ryskin, {\sl Phys. Lett. {\bf
B266}}, 482 (1991); J. D. Bjorken, preprints

\itemitem{} SLAC-PUB-5545 and -5673,
Stanford (1991); J. P. Blaizot and A. Krzywicki, preprint LPTHE Orsay
92/11.

\item{154.} B. M\"uller and S. Schramm, {\sl Phys. Rev. {\bf C43}},
2791 (1991); B. M\"uller, {\sl Nucl. Phys. {\bf A544}}, 95c

\itemitem{} (1992).
\bigskip}

\end